%% file: thimble-review.tex
\documentclass[rmp,aps,twocolumn,eqsecnum,amsmath,notitlepage,nofootinbib]{revtex4-1}

\usepackage{graphics,setspace,epsfig,color}
\usepackage[letterpaper, dvips,width=7.5in,height=8.5in,includemp=false]{geometry}

\usepackage[vcentermath]{}
\usepackage{epsf}
\usepackage{amscd}
\usepackage{amsmath}
\usepackage{amsfonts}
\usepackage{braket}
\usepackage{bm}
\usepackage{xcolor}
\usepackage{multirow}
\usepackage{booktabs}
\usepackage{graphicx}
\usepackage{caption}
\usepackage{subcaption}
\usepackage{physics}

\newcommand{\partialslash}{\partial\!\!\!\slash}

\newcommand{\CC}{{\cal C}}
\newcommand{\CaD}{{\cal D}}
 \newcommand{\CM}{{\cal M}}
\newcommand{\CG}{{\cal G}}\newcommand{\CO}{{\cal O}}

\newcommand{\beq}{\begin{equation}}
\newcommand{\eeq}{\end{equation}}
\newcommand{\bea}{\begin{eqnarray}}
\newcommand{\eea}{\end{eqnarray}}

\newcommand{\nn}{\nonumber}
\newcommand{\C}{\mathbb{C}}
\newcommand{\R}{\mathbb{R}}
\newcommand{\av}[1]{\langle#1\rangle}
\def\beqs#1\eeqs{\beq\begin{split} #1 \end{split}\eeq}
\def\dd#1#2{\frac{d #1}{d #2}}

\let\Re\relax 
\DeclareMathOperator\Re{Re}
\let\Im\relax 
\DeclareMathOperator\Im{Im}

\newcommand{\eq}[1]{Eq.~\ref{#1}}
\newcommand{\fig}[1]{Fig.~\ref{#1}}

\newcommand{\sect}[1]{section~\ref{#1}}

\usepackage[colorlinks=true,backref=true, linktocpage=true,
citecolor=blue,urlcolor=blue,linkcolor=blue,pdfpagemode=UseOutlines]{hyperref}

\hypersetup{%
  bookmarksnumbered=true,
  pdftitle = {},
  pdfsubject = {},
  pdfauthor = {},
  pdfkeywords = {}
}

\graphicspath{{./figures2/}}

\begin{document}

\title{Complex paths around the sign problem}
\author{Andrei Alexandru}
\email{aalexan@gwu.edu }
\affiliation{Department of Physics\\The George Washington University\\
Washington, DC 20052}
\author{G\"ok\c ce Ba\c sar}
\email{basar@unc.edu}
\affiliation{Department of Physics \\
University of North Carolina\\Chapel Hill, NC 27599}
\author{Paulo F. Bedaque}
\email{bedaque@umd.edu}
\affiliation{Department of Physics \\
University of Maryland\\College Park, MD 20742}
\author{Neill C. Warrington}
\email{ncwarrin@umd.edu}
\affiliation{Institute for Nuclear Theory \\
University of Washington\\Seattle, WA 98195}

\begin{abstract}
The Monte Carlo evaluation of path integrals is one of a few general purpose methods to approach strongly coupled systems. It is used in all branches of Physics, from QCD/nuclear physics to the correlated electron systems. 
However, many systems of great importance (dense matter inside neutron stars, the repulsive Hubbard model away from half-filling, dynamical and non-equilibrium observables) are not amenable to the Monte Carlo method as it currently stands due to the so-called ``sign-problem". We review a new set of ideas recently developed to tackle the sign problem based on the complexification of field space and the Picard-Lefshetz theory accompanying it. The mathematical ideas underpinning this approach, as well as the algorithms so far developed, are described together with non-trivial examples where the method has already been proved successful. Directions of future work, including the burgeoning use of machine learning techniques, are delineated.
\end{abstract}

\maketitle

\tableofcontents
\section{ The sign problem }
 \input intro

\section{Cauchy theorem, homology classes and holomorphic
  flow}\label{sect:cauchy}

 \input cauchy

\section{Algorithms on or near  thimbles }\label{sect:algo}
\input algo

\section{Other manifolds and the algorithms that can find
  them }\label{sect:others}

\input glory

\section{Conclusion and prospects }\label{sect:conclusion}
\input conclusions

\begin{acknowledgments}
\input{acknowledgement}
\end{acknowledgments}

\input{appendix-jacobian}

\input{appendix-thimbles}

\bibliography{review}

\end{document}

%% file: intro.tex
\def\ncfg{{\tt n}}

Monte-Carlo methods have been used with great success to study problems ranging 
from classical systems of particles 
to studies of hadrons using lattice quantum chromodynamics.
The usual setup  is to   a formulate the problem -- classical or quantum -- in a way analogous  to
 a classical statistical system. Observables are then given by a multidimensional integrals involving a Boltzmann factor which is computed numerically by importance sampling.
There are, however, important systems of great interest that cannot yet be solved using 
standard Monte-Carlo methods. These are the systems where the statistical weights 
become either complex or whose signs oscillate.
Roughly speaking, we say that the system suffers from a {\em sign problem}
when the phase fluctuations increase as the size of the system is increased. These fluctuations lead to delicate cancellations that preclude a stochastic evaluation of the integral.
This occurs in the study of neutron matter found in neutron stars, the repulsive Hubbard model away from half-filling and all field theoretical/many-body observables in real time. Not surprisingly, solving the sign problem is of central importance in many fields of Physics and
a number of approaches have been proposed to either solve or alleviate this problem.
Some are more generic, and some are problem specific, but all approaches have fallen short of meaningfully addressing the Physics of the systems mentioned above. 

In this review we 
will focus on a novel set of related methods relying on the analytical properties of the 
configuration weights. The fundamental idea is to express the partition sum
as an integral over real degrees of freedom and complexify each variable. The partition
sum is originally an integral over the real manifold in this
enlarged configuration space, however, as we will discuss, we can deform the multidimensional integration contour---without 
changing the value of the partition function---to a manifold
that has better numerical properties. In particular, the phase fluctuations
are either eliminated or significantly reduced. We will describe the geometry of the complex field space, its critical points, and the algorithms used to both find suitable manifolds and to integrate over them. All of these steps will be exemplified in simple field theories, usually in lower number of dimensions, that contain, however, all the properties of the theories of the greater physical interest. 

 \subsection{Field Theory/Many-Body Physics as a path integral  }
 
 The expectation value of any observable $\mathcal{O}$ in field theory can be calculated by the path integral
 \footnote{Similar expressions are obtained for the partition function $Z={\rm tr} e^{-\beta H}$ of non-relativistic quantum systems by discretizing both space and time, then using the Trotter formula.}
\beq\label{eq:pathintegral}
\av{O} = \frac1Z \int D\phi\, e^{-S_E(\phi)} O(\phi), \qquad Z=  \int D\phi\, e^{-S_E(\phi)} .
\eeq Here, $\phi$ is the generic name of the fields in the theory and $S_E$ is the  euclidean (imaginary-time) action evaluated over a euclidean ``time" $\beta$, equal to the inverse temperature of the system \footnote{There is no assumption that the theory is relativistic. In fact, non-relativistic systems in the second quantized form are frequently studied within this formalism. }. The path integral in \eq{eq:pathintegral} is an integral over an infinite dimensional space. In order to evaluate it numerically (and to properly define it), we consider a discretized version where spacetime is replaced by a finite lattice.
After discretization, the path integral becomes a finite dimensional integral, albeit one over a very large number of dimensions, proportional to the number of spacetime points composing the lattice.
This is equivalent to a classical statistical mechanics problem in four spatial dimensions, where the state of the system is described by the field $\phi$ defined on the entire four dimensional grid, and the probability of each state is controlled
by the Boltzmann factor $\exp[-S_E(\phi)]$. Using Monte-Carlo methods, a set of $\ncfg$ configurations $\{\phi^{(1)},\ldots,\phi^{(\ncfg)}\}$
is generated with the probability distribution $\exp[-S_E(\phi)]/Z$. The observables and their errors are then estimated using
\beq\label{eq:estimator}
\av{O} = \frac1{\ncfg} \sum_a O(\phi^{(a)}) \,,\quad \epsilon_O = \sqrt{\frac1{\ncfg(\ncfg-1)} \sum_a [O(\phi^{(a)})-\av{O}]^2}\,.
\eeq Numerous algorithms have been developed to obtain configurations $\phi^{(a)}$ distributed according to $e^{-S_E[\phi]}$ in an efficient way. The cost of the sampling process increases with a moderate power of the spacetime volume $V$ (between $1$ and $2$), despite the fact that the Hilbert space dimension of the corresponding quantum system grows exponentially with the space volume. This is the great advantage of Monte Carlo methods over direct diagonalization procedures.

\subsection{Physical systems with sign problems}

    Many theories of interest in theoretical physics have sign problems in all currently known formulations. 
   In fact, systems that cannot be fully understood because a sign problem hinders the use of Monte Carlo simulations are pervasive in all subfields of Physics (and Chemistry). Among those some have become ``holy grails" in their respective field, problems whose solutions would have a revolutionary impact.

For instance, in nuclear physics, QCD at finite baryon density has a sign problem. This prevents the  understanding from first principles of both neutron stars and supernovae. Extensive work has been expended to evade this sign problem (see, for isntance, the following reviews and the references therein \cite{Aarts_2016,Philipsen:2007aa,forcr2010simulating,10.1143/PTP.110.615,KARSCH200014}. 
Quantum Monte Carlo (QMC) studies of nuclei using ``realistic nucleon-nucleon interactions" also suffer from the sign problem \cite{Carlson_2015,lhde2015nuclear,Wiringa_2000}. The ``constrained path algorithm" \cite{PhysRevLett.74.3652,PhysRevB.55.7464} is a widely-used approximate method to address these sign problems \footnote{The constrained path algorithm is a generalization of the ``fixed-node approximation", a similar approximate technique for avoiding the sign problem \cite{doi:10.1063/1.431514}}.  Lattice Field Theory studies of nuclei have similar behavior; sign problems appear in studies of nuclei with different proton and neutron numbers, and when repulsive forces become sufficiently large \cite{Lee_2004,Lee_2009,PhysRevLett.119.222505,Epelbaum_2014}. Furthermore, lattice and QMC studies of nuclear matter encountered in astrophysics suffer from the sign problem. This includes spin polarized neutron matter \cite{Fantoni_2001,PhysRevC.83.065801,Gandolfi_2014}\footnote{Unpolarized neutron matter, however, can be formulated free of the sign problem \cite{PhysRevLett.92.257002,Lee_2005}} and lattice EFT studies of nuclear matter beyond leading order \cite{lu2019ab}\footnote{''Wigner $SU(4)$" symmetric approximations to pionless EFT have no phase oscillations and have been profitably used \cite{PhysRev.51.106,Lee_2007,Lu_2019}.}.

Many cold atom systems, when formulated with lattice or QMC methods, exhibit sign problems as well. Both spin and mass imbalanced spin 1/2 fermions have a sign problem \cite{PhysRevLett.110.130404,PhysRevA.89.063609}. This sign-problem makes it prohibitively difficult, for example, to conclusively demonstrate the existence of a number of conjectured phases (like the  ``LOFF" phases) in more than $1+1$ dimensions.  Bosonic non-relativistic systems exhibit sign problems as well. This includes bosons under rotation \cite{berger2020thermodynamics} and coupled to spin-orbit interactions \cite{PhysRevA.101.033617}. For a review see \cite{Berger:2019odf}.


A wide variety of lattice-supersymmetric models suffer from a sign problem too (for a review see Ref.~\cite{Schaich:2018mmv}). In particular, first-principles tests of the gauge-gravity duality conjecture, even in the 
simplest case of reproduce supergravity black hole thermodynamics from D0-brane quantum mechanics, can only claim to be bona-fide controlled tests of the duality if the phase fluctuations are under control \cite{Hanada:2011fq,Berkowitz:2016jlq}.

Sign problems are found in condensed matter physics as well. 
A particularly well-known example is the Hubbard model away from half filling  \cite{10.2307/2414761,PhysRevB.41.9301,PhysRevB.40.506} thought to model essential characteristics of high $T_c$ superconductors. Path integral formulations of fullerene exhibit the sign problem as well \cite{ostmeyer2020semimetalmott}. Furthermore, some models of frustrated magnetism on triangular and \emph{kagom{\'e}} lattices, of interest for their conjectured spin-liquid ground states, exhibit the sign problem \cite{PhysRevB.50.3108,frustrate}. As a result there is uncertainty in the zero-temperature properties of these models. 

\subsection{Reweighting and the sign problem}

The standard workaround for sampling complex actions is to use {\em reweighting}. The idea is to split the integrand
into a positive part that is used for Monte-Carlo sampling, usually the absolute value of the integrand, and a fluctuating 
part that is included in observables. Using the absolute value as a sampling weight, we have the following identity
\beq
\av{O} = \frac{\av{O e^{-i \Im S_E(\phi)}}_0}{\av{e^{-i \Im S_E(\phi)}}_0},
\qquad
\av{O}_0=\int D\phi\, \frac{e^{-\Re S_E(\phi)}}{Z_0} O(\phi) 
\eeq
and
$
Z_0\equiv \int D\phi\, e^{-\Re S_E(\phi)}\,.
$
The idea, then, is to use the {\em phase quenched} action $\Re S_E$ to sample configurations, and take into account
the imaginary part of the action when computing observables.
From a numerical point of view, this procedure works when the phase fluctuations are mild and 
we can estimate the phase average, $\av{e^{-i \Im S_E(\phi)}}_0$, with enough accuracy; this means that the error
estimate for this average should be significantly smaller that its mean. Since the magnitude of the phase for each configuration
is one, to resolve the mean accurately we require a number of configurations $\ncfg \gg 1/\av{e^{-i \Im S_E(\phi)}}_0^2$.
When the average phase is very small, reweighting requires a very large number of samples and becomes impractical. For many systems at finite
density, the phase average goes to zero exponentially fast in the spatial volume/inverse temperature. This is because the phase average is the ratio of two partition functions:
\beq\label{eq:average-sign-free-energy}
\av{e^{-i \Im S_E(\phi)}}_0 = \frac{Z}{Z_0} = \frac{e^{-\beta f V}}{e^{-\beta f_0 V}} = e^{-\beta V \Delta f } \,,
\eeq
where $\Delta f = f-f_0 >0$ is the difference in the free energy density between the original system and the {\em phase quenched} system.
In this case, the numerical effort grows exponentially as we increase the volume and/or lower the temperature. This is what is usually defined to be 
the {\em sign problem}. An even worse problem arises when calculating real time correlation functions. In that case, we are interested in integrals of the form
\beq
\langle \mathcal{O} \rangle
=
\int D\phi\  e^{i S(\phi)} \mathcal{O} ,
\eeq where $S$ is the real time (Minkowski space) action of the system\footnote{In thermal equilibrium at non-zero temperature, real time correllators can be computed from path integrals defined in the closed-time contour in complex time \cite{Schwinger:1960qe,Keldysh:1964ud}. See section \ref{sec:realtime}.}. Since there is no damping of the magnitude of the integrand and the value of the field $\phi(t,\mathbf{x})$ (for any $t, \mathbf{x}$) grows, the average phase is strictly zero, even for small sized systems. A similar argument applies to observables, like parton distribution functions, defined on the light cone.

We should note that, the existence of a sign problem does not necessarily preclude numerical study. There are cases where the sign problem is mild enough that most relevant information about the system
in the region of interest can be extracted before the sign fluctuations become an obstacle. For example, when studying the
phase diagram of a simple heavy-dense quark model for QCD (see below), the endpoint of the first order phase transition can be studied via
reweighting for system sizes as large as $100^3$ even tough the model has a sign problem~\cite{Alford:2001ug}. We mention this study to point out that, from a practical point of view, methods which merely reduce sign fluctuations, without completely eliminating them, are also important.

\subsection{The absence of a general solution}

It is of theoretical, if not practical, interest to know if a generic solution to the sign problem exists. 
If one takes an exponentially vanishing average sign in the system size as the definition of the sign problem, then there are definitely models in which the sign problem can be solved. 
For instance, for many systems it is possible to rewrite the path integral using a different
set of states and obtain an expression free of phase fluctuations. This was accomplished, for example, for 
the two-component scalar theory using dual variables~\cite{Endres:2006xu,Gattringer:2012df}, and by reorganizing the summation over configurations for the heavy-dense system mentioned 
earlier~\cite{Alford:2001ug,Alexandru:2017dcw}. Similarly, there is a class of fermionic models that, when formulated in terms of fermion bags \cite{Ayyar:2017xmi,Chandrasekharan:2013rpa,Huffman:2019efk,Chandrasekharan:1999cm,Alford:2001ug,Huffman:2013mla,Huffman:2016nrx,Hann:2016xsw,Chandrasekharan:2012fk} have strictly positive Boltzmann weights even though other formulations have a severe sign problem. As it turns out, a solution of this kind is unlikely to work for all systems.

There is an often-cited, general argument implying that a generic solution to the sign problem, applicable to all systems, is extremely unlikely to exist.
It relies on the  NP$\not=$P conjecture from computational theory. NP decision problems are problems 
that can be solved on a {\em non-deterministic} Turing machine in a time that  increases only polynomially with the system size, 
whereas P problems are the ones that can be solved in polynomial time in a {\em deterministic} way. While no proof exists, it is widely believed that there are NP problems 
that are not P. In connection to this question, an important subset of NP problems are the NP-hard or NP-complete problems. 
If any of these NP-hard problems can be solved in polynomial time on a classical computer, then all NP problems can, 
invalidating the conjecture.
There are spin glass-like systems with a sign problem that can be mapped into NP-hard problems~\cite{Troyer:2004ge}.
Using the chain of arguments above, a generic solution to the sign problem that would solve this problem,
would imply $NP=P$, which is considered highly unlikely.

\subsection{A brief survey of methods to deal with sign problems}

As mentioned above, some of the most physically interesting models in particle, nuclear and condensed matter physics have sign problems. 
Given the interest in these problems, it is not surprising that a variety of approaches have been tried to either
solve or circumvent the sign problem. In this review we will focus on Lefschetz thimble inspired methods,
but we want to point out
some approaches attempted through the years 
  to understand the phase diagram of QCD and other relativistic theories.
  
A first set of methods uses simulations in the parameter region where the action is real; the result is then extrapolated in the region of interest.
One version of this idea is to rely on results from imaginary chemical potential. Monte Carlo simulations
can be used either directly to infer features of the phase diagram for real chemical potential or to compute 
observables and fit them using a polynomial ansatz or a Pad\'e approximations and then analytically continue 
these functions to real values of $\mu$~\cite{deForcrand:2003vyj,deForcrand:2002hgr,DElia:2002tig,DElia:2004ani,Cea:2014xva,Bonati:2015bha,Bellwied:2015rza,Borsanyi:2020fev}.
Another approach is to compute the derivatives of thermodynamic observables with respect to $\mu$
at $\mu=0$, then use Taylor expansions to extend these results to $\mu>0$~\cite{deForcrand:1999ih,Miyamura:2002mpl,Kaczmarek:2011zz,Endrodi:2011gv,Bonati:2018nut,Bazavov:2018mes}.
Yet another method is to use multiparameter reweighting by combining simulations
from different temperatures at $\mu=0$ to determine the phase transition line and critical point in QCD~\cite{Fodor:2001pe}.

Another class of methods attempt to alleviate the sign problem by a clever rewriting the path integral in terms of new variables. One possibility
is to reorganize the sum over the configurations in subsets that have either only positive sign contributions
to the partition function, thus solving the sign problem, or a much reduced sign problem~\cite{Rossi:1984cv,Karsch:1988zx,Chandrasekharan:1999cm,Alford:2001ug,Bloch:2013ara,Alexandru:2017dcw}.
Another direction is to reformulate the problem in terms of dual variables in which the sign problem
is absent~\cite{Endres:2006xu,Gattringer:2012df}. 
It turns that for QCD, the use of the canonical ensemble partition function (as opposed to the grand canonical ensemble) makes the
sign fluctuations milder and it can be used to investigate small enough
systems~\cite{Barbour:1988ax,Hasenfratz:1991ax,deForcrand:2006ec,Kratochvila:2005mk,Li:2010qf,Li:2011ee,Alexandru:2005ix,Alexandru:2010yb}. Finally, Fermi bags are enough to completely eliminate the sign problem in some low dimensional models \cite{Ayyar:2017xmi,Chandrasekharan:2013rpa,Huffman:2019efk,Chandrasekharan:1999cm,Alford:2001ug,Huffman:2013mla,Huffman:2016nrx,Hann:2016xsw,Chandrasekharan:2012fk}. These methods are very model dependent and require insight to be applied in each new class of models.

Recently a proposal
based on the density of states method was explored as a way to alleviate sign fluctuations~\cite{Fodor:2007vv,Langfeld:2014nta,Garron:2016noc,Garron:2017fta,Gattringer:2015lra}. 

Finally, there is a significant effort to simulate QCD at finite density using the complex Langevin approach~\cite{Parisi:1984cs,Klauder:1983nn}\footnote{See \cite{Berger:2019odf} for a recent review of complex Langevin approach.}, based on the idea of stochastic quantization \cite{Parisi:1980ys}.
This method shares with the thimble methods its starting point: the configuration space of $N$ real degrees 
of freedom is extended to a $N$ dimensional complex one. The important difference is that complex Langevin approach
sets up a stochastic process that moves freely in this enlarged space of $2N$ real degrees of freedom, 
whereas the methods we discuss in this review sample an $N$ dimensional manifold. Results show that, while instabilities are present in complex Langevin QCD simulations, for heavy quark masses
credible results can be obtained for temperatures above the deconfinement transition. In the hadronic
phase, the simulations become unstable and unreliable~\cite{Fodor:2015doa,Sexty:2013ica,Aarts:2013uxa,Seiler:2012wz,Aarts:2011ax,Aarts:2009uq,Aarts:2008rr,Aarts:2008wh}. 


%% file: cauchy.tex
 \subsection{Deformation of domain of integration: a multidimensional Cauchy theorem}
 \label{sec:deformation}
 
\begin{figure}[t]
\centering
  \includegraphics[width=.55\linewidth]{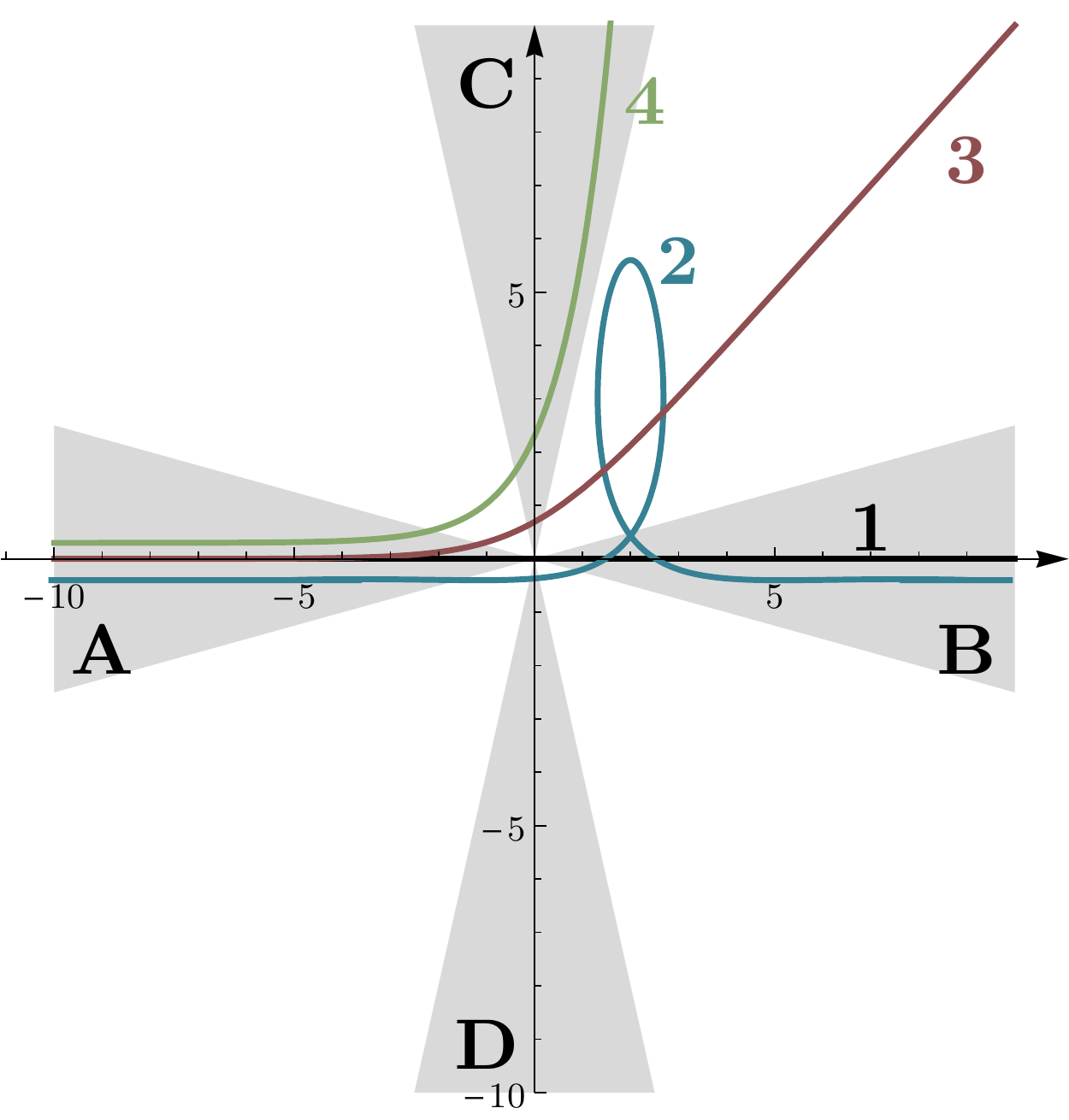}
  \caption{Several contours of integration for the integral in \eq{eq:z4}. Contour $1$ (the real line) and $2$ produce the same result. Contour $4$ a different result while the integral over contour $3$ is divergent. The gray areas show direction in the complex plane (``good" regions) where the integrand vanishes fast enough so the integral converges.}
  \label{fig:cauchy}
\end{figure}

 The well known Cauchy theorem for functions of one complex variable states that for an analytic function $f(z)$ the integral over a closed loop vanishes:
 \beq
 \oint_C f(z) = 0.
 \eeq This can be used to ``deform" the contour  of integration from, say, the real line, to a different contour on the complex plane, as long as the initial and final points of the contours coincide. In many applications the contour starts and/or ends at a point on the infinity and the issue becomes whether moving these ending points may cross a ``singularity of $f(z)$ at infinity". For instance, take the integral
 \beq\label{eq:z4}
 \int d\phi \ e^{-\phi^4}
 \eeq over different contours on the complex plane starting/ending at different points at the infinity.  Since there are no singularities at any finite values of $z$, Cauchy's theorem allows us to deform the contour of integration as long as no singularity ``at infinity" is crossed.
 The integral in \eq{eq:z4} is well-defined (it converges) if and only if the initial and final asymptotic directions of the contour are in the regions $A, \ldots, D$ shown in \fig{fig:cauchy}. The integral over two different contours whose ends lie on the same regions have, on account of Cauchy's theorem, the same value.  For instance, the real line, contour $1$, is equivalent to contour $2$ since both start in region $A$ and end in region $B$. The integral over contour $3$ is not even well-defined as it diverges, while the value for the integral over contour 4 is different from the value on contours 1 or 2. In fact,
 imagine starting from the real line and continuously deforming it towards contour $4$. At some point the integral will cease to be well-defined as its end point leaves region $B$ and the integral becomes divergent. As the end point enters region $C$ the integral  becomes finite again but acquires a different value than on the real line.

In fact, there are only three independent classes of contours (known as ``homology classes") on which the integral in \eq{eq:z4} may be evaluated: those that start in region $A$ and end in region $B$, $C$ or $D$, denoted $A\rightarrow B$, $A\rightarrow C$ and $A\rightarrow D$, respectively. Any other contour with different a asymptotic behavior, for instance $B\rightarrow C$, can be expressed as a linear combination of contours (with integer coefficients) belonging to one of these three classes. Cauchy's theorem guarantees that any contour that lies in one of these classes can be smoothly deformed to some other contour in the same class without changing the value of the integral. In contrast, as explained above, it cannot be deformed to a contour that lies in a different class.  In short, all possible domains over which the integral \eq{eq:z4} is well-defined can be classified as a linear combinations of three discrete classes of contours. Each class contains a continuous family of ``equivalent" contours that can be smoothly deformed to one another without changing the value of the integral. As we will see below, the reason that there are three classes is  that the function $\phi^4$ in the exponent is a quartic polynomial which in general has three saddle points.\footnote{Note our simple example actually a degenerate case where all three saddle points are at $\phi=0$, but it is easy to lift the degeneracy by adding a term $\epsilon \phi$ to the exponent.}
  
 

\begin{figure}
\includegraphics[width=0.85\linewidth]{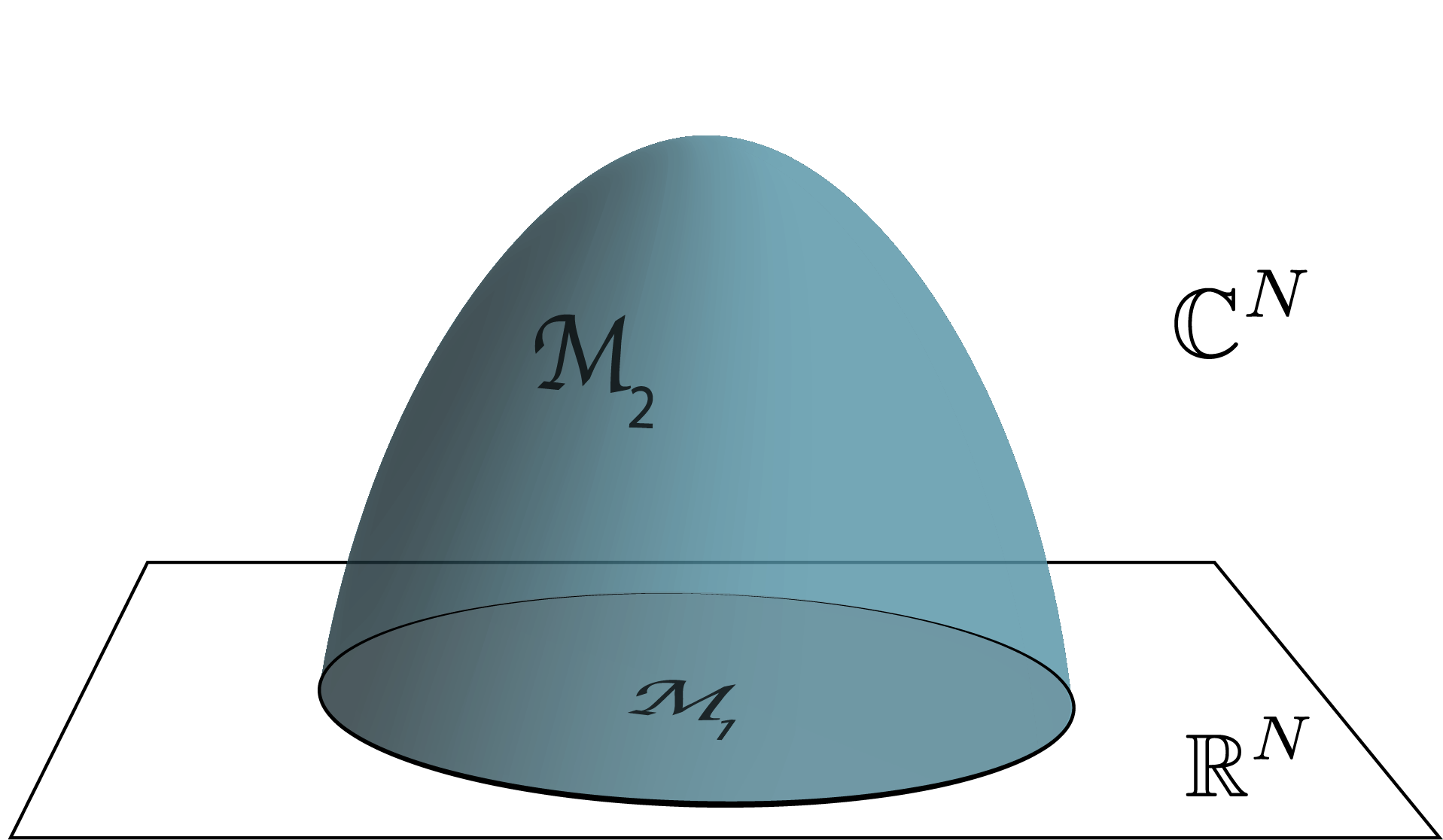}
\caption{Above is a schematic of a multi-dimensional deformation. The original domain of integration, ${\cal M}_1 \subset \mathbb{R}^N$, is deformed to ${\cal M}_2 \subset \mathbb{C}^N$. This deformation sweeps out a manifold $\mathcal{B} \subset \mathbb{C}^N$ whose boundary is $\partial \mathcal{B} = -{\cal M}_1 \cup {\cal M}_2$. }
\label{fig:deformation-process}
\end{figure}
 
All the observations above generalize to higher dimensions. Instead of integrals over one dimensional paths we will consider 
integrals over $N$-cycles, orientable manifolds with no boundary with real dimension $N$ immersed in the $2N$ dimensional space. The  integral over a cycle $\mathcal{M}$ is defined by
\bea\label{eq:int_def}
 \int_{\mathcal{M}}  \!\!\!\!\!\! f(\phi) d\phi_1 \wedge \cdots \wedge d\phi_N= \int_{\mathbb{M}}  \!\!\! \!\!\!\ f(\phi(\zeta))  \det J(\zeta)d\zeta_1 \dots d\zeta_N, \nn \\
 \eea
 where $\phi_i=\Phi_i(\zeta_1, \dots ,\zeta_N)$ is a parametrization of the $N$-dimensional manifold $\mathcal{M}$ by $N$ real coordinates $\zeta_1, \dots \zeta_N$, $\mathbb{M}$ is the region  of $\R^N$ used to parametrize 
$\mathcal{ M}$ 
 and $\det J(\zeta)=   \frac{\partial (\phi_1 \dots \phi_N)}{\partial(\zeta_1 \dots \zeta_N)} $ is the determinant of the Jacobian of the parametrization, which is in general a complex number. 
 $\phi$  stands for all $\phi_1, \dots, \phi_N$ (and similarly for $\zeta$).

 Assume that we have two such cycles $\mathcal{M}_1$ and $\mathcal{M}_2$ that can be smoothly deformed into one another. The space swept by
the deformation will be denoted with ${\cal B}$ and the two cycles form the boundary $\partial{\cal B}=\mathcal{M}_1-\mathcal{M}_2$ where the minus sign means oriented in opposite way  (see \fig{fig:deformation-process}).
 By Stokes' theorem we have  \footnote{Readers not familiar with the formalism of differential forms may take the right side of \eq{eq:int_def} as the definition of an integral over $N$-dimensional manifolds embedded in $\C^N$. We will  use this definition  extensively in this paper.}:
 \beq
 \int_{\partial\mathcal{B}} f(\phi) \  d\phi_1 \wedge \dots \wedge d\phi_N
 =
 \int_{\mathcal{B}} df(\phi) \wedge d\phi_1 \wedge \dots \wedge d\phi_N,
 \eeq 
 where $df = \frac{\partial f}{\partial \phi_i } d\phi_i + \frac{\partial f}{\partial \bar \phi_i } d\bar \phi_i $ ($\bar \phi$ is the complex conjugate of $z$). Since $f(z)$ is assumed to be holomorphic we have $ \frac{\partial f}{\partial \bar \phi_i} =0$. In the sum $\frac{\partial f}{\partial \phi_1}  d\phi_1 + \dots \frac{\partial f}{\partial \phi_N}  d\phi_N$ every term is proportional to  one of the terms in $d\phi_1 \wedge \dots \wedge d\phi_N$ so $df\wedge d\phi_1 \wedge \dots \wedge d\phi_N=0$ since $d\phi_i\wedge d\phi_i=0$. We arrive then at
 \beqs\label{eq:multi-cauchy}
   \int_{\partial\mathcal{B}} & 
   f(\phi)  \  d\phi_1 \wedge  \dots \wedge d\phi_N =\\
   &= \int_{\mathbb{M}_1-\mathbb{M}_2} 
  f(\Phi(\zeta))\ \det J(\zeta) \ d\zeta_1 \dots d\zeta_N 
  = 0 \, . 
\eeqs 
 which is the generalization of the Cauchy theorem we are interested in \footnote{We thank Scott Lawrence for a discussion on this point.}. This theorem can be used to deform the manifold of integration without altering  the value of the integral  just as we discussed above for the one dimensional case. In fact, our discussion of contour deformation readily generalizes to the multidimensional case.
 For manifolds approaching the infinity along certain ``directions" (in reality, $N$-dimensional planes)
the integral is convergent and well-defined (``good regions"); for others it is not.  Furthermore, it can be shown, assuming the integrand is well behaved in a sense discussed below, that the manifolds for which the integral converges are separated in discrete equivalence classes: those with the same asymptotic properties lead to the same integral. A continuous deformation of manifolds of integration from one equivalent class to another, that is, from one ``good region" to another necessarily goes through manifolds where the integral diverges. Such deformations are the analogue of deformations crossing a ``singularity at the infinity" in the one-dimensional case.  All this is in close analogy to the familiar one-dimensional case. A detailed discussion of the mathematical details can be found in \cite{pham}.

  \subsection{Holomorphic gradient flow}
  We will be interested in deforming integrals from $\R^N$ (the real cycle) to some other $N$-cycle  without altering the value of the integral but alleviating the sign problem in integrals of interest in field theory, which are typically of the form
  \beq
  \int_{\R^N} e^{-S(\phi)} \mathcal{O}(\phi)d\phi_i,
  \eeq where $S$ is the action of the theory and $\mathcal{O}$ some observable.
  One way of performing this deformation is with the help of the holomorphic flow. The holomorphic flow is defined for every action $S$ by the differential equations:
  \beq\label{eq:flow}
  \frac{d \phi_i}{dt} =\overline{ {\frac{\partial  S}{\partial \phi_i}}  }.
  \eeq  For every point $\phi$ in $\R^N$ and a fixed flow time $T$, the solution of \eq{eq:flow} with the initial condition $\phi(t=0) = \zeta$ defines a point $\tilde\phi=\mathcal{F}_T(\zeta)$ in $\C^N$. By flowing all points of $\R^N$ in this manner we obtain the flowed manifold $\mathcal{M}_T=\mathcal{F}_T(\R^N)$\footnote{Other flows to generate manifolds were proposed in \cite{Tanizaki:2017yow}.}.
  
The holomorphic flow has two important properties:
  \bea
\frac{d}{dt} S_R &=& 
\frac{1}{2} \left[ 
\frac{dS}{dt} +   \overline{ \frac{d S}{dt} }
\right]
=
\frac{\partial S}{\partial \phi_i}  \overline{\frac{\partial  S}{\partial \phi_i}}| \geq 0, \\
\frac{d}{dt} S_I &=&
\frac{1}{2i} \left[ 
\frac{dS}{dt} -   \overline{ \frac{d S}{dt} }
\right]
=
 \frac{1}{2i} \left[
\frac{\partial S}{\partial \phi_i}  \overline{ \frac{\partial  S}{\partial \phi_i} } - \overline{ \frac{\partial  S}{\partial \phi_i} }\frac{\partial  S}{\partial \phi_i}  
\right] = 0, \nn
  \eea that is, the imaginary part  $S_I$ is constant along the flow while the real part of the action $S_R$ increases monotonically (that is why \eq{eq:flow} is also called {\it upward} flow).\footnote{This can also be seen by noting that the holomorphic flow is the gradient flow of $S_R$ and the hamiltonian flow for the ``hamiltonian" $S_I$.}  The fact that $S_R$ increases along the flow means that the integrand    vanishes along asymptotic directions even faster in the flowed manifold $\mathcal{M}_T$ than in $\R^N$, leading to  the convergence of the integral  at all $T$. By the arguments exposed above, this means that $\mathcal{M}_T$ is equivalent to $\R^N$ for the purpose of computing the integral, that is, it is in the same homology class as $\R^N$, as in the one dimensional example explained in the beginning of this section.

   \subsection{Lefschetz Thimbles and Picard-Lefschetz theory}\label{sect:thimbles-and-pl}

Even though $\mathcal{M}_T$ is equivalent to $\R^N$, evaluating the path integral on $\mathcal{M}_T$ rather than $\R^N$ is computationally advantageous in controlling the sign problem. Before we explain why this is, we first introduce the necessary mathematical background (for a different perspective, see Appendix \ref{app:thimble-alt}). 

We begin by focusing on the stationary points of the flow, namely the critical points of the action $\phi^c$ where $\partial S/\partial\phi_i |_{\phi^c }=0$. The \emph{Lefschetz thimble} ${\cal T}$ attached to a critical point $\phi_c$ is defined as the set of initial conditions $\phi(0) \in \C^N$ for which the {\it downward} flow
  \beq\label{eq:reverse-flow}
  \frac{d \phi_i}{dt} = - \overline{ {\frac{\partial  S}{\partial \phi_i}}  }.
  \eeq  
asymptotically approaches the critical point. Similarly, the \emph{dual-thimble} $\mathcal{K}$ is the set of all point for which the upward flow asymptotes to $\phi_c$. For a constructive definition for $\mathcal{T}$, we begin by linearizing the flow around $\phi^c$:
  \beq\label{eq:linear_flow}
    \frac{d \phi_i}{dt} = \underbrace{
    \left.\overline{ \frac{ \partial^2 S}{\partial\phi_i\partial\phi_j}}\right|_{\phi=\phi^c} 
    }_{\overline H_{ij}}   (\bar\phi_j- \bar\phi_j^c)
  \eeq 
  whose solution can be written as
  \beq\label{eq:linear_sol}
  \phi(t)-\phi^c = \sum_{a=1}^N c_a  \rho^{(a)} e^{\lambda_a t},
  \eeq where $c_a$ are real and $\rho^{(a)}$ are the solutions to the modified eigenvector problem (``Takagi vectors"  \cite{192483})
  \beq\label{eq:takagi}
  H_{ij} \rho^{(a)}_j = \lambda_a \bar\rho^{(a)}_i .
  \eeq 
  The modified eigenvalues $\lambda_a$ can be chosen to be real, and then  the eigenvalues/eigenvectors come in pairs $(\lambda_a, \rho^{(a)}), (-\lambda_a, i\rho^{(a)})$. The set of $N$ vectors $\rho^{(a)}$ which define the directions around a critical point where the flow moves away from the critical point forms a basis (with real coefficients) for the tangent space of ${\cal T}$ at $\phi^c$. Likewise the set of $N$ vectors $i\rho^{(a)}$ which define the directions around a critical point where the flow moves towards the critical point forms a basis for the tangent space of ${\cal K}$ at $\phi^c$. These two tangent spaces together span the tangent space of $\C^N$. 
With this knowledge, in the infinitesimal neighborhood of the critical point, we can solve for the \emph{vanishing cycle} $v(\epsilon)$ as $S(\phi)-S(\phi^c)\approx z_i H_{ij} z_j=\epsilon$ which is an $N-1$ dimensional surface in the tangent space of ${\cal T}$. The thimble can be constructed by taking the vanishing cycle as the initial condition and flowing by upward flow: ${\cal T}=\cup_{0\leq T<\infty} {\cal F}_T (v(\epsilon) )$, when $\epsilon\rightarrow 0$. In other words we can build the thimble slice by slice by using the flow.  We can further use the fact that the flow defines a one-to-one map between the initial point and the flowed point and instead consider an infinitesimally small $N$ dimensional ball, ${\cal B}$, in the tangent plane. ${\cal B}$ is already a small portion of the thimble near $\phi^c$. If we take ${\cal B}$ as the initial condition, its image under upward flow with $T\rightarrow \infty$ is the thimble: ${\cal T}={\cal F}_{T\rightarrow\infty}(\cal B)$. This is the main idea behind the ``contraction algorithm" that is a method to simulate path integrals on a given thimble (see section \ref{sec:contraction}).

\begin{figure}[t]
  \centering
  \includegraphics[width=.7\linewidth]{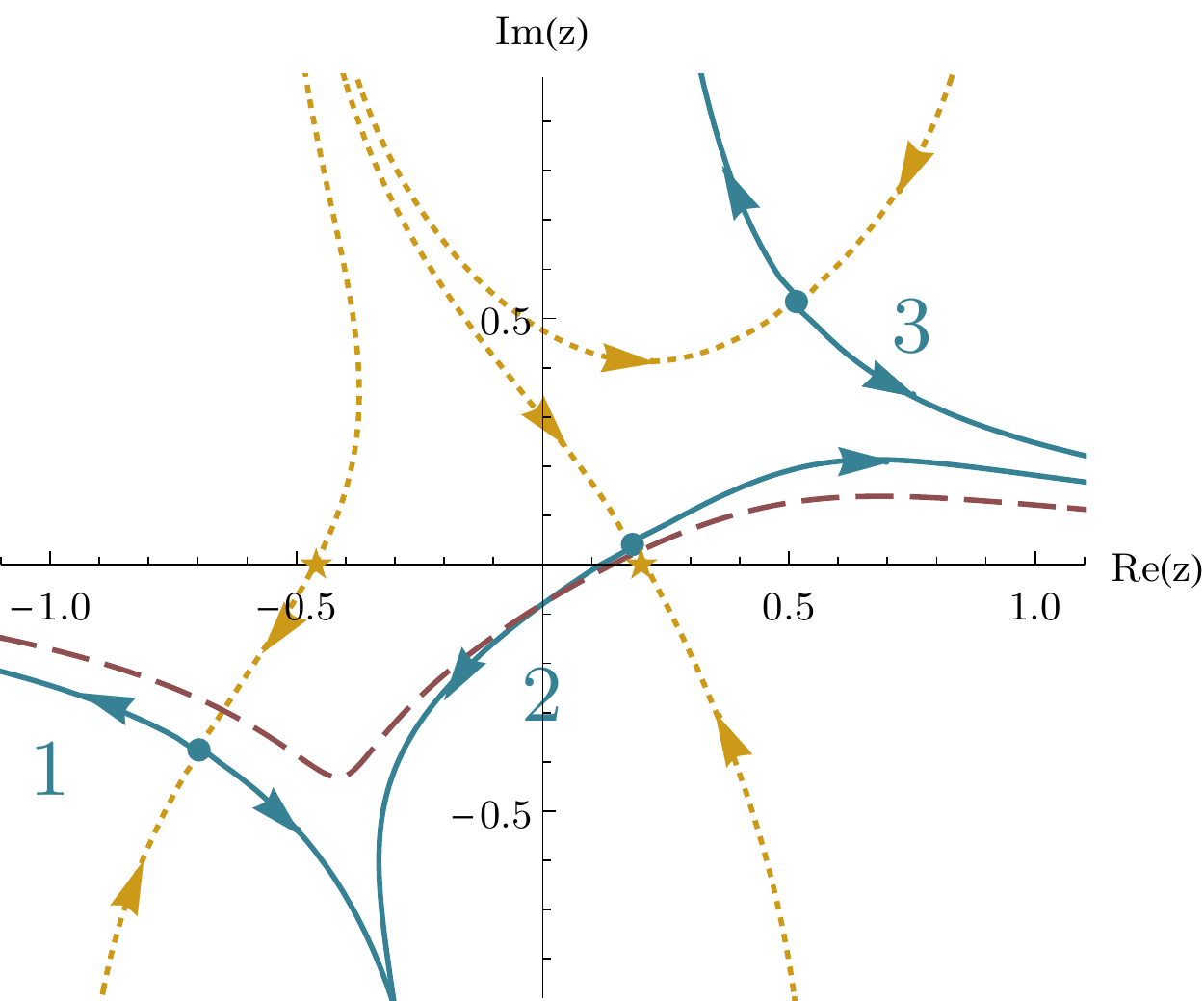}
  \caption{ Thimbles (blue), dual thimbles (yellow), critical points (blue dots), their pre-image under the flow (orange stars), and the flowed real line (dashed red) for $G=1.1 e^{i 0.05},  p=1,  \mu=0.3$ and $m=i 0.1$. The arrows indicate the direction of the upward flow.}
  \label{fig:1D-thimbles}
\end{figure}

For a concrete illustration of these ideas, consider \fig{fig:1D-thimbles}, where the action is taken to be $S(\phi)=\phi^2/G - \log\left[ (p^2+i \mu)^2 +(\phi+m)^2 \right]$. $S$ can be thought of as a toy model for the action of a fermionic model coupled to an auxiliary field $\phi$, after the fermions have been integrated out. Notice that $e^{-S}$ is a holomorphic function, even though $S$ is not; this is a feature common to theories with fermions. This theory has three critical points, attached to which are thimbles and dual-thimbles. Only thimbles $1$ and $2$
contribute to the integral. The real line, evolved by the holomorphic flow by a time $T=1.0$ is shown as the dashed red line. Notice how it approximates the union of the two contributing thimbles.

%
 
In section \ref{sec:deformation} we stated that the domain of integration of an integral of the form \eqref{eq:int_def} is naturally identified by a set of equivalence classes of $N$-cycles identified by their asymptotic behavior. The thimbles are representatives of these equivalence classes, each thimble representing a different class\footnote{In this review we only consider integration domains with no boundaries. The generalization of thimbles with boundaries are studied extensively in \cite{delabaere2002}. }. More concretely, let us assume that there are finitely many critical points, $\phi^c_\alpha$ indexed by $\alpha$ and $\Im S(\phi^c_\alpha)\neq \Im S(\phi^c_\beta)$ for $\alpha\neq \beta$ \footnote{These assumptions ensures that no two critical point is connected by flow since the flow conserves the imaginary part which is known as the \textit{Stokes phenomenon}. We will discuss Stokes phenomenon briefly in chapter \ref{sect:algo}.}. Attached to each critical point there exists a thimble, ${\cal T}_\alpha$, and a dual thimble ${\cal K}_\alpha$. As explained above, different thimbles do not intersect each other (they carry different values of $\Im S$) and ${\cal T}_\alpha$ intersects ${\cal K}_\beta$ if and only if $\alpha=\beta$.  In other words $\langle {\cal K}_\alpha, {\cal T}_\beta \rangle=\delta_{\alpha\beta}$ where $\langle,\rangle$ denotes the intersection number between two cycles. The intersection occurs at $\phi^c_\alpha$.  Since $\Re S$ is bounded from below on a thimble, the integral \eqref{eq:int_def} is guaranteed to be well-defined when evaluated on a thimble ${\cal T}_\alpha$. In fact, the set of all thimbles forms a complete basis for the space of equivalence classes of ``good domains" (i.e. the homology group) and any domain, say ${\cal M}$, over which \eqref{eq:int_def} is well-defined is equivalent to a unique linear combination of thimbles \cite{pham}:
\beq \label{eq:thimble-decomp}
{\cal M} \equiv \sum_\alpha n_\alpha({\cal M}) {\cal T}_\alpha, \quad n_\alpha({\cal M})=\langle {\cal K}_\alpha, {\cal M}\rangle\,.
\eeq
Here the integer coefficients $n_\alpha$  are given by the number of intersections between ${\cal M}$ and the dual thimble ${\cal K}_\alpha$. The sign depends on the relative orientations of ${\cal K}_\alpha$ and ${\cal M}$. Notice that some of the $n_\alpha $ may vanish; it is said then that those particular thimbles do not contribute to the integral. A simple example of this is shown in \fig{fig:1D-thimbles}.

Thimbles are the multi-dimensional generalization of the concept of ``steepest descent" or ``stationary phase" contour from the theory of complex functions of one variable. Naturally, they are useful in studying the semi-classical expansion of path integrals in field theory \cite{Cherman:2014ofa,Dunne:2015eaa} and their asymptotic analysis (see \cite{Aniceto:2018bis} for a recent review of the new developments related to ``resurgent transseries"). Also, thimbles have been used in attempts at defining ill defined path integrals by defining the relevant partition function as an integral over one or more thimbles instead of over $\R^N$ \cite{witten2010new,Witten:2010cx,Harlow:2011ny}. For our purposes, the relevant property of the thimbles is that the imaginary part of the action and, consequently, the phase of the integrand of the partition function, is constant on the thimble. Therefore instead of evaluating the path integral on $\R^N$ where the phase is a rapidly oscillating function, evaluating it in on the equivalent thimble decomposition where the phase is piecewise constant can provide significant practical advantage. This fact by itself, however, is not quite enough to solve the sign problem. As can be seen from \eq{eq:int_def}, the phase of the integrand depends also on the phase of the Jacobian (the ``residual phase"). The Jacobian will have a rapidly oscillating phase if the shape of the manifold of integration oscillates quickly along real and imaginary directions. For theories in the semi-classical regime this does not happen because the parts of the thimble with significant statistical weight are close to the critical point.  Experience shows that the residual phase in many strongly coupled models introduces a very mild sign problem (see below for many examples) \footnote{One can construct examples of extremely strongly coupled theories where the residual phase introduces a severe sign problem \cite{Lawrence:2020irw}}.

\begin{figure}[t]
\centering
  \includegraphics[width=0.9\linewidth]{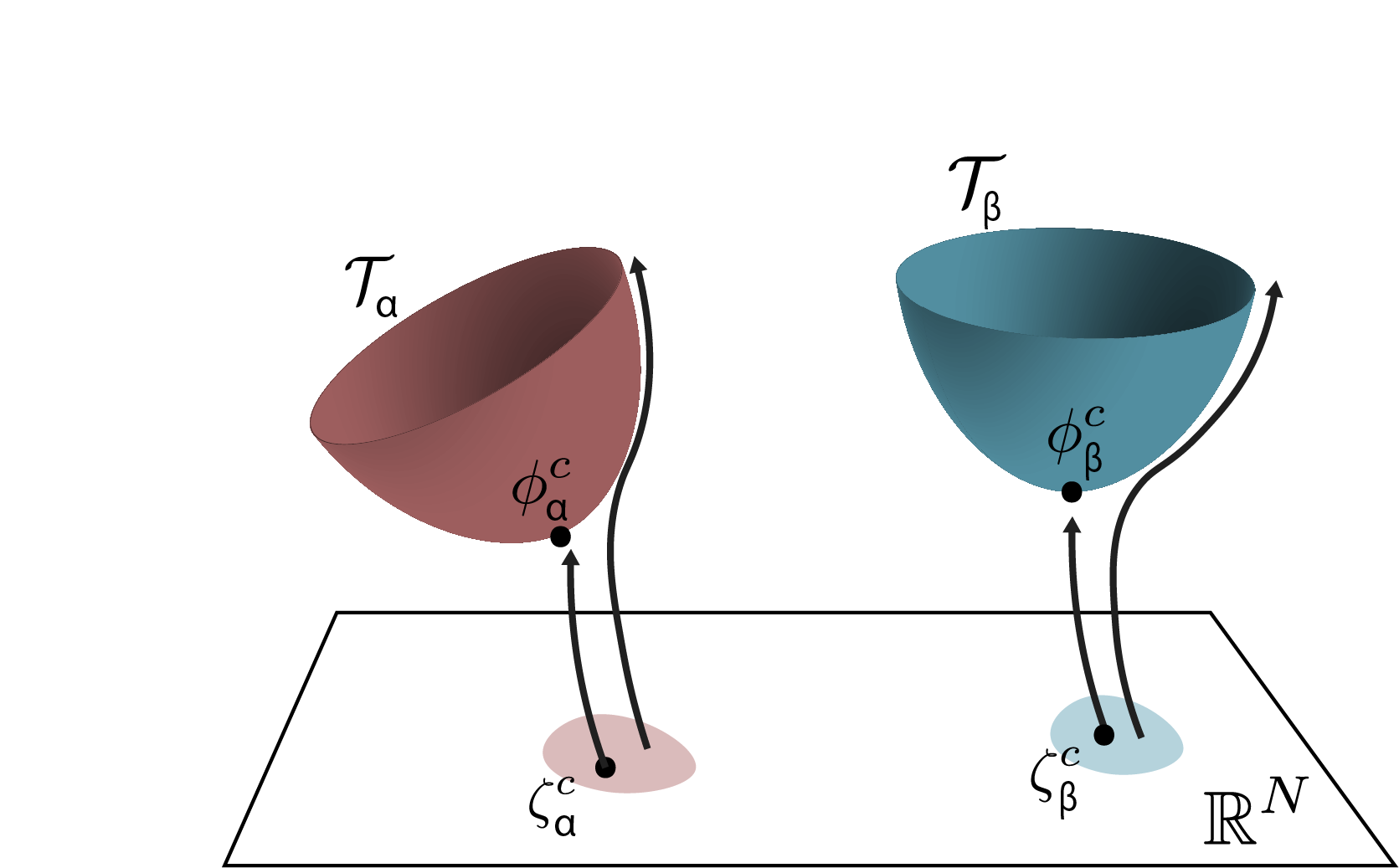}
  \caption{ The points $\zeta^c \in \R^N$ flow to the critical points $\phi^c\in\C^N$. The points in the neighborhood of each $\zeta^c$ approach the thimble but eventually veer off. In this figure we show two such thimbles ${\cal T}_\alpha$ and ${\cal T}_\beta$ for illustrative purposes. }
  \label{fig:approach-thimble}
\end{figure}

 An important question that naturally arises then is: which thimble, or combination of thimbles, is equivalent to the $\R^N$?  We can answer this question by considering the manifold $\mathcal{M}_T$ obtained by taking every point of $\R^N$  as an initial condition and flowing them by a ``time" $T$. Since the real part of the action grows monotonically with $T$ the integral remains convergent at all $T$ and, by the arguments above, the value of the integral remains the same. Since $\R^N$ and the dual thimble of any critical point are $N$ dimensional spaces they will generically intersect on isolated points, if they intersect at all. If we call each of those points $\zeta^c$ we have $\phi^c=\mathcal{F}_{T\rightarrow\infty}(\zeta^c)$. Starting from one of these intersection points $\zeta^c$ the flow leads to the critical point on a trajectory lying on the dual thimble $\mathcal{K}$ (see \fig{fig:1D-thimbles} and \fig{fig:approach-thimble}). The trajectory starting at points near $\zeta^c$ initially approaches the critical point but then veers along the unstable directions of the critical point slowly approaching the thimble (see \fig{fig:approach-thimble}). Points in $\R^N$ far from the intersection points take a more direct route towards infinity (or some other point where the action diverges). Therefore, all points in $\R^N$ flow, at large times, to points near a set of thimbles that, together, are equivalent to $\R^N$ (or to points where the action diverges).
Furthermore every thimble is counted as many times as there are intersection points between the corresponding dual thimble. Consequently the thimble decomposition of $\R^N$ can explicitly be obtained as the limit, 
  \beq\label{eq:RN-thimble-decomposition}
{\cal M}_{T\rightarrow \infty} = \sum_{\alpha} n_\alpha (\R^N) {\cal T}_\alpha, \quad \text{where}\quad{\cal M}_{T=0}=\R^N\,.
    \eeq 
 It is worth stressing that even though the thimble decomposition is obtained as the infinite flow time limit, the value of the integral remains unchanged during the deformation and ${\cal M}_T$ is equivalent to $\R^N$ for any finite value of $T$:
 \bea
 Z=\int_{\R^N} d\phi\ e^{-S(\phi)} &=&\int_{{\cal M}_T} d\phi\ e^{-S(\phi)} 
 \nn\\
 &=& \sum_{\alpha} n_\alpha(\R^N) \int_{\mathcal{T}_\alpha} d\phi\ e^{-S(\phi)}
 \eea
 
It should be noted that in theories where more than one thimble contribute to the partition function, there is a possibility that the contributions from different thimbles come with  phases $\exp(-i \Im S_{\text{eff}})$ (constant over each separate thimble) which induces a sign problem. This kind of sign problem is not helped by integrating over thimbles. However, in order for the contributions from different thimbles to (nearly) cancel an (approximate) symmetry is required relating the contribution of different thimbles. Monte Carlo methods can be adapted to situations like that by sampling points related by the symmetry at the same time. 
  
  In field theories, where the dimensionality of the integral is large, it is extremely difficult to find the thimbles -- it is in fact equivalent to classifying all {\it complex} solutions of the equations of motion-- and even harder to find their intersection numbers $n_\alpha$.  
  The discussion of the previous paragraph will be useful however, in establishing an algorithm to solve this  problem numerically and ``on-the-fly" during a Monte Carlo run. It also clarifies the fact that there is nothing special about thimbles as opposed to other manifolds obtained from flowing $\R^N$ by a finite time $T$. These other manifolds do not improve the sign problem as much as the thimbles do but still give the correct result for the integral  and can be advantageous for numerical/algorithmic reasons. 

%% file: algo.tex
%
%
%
%

\subsection{Single Thimble Methods}
  
Early simulations using complex manifolds focused on sampling the path integral contribution
from the ``main'' thimble, the thimble associated to the critical point with the smallest value of $S_R(\phi_c)$. This was based on the hope that in the relevant continuum/thermodynamic
limits the path integral would be dominated by the contribution of a single thimble or that a regularization
can be defined for relevant QFTs in term of a single thimble path integral~\cite{Cristoforetti:2012su,DiRenzo:2020vou}. Although there is no evidence that this conjecture is valid, algorithms to sample a single thimble are obvious stepping stones towards multi-thimble integration.
We will discuss in this section the algorithms proposed to sample the integral along a single thimble:
the contraction algorithm, a Metropolis based algorithm~\cite{Alexandru:2015xva}, a Hybrid Monte-Carlo
algorithm~\cite{Fujii:2013sra}, and the Langevin algorithm~\cite{Cristoforetti:2012su}.

As discussed earlier finding the thimble decomposition for the path integral is a very hard
problem which was only attempted for quantum mechanical systems~\cite{Fujii:2015bua}.
However, in many cases it is feasible to find the ``main'' thimble even for realistic systems 
using the symmetry of the problem. The problem of finding the critical point is usually reduced
to a ``gap'' equation to be solved analytically or numerically. 
For the algorithms discussed in this section, we assume that we have identified this critical point
and we want to sample configurations on the corresponding thimble.

Another important challenge facing any algorithm for the Monte Carlo evaluation of integrals over thimbles is 
to restrict sampling to the thimble manifold. For most systems there is no known method 
that can identify points
on the thimble based on the local behavior of the action.
Rather, a point has to be transported though the reverse flow (\eq{eq:reverse-flow}) to decide whether it approaches the critical point or not.
The thimble attached to this
point can then be constructed by integrating the upward flow equations starting in the neighborhood
of the critical point. As the thimble on the neighborhood of the critical point is approximated by the tangent space spanned by the Takagi vectors with positive eigenvalues (in~Eq.~\ref{eq:takagi}) we can take points on the tangent plane (close enough to the critical point) as the initial conditions of the holomorphic flow \eq{eq:flow} to find points lying on the thimble.  This ``backward-and-forward" procedure then allows us to find points on the thimble nearby other points on the thimble, as required by Monte Carlo procedures, at the expense of integrating the flow equations.
This process provides a map between the $N$ dimensional neighborhood of the critical point to the thimble
attached to it. It is an essential ingredient for all single thimble algorithms discussed here.
For a given parametrization of the tangent space near the critical point~$\phi^c$: 
\beq
\phi_n = \phi^c+\sum_{a=1}^N\zeta_a \rho^{(a)}\,,\quad \zeta_a\in \R \,,
\eeq
integrating the upward flow for a time $T$ produces a map $\phi_n\to\phi_f=\mathcal{F}_T(\phi_n)$.
Here $\phi_n$ is a point {\em near} $\phi^c$ and $\phi_f$ is moved {\em far} by the flow. For large enough
$T$, this will map a small neighborhood of the critical point into a manifold very close to the thimble and the larger the value of $T$, the closer the manifold generated by the $\phi_n\rightarrow\phi_f$ mapping is to the thimble.
 As a practical method of determining an appropriate value for $T$,
simulations can be carried out for increasing values of $T$ until the results converge.

Having chosen an appropriate $T$, we have now the means to parametrize
the thimble using the tangent plane close to the critical point. We can then approximate the integral over the
thimble as
\beq
\int_{\cal T} d\phi_f\, e^{-S(\phi_f)} \approx \int_{U} d\phi_n\, \det J(\phi_n)\, e^{-S(\phi_f(\phi_n))} \,,
\eeq
where $J_{ij}=\partial (\phi_f)_i/\partial (\phi_n)_j$ is the Jacobian of the map and 
$U$ is the region around $\phi^c$ in the tangent plane that is mapped to the manifold approximating the
region of the thimble that dominates the integral.
For the special case where the tangent plane is in the same homology class as the thimble, the region $U$ can
be extended to the entire tangent plane and the relation above becomes exact for all flow times $T$. 
For the case when the tangent plane is not in the same homology class, the relation only becomes exact in the 
limit of large $T$. 
In practice the region $U$ is generated implicitly in the simulations: we start in the neighborhood of the
critical point and the proposed updates move smoothly, or in small discrete steps, through the configuration
space and the potential barriers force the simulation to stay in the relevant region. To fix terminology
we will refer to the region $U$ in the tangent plane as the {\em parametrization} manifold and the image 
under the map ${\cal F}_T(U)$ as the {\em integration} manifold.

The goal of the algorithms presented here is to sample the integration manifold according to the Boltzmann 
factor $\exp(-S)$. Since the action and the integration measure are complex, we need to use a modified
Boltzmann factor for sampling. The probability density we will sample corresponds to 
\beq
P_0(\phi_f) |d \phi_f| = \frac1{Z_0} e^{-\Re S(\phi_f)} |d\phi_f|\,,\quad Z_0\equiv \int_{\cal T} |d\phi_f| e^{-\Re S(\phi_f)} \,.
\eeq
The final result for observables will have to include the phase
\beq
\av{{\cal O}} = \frac{ \av{{\cal O} e^{i\varphi}}_0}{\av{e^{i\varphi}}_0}\,,\quad e^{i\varphi}\equiv e^{-i \Im S(\phi_f)} \frac{d\phi_f}{|d\phi_f|}
\eeq
Since we are sampling the configurations from a single thimble, or from a manifold that is very close to it, the imaginary
part of the action is constant (or nearly so.) The only fluctuation come from the {\em residual phase} associated with
the phase of the measure $d\phi_f$. If we view this as an integral over the parametrization manifold, then the
probability measure is
\beq
P_0(\phi_n) = \frac1{Z_0} e^{-\Re S_\text{eff}(\phi_n)}\,,\quad S_\text{eff} = S(\phi_f(\phi_n)) - \ln\det J(\phi_n) \,.
\eeq
The complex phase in this case is $\exp(-i \Im S_\text{eff})$ and the fluctuations of this phase are dominated
by the Jacobian phase which correspond to the residual phase. Note that to compute the effective action
for a point $\phi_n$ in the parametrization space, we have to integrate the upward flow differential equation 
with initial condition $\phi_n$ for a time $T$ to get $\phi_f$. Then $S(\phi_f)$ is the action contribution.
The other contribution comes from the Jacobian. As explained in Appendix~\ref{app:jacobian} the Jacobian matrix can
be computed by integrating the matrix differential equation
\beq
\dd Jt = \overline{H(\phi(t)) J(t)}\,,
\label{eq:vectorflow}
\eeq
where $H(\phi(t))$ is the Hessian of $S$ along the flow and the initial condition $J(0)$ is a matrix whose columns 
form an orthonormal basis in the tangent to the parametrization space at $\phi_n$. This equation flows a basis in the tangent space
at $\phi_n$ to a basis in the tangent space at $\phi_f$.
Since our parametrization space is a hyperplane the basis for the tangent space at $\phi_n$ can be chosen to be the same
at all points in $U$, for example the positive Takagi vectors or any other basis spanning this tangent space. 

This equation can also be used to map a single infinitesimal displacement represented by a vector $v_n$ in the tangent space
at $\phi_n$ to a displacement represented by a vector $v_f$ in the tangent space on the thimble at $\phi_f$. 
In the equation above $J(t)$ is then replaced with $v(t)$ the column vector representing the displacement. The initial
condition is $v(0)=v_n$ and the final result, $v(T)=v_f$, is a vector in the tangent space at $\phi_f$. 
Because of this we will sometime call this equation the {\em vector flow}. 

\subsection*{Contraction Algorithm}
\label{sec:contraction}

Several sampling algorithms are based on the mapping between the tangent plane and the (approximate) thimble. The most straightforward is the 
{\em contraction algorithm}~\cite{Alexandru:2015xva,Alexandru:2016san}, which is generates configurations
in the parametrization manifold based on the probability $P_0$ using the Metropolis method~\cite{Metropolis:1953} 
based on the effective action $\Re S_\text{eff}$. The basic process is detailed below.

\begin{enumerate}

\item After a critical $\phi^c$ point is identified, the tangent space of its thimble is compute by solving~\eq{eq:takagi} 
and finding the $\rho^{(a)}$ corresponding to positive $\lambda^{(a)}$.

\item Start with a point $\phi_n=\phi^c + \sum_{a=1}^N \zeta_a \rho^{(a)}$ on the tangent 
space.~
Evolve $\phi_n $  by the holomorphic flow by a time $T$ to find $\phi_f$, 
compute the Jacobian $J(\phi_n)$ by integrating the flow equation for the basis, and
then compute the action $S_\text{eff}(\phi_n)$.

\item Propose new coordinates $\zeta'=\zeta+\delta\zeta$, where $\delta\zeta$ is a random vector chosen with 
symmetric probability function, that is $P(\delta\zeta)=P(-\delta\zeta)$. Evolve $\phi_n' = \sum_a \zeta'_a \rho^{(a)}$ 
by the holomorphic flow by a time $T$ to find $\phi_f'$, compute $J(\phi_n')$, and $S_\text{eff}(\phi_n')$.
 
\item Accept/reject $\zeta'$ with probability $\min\{ 1, e^{-S_\text{eff}'+S_\text{eff}}\}$.
\item Repeat from step 3 until a sufficient ensemble of configurations is generated.
\end{enumerate}
  
  
To make the updating effective, we have to account for the fact that the map ${\cal F}_T$ is highly anisotropic.
If we consider the flow close to the critical point, we see that displacements in the direction of the Takagi
vector $\rho^{(a)}$ are mapped into vectors that have their magnitude increased by $\exp( \lambda^{(a)} T)$.
Even small differences in the eigenvalues $\lambda^{(a)}$ lead to large differences as $T$ increases.
If the parametrization space proposals $\delta\zeta$ are isotropic then the update process becomes inefficient.
Ideally we would like to generate proposals that are isotropic on the integrations manifold, but since the
map changes from point to point, this requires care to ensure that the detailed balance is preserved. As it
turns out this is possible but we will discuss this point later. An easy fix for this problem is to adjust
the size of displacement for proposal based on the flow around the critical point. The proposal is then
$\delta\zeta_a = \exp(-\lambda^{(a)}T) \delta$ with $\delta$ a random variable chosen with uniform probability
in the interval $[-\Delta,\Delta]$. The step size $\Delta$ is tuned to get reasonable acceptance rates.
If the distortions induced by the map ${\cal F}_T$ vary little from $\phi^c$ to the points sampled by the process, 
then this algorithm is effective.
  
By far the most computationally expensive part of the contraction algorithm---and most other thimble algorithms---is 
the computation of the Jacobian (even for most bosonic systems the cost scales with $N^3$ and $N$ is proportional 
to the spacetime volume.) Methods to deal with this problem are discussed in \sect{sec:algo-jacobians}.

Another Metropolis based method was proposed to sample single thimble configurations~\cite{Mukherjee:2013aga} and
was tested for a single plaquette $U(1)$ problem. In this proposal the Jacobian is not included in the sampling 
and it is to be included via reweighting in the observable measurement. This reweighting will fail for most systems
that have more than a few degrees of freedom since for this systems the Jacobian fluctuates over many orders of magnitude.


\subsection*{HMC on thimbles}

A more sophisticated algorithm based on Hybrid Monte Carlo~\cite{Duane:1987de} was proposed and tested for
the $\phi^4$ model~\cite{Fujii:2015bua}. In principle, a straightforward extension of HMC could be applied to
the action $\Re S_\text{eff}$ on the parametrization manifold. The problem with such an approach is that it would
require the calculation of the derivatives of $\det J$, or some related quantity, which is quite cumbersome.
Of course this could be side-stepped by neglecting the Jacobian in the sampling~\cite{Ulybyshev:2019fte}, but this
requires reweighting it in the observables which fails for large systems. The proposal is then to use HMC as defined
by the Hamiltonian in the larger $\C^N$ space, where the motion is confined to be on the thimble via forces of
constraint~\cite{Fujii:2015bua}. This has the advantage that the Jacobian is accounted for implicitly, but the
algorithm requires solving implicit equations to project back to the thimble. For the cases where the thimble
is relatively flat/smooth, these equations can be solved robustly via iteration, as is the case with the $\phi^4$
system in the parameter range investigated.

The basic idea is to integrate the equations of motion generated by the Hamiltonian
\beq
{\cal H}(\pi, \phi_f) = \frac12 \pi^\dagger\pi + \Re S(\phi_f) \,,
\eeq
subject to the constraint that $\phi_f \in {\cal T}$.
Forces of constraint perpendicular to the thimble keep the system confined
on its surface. The momentum $\pi$ is in the tangent space at $\phi_f$, so it is a {\em real}
linear combination of columns of $J(\phi_n)$. The perpendicular force has to be a {\em real} linear combination of
the columns of $iJ(\phi_n)$, since this forms a basis in the space perpendicular (according to the scalar product $\braket{v}{w} \equiv \Re v^\dagger w$) to the thimble.

For a practical implementation we need to provide an integrator for these equations of motion
for finite time steps. A symplectic integrator for this problem is provided by the following method
\beqs
&\pi_{1/2}=\pi-\partial_{\phi_f} \Re S(\phi_f) \frac{\Delta t}2 + i J(\phi_n) \lambda \,,\\
&\phi_f' = \phi_f + \pi_{1/2} \Delta t \,,\\
&\pi' = \pi_{1/2} -\partial_{\phi_f} \Re S(\phi_f') \frac{\Delta t}2 + i J(\phi_n') \lambda' \,.
\eeqs
The map $(\pi,\phi_f)\to(\pi',\phi_f')$ is symplectic and time reversible, thus satisfying the requirements
for HMC. Note that this map requires the determination of $\lambda$ and $\lambda'$, two sets of $N$ real
numbers which encode the effect of the constraint forces acting perpendicular on the thimble. 
$\lambda$ is determined by the requirement that $\phi'_f \in {\cal T}$ and $\lambda'$ by requiring that
$\pi'$ is in the tangent space at $\phi_f'$. For small enough $\Delta t$, these requirements
lead to unique ``small"  solutions (which vanish in the $\Delta t\rightarrow 0$ limit) for $\lambda$s. A solution for $\lambda'$ can be computed in a straightforward way, via the projection
method we discuss below. Computing $\lambda$ is more difficult and the current proposal is to 
use an iterative method~\cite{Fujii:2015bua}. This iteration is guaranteed to converge for small enough $\Delta t$, but
for a fixed size $\Delta t$ no guarantees can be made even for the existence of a solution.
  
With these ingredients in hand, the basic steps of HMC are the following:
\begin{enumerate}
\item At the beginning of each ``trajectory'' an isotropic gaussian momentum $\pi$ is generated in
the tangent space at $\phi_f$, $P(\pi)\propto \exp(-\pi^\dagger\pi/2)$.
\item The equations of motion are integrated by repeatedly iterating the integrators steps above for
a $t/\Delta t$ times, where $t$ is the ``trajectory" length.
\item At the end of trajectory the proposed $(\pi',\phi_f')$ are accepted with a probability
determined by the change in Hamiltonian $P_\text{acc}=\min\{1, \exp(-{\cal H}+{\cal H'})\}$.
\end{enumerate}
 
One important ingredient for this and other algorithms we will discuss later, is the projection
to the tangent space at $\phi_f$. If we have the Jacobian matrix in hand $J(\phi_n)$, its columns form a real basis of
the tangent space and the columns of $iJ(\phi_n)$ form a basis for the orthogonal space. Every vector $v\in\C^N$
can then be decomposed in its parallel, $P_\parallel(\phi_f) v$, and perpendicular component, $P_\perp(\phi_f) v$,
using standard algebra.
This step is required to find $\lambda'$ in the symplectic integrator. It can also be used to find the
starting momentum, at the beginning of the trajectory: we generate a random vector in $\C^N$ with
probability $P(\tilde\pi)\propto \exp(-\tilde\pi^\dagger \tilde\pi/2)$ and then project it to the
tangent plane $\pi = P_\parallel(\phi_f) \tilde\pi$.

The projection discussed above can be readily implemented when we have the Jacobian matrix $J(\phi_n)$.
However, calculating this matrix is an expensive operation that is likely to become a bottle-neck for
simulations of systems with large number of degrees of freedom. One solution for this problem is
the following~\cite{Alexandru:2017lqr}:
we use the map $v\to J(\phi_n)v$, that maps the tangent space at $\phi_n$ on the parametrization manifold
to the tangent space at $\phi_f$ on the thimble. 
This calculation can be implemented efficiently, by
solving the vector flow equation, Eq.~\ref{eq:vectorflow}, for a single vector~$v$. We extend
this to arbitrary vectors that are not included the tangent space. For a generic vector $v$ we split it into 
$v_1=P^{(0)}_\parallel v$ and $i v_2 = P^{(0)}_\perp v$. Here $P^{(0)}_\parallel$
is the projection on the tangent space of the parametrization manifold, the space spanned by the Takagi vectors,
and $P^{(0)}_\perp$ its orthogonal complement. Both $v_1$ and $v_2$ belong to the tangent space at $\phi_n$,
so $J(\phi_n) v_{1,2}$ can be computed using the vector flow equations. This defines then a map from any vector
$v$ to $J(\phi_n)v=J(\phi_n)v_1 + i J(\phi_n) v_2$, which requires two integrations of the vector flow. Using this map
we can then compute $J^{-1}(\phi_n) v$ using an iterative method, such as BiCGstab. It is then straightfoward to
prove that $P_\parallel(\phi_f) v = J(\phi_n) P_\parallel^{(0)} J(\phi_n)^{-1} v$.

\subsection*{Langevin on thimbles}

The Langevin algorithm was proposed as possible sampling method for single thimble 
manifolds~\cite{Cristoforetti:2012su,Cristoforetti:2012uv,Cristoforetti:2013wha}.
The idea is to sample the thimble manifold ${\cal T}$ with probability density proportional
to $\exp(-\Re S)$ with respect to the Riemann measure induced by embedding ${\cal T}$ in $\C^N$.
The residual phase of the measure is taken into account via reweighting. The imaginary
part of the action is constant over the thimble and will not contribute to averages. 
 
The Langevin process simulates the evolution of the system via a drift term due to the
action and a brownian motion term. The discretized version of the process is given by
the following updates:
\beq
\phi_f' = \phi_f - \partial_{\phi_f} \Re S(\phi_f) \Delta t + \eta \sqrt{2\Delta t} \,
\eeq
where the vector $\eta$ is a random $N$ dimensional vector, in the tangent space at the thimble at $\phi_f$.

Two details are important here: how the vector $\eta$ is chosen and how the new configuration $\phi_f'$
is projected back to the thimble. The proposal is to chose $\eta$ isotropically at $\phi_f$ by generating
a gaussian $\tilde\eta$ unconstrained in $\C^N$ and then projecting it to the tangent space at $\phi_f$ using a
procedure similar to the projection outlined in the section above, $\eta=P_\parallel(\phi_f)\tilde\eta$. 
This ensures an isotropic proposal in
the tangent space and the norm of the vector is adjusted such that it follows the $\chi^2$-distribution with
$N$ degrees of freedom~\cite{Cristoforetti:2013wha}.

 \begin{figure*}[t!]
\includegraphics[scale=0.5]{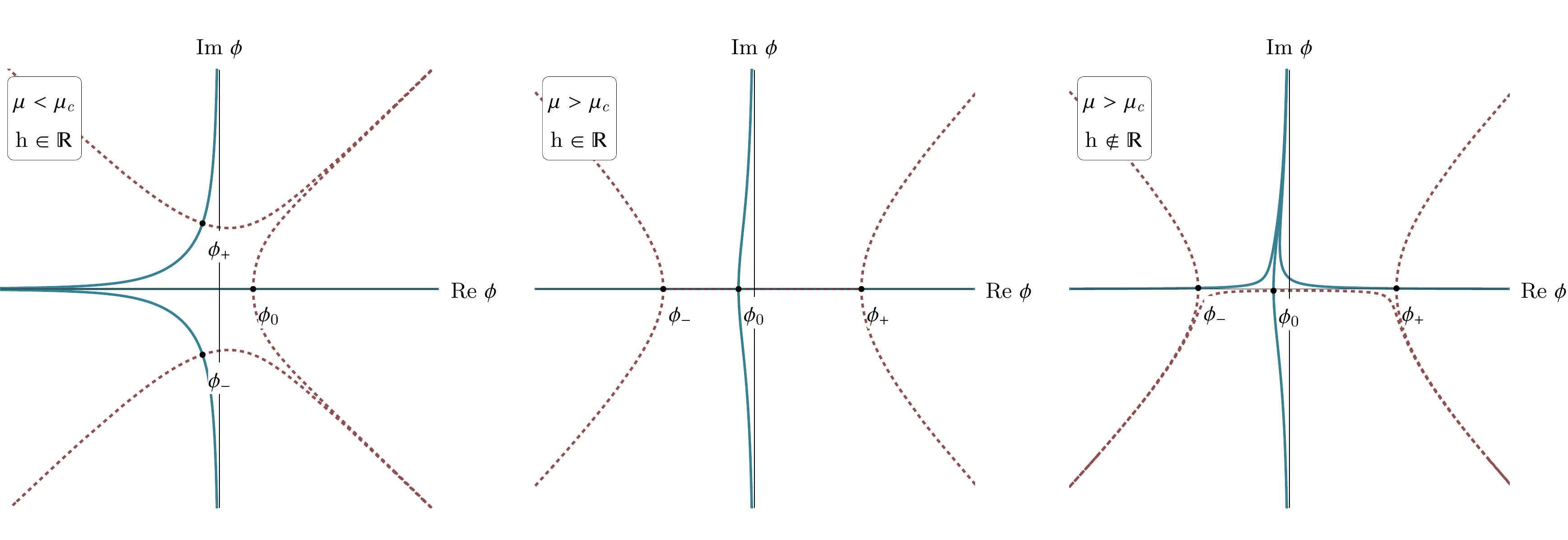}
\caption{Projections of the thimbles (blue) $\mathcal{T}_0, \mathcal{T}_+$, and $\mathcal{T}_-$ and dual thimbles (red) $\mathcal{K}_0, \mathcal{K}_+$, and $\mathcal{K}_-$ onto the 1-complex-dimensional subspace constant fields. The intersection of the original domain of integration with this subspace corresponds to the real line. The typical arrangement of thimbles varies with the chemical potential. 
}
\label{fig:1d-projection-thimbles}
\end{figure*}

At every step we start with $\phi_f$ on the thimble and we move along the tangent direction, since both
the drift and the random vector lie in the tangent plane. Unless the thimble is a hyperplane, this shift
will take us out of the thimble. A projection back to the thimble is required. The methods proposed rely
on evolving the new configuration in the downward flow toward the critical point, projecting there 
to the thimble and flowing back~\cite{Cristoforetti:2012su,Cristoforetti:2012uv}. This proposal was
found to be unstable~\cite{Cristoforetti:2012uv}. The only simulations that we are aware of that employ
this algorithm involve simulations on the tangent plane to the thimble~\cite{Cristoforetti:2013wha}. In this
case the updates do not require any projection since the manifold is flat. To make this algorithm
practical for the general case a robust projection method is needed.

A final note about Langevin algorithm: for a finite $\Delta t$ the method is not exact. Simulations
have to be carried out for decreasing $\Delta t$ and then extrapolated to $\Delta t=0$ to remove
the finite step-size errors. For other Langevin methods, an accept/reject step can be used to remove
the finite step-size errors, but this has not been developed for thimble simulations. 

While both Langevin method and HMC algorithm perform updates directly on $\phi_f$ with
drift (or force) term evaluated locally, it is worth emphasizing that the updates still require 
the integration of the flow equations.
This is because the projection of the shift to the tangent plane to the thimble 
and the required projection back to the manifold after the update, can only be currently done by connecting
$\phi_j$ with its image under the flow $\phi_n$ in the infinitesimal neighborhood of the critical point.
The advantage of these methods over Metropolis, assuming that a practical projection method is available,
is that the updates can lead to large change in action leading to small autocorrelation times in the Markov chain.

\subsection*{Case study: bosonic gases}
We presently consider the relativistic Bose gas at finite density for an application of these algorithms to bosonic systems with sign problems. The continuum Euclidean action of this system is 
\begin{align}
\label{eq:bose-gas-continuum}
S = \int d^4x\,\big[&\partial_0\phi^* \partial_0\phi + \nabla\phi ^*\cdot\nabla\phi + 
(m^2-\mu^2)\lvert\phi\rvert^2 \nonumber \\&+ \mu\underbrace{(\phi^* \partial_0\phi - \phi\partial_0\phi^*)}_{j_0(x)}  + 
\lambda {\lvert \phi \rvert}^4 \big]~,
\end{align}
where $\phi = (\phi_1 + i \phi_2)/\sqrt{2}$ is a complex scalar field. This action encodes the properties of a two-component system of bosons with a contact interaction and an internal global $U(1)$ symmetry which breaks spontaneously at high density. In Euclidean space, the current $j_0$ is complex and causes a sign problem \footnote{This is most readily seen in Fourier space in the continuum: \\ $\int{d^4x~j_0(x)} = (2\pi)^{-4}\int{d^4p ~(-2i p_0) |\phi(p)|^2}$ is purely imaginary.}.


This system was studied with the contraction algorithm in~\cite{Alexandru:2016san}, HMC method~\cite{Fujii:2013sra}, 
and the Langevin process~\cite{Cristoforetti:2013wha}. The following lattice discretization of \eq{eq:bose-gas-continuum} was used
%
%
\beqs
\label{eq:lat_act}
S =  \sum_{x,a} \Bigg[&\Big(4+\frac{m^2}{2}\Big)\phi_{x,a}\phi_{x,a}-\sum_{\nu=1}^{3}\phi_{x,a}\phi_{\smash{x+\hat{\nu},a}} 
\\&- \cosh\mu\ \phi_{x,a}\phi_{x+\hat{0},a}  +  i\sinh\mu\ \epsilon_{ab}\phi_{x,a}\phi_{\smash{x+\hat{0},b}}  \\
     &+ \frac{\lambda}{4}  \big(\phi_{x,a}\phi_{x,a})^2 - h (\phi_{x,1}+\phi_{x,2}) \Bigg]~,
\eeqs
where $\epsilon_{ab}$ is the antisymmetric tensor and $\epsilon_{12} = 1$. This lattice action will be used for the remainder of this discussion. The final term must be included in the lattice theory to obtain a well-defined thimble decomposition and we take $h$ small.

To apply the contraction algorithm, it is first necessary to find critical points (extrema) of the action \eq{eq:lat_act}. 
Restricting attention to those critical points which are constant in spacetime, the following extremum condition is obtained:
\beq
\label{eq:cubic}
(2+m^2)\phi -2 \cosh\mu \phi + 2 \lambda |\phi|^2 \phi =h~.
\eeq
Three extrema exist and we denote them $\phi_0, \phi_{+}, \phi_{-}$. The corresponding Lefschetz thimbles will be denoted $\mathcal{T}_0, \mathcal{T}_{+}, \mathcal{T}_{-}$.
Depending on the parameters of the theory, different combinations of thimbles contribute to the path integral. To this end, the one-dimensional projections of $\mathcal{T}_0,\mathcal{T}_+,\mathcal{T}_-$ depicted in \fig{fig:1d-projection-thimbles} are useful.

For $\mu < \mu_c = \text{cosh}^{-1}(1+m^2/2)$, only $\mathcal{T}_0$ contributes to the path integral. This is because $S_R(\phi_{\pm}) < S_R(\phi)$ for any $\phi$ on the original integration manifold, and therefore no point can flow to $\phi_{\pm}$ by the upward flow. This is sufficient to eliminate $\mathcal{T}_{\pm}$ as contributing thimbles.

For $\mu > \mu_c$, the contributing thimbles changes. As seen in the center of \fig{fig:1d-projection-thimbles}, when $h\in \mathbb{R}$, there are flow trajectories connecting both $\phi_-$ and $\phi_+$ to $\phi_0$. This feature, called \textit{Stokes phenomenon}, introduces complications into the decomposition of the path integral into an integer linear combination of thimbles. We avoid Stokes phenomenon altogether by simply introducing a complex $h$; for a detailed discussion of our procedures see \cite{Alexandru:2016san}.

\begin{figure}[t!]
\includegraphics[scale=0.55]{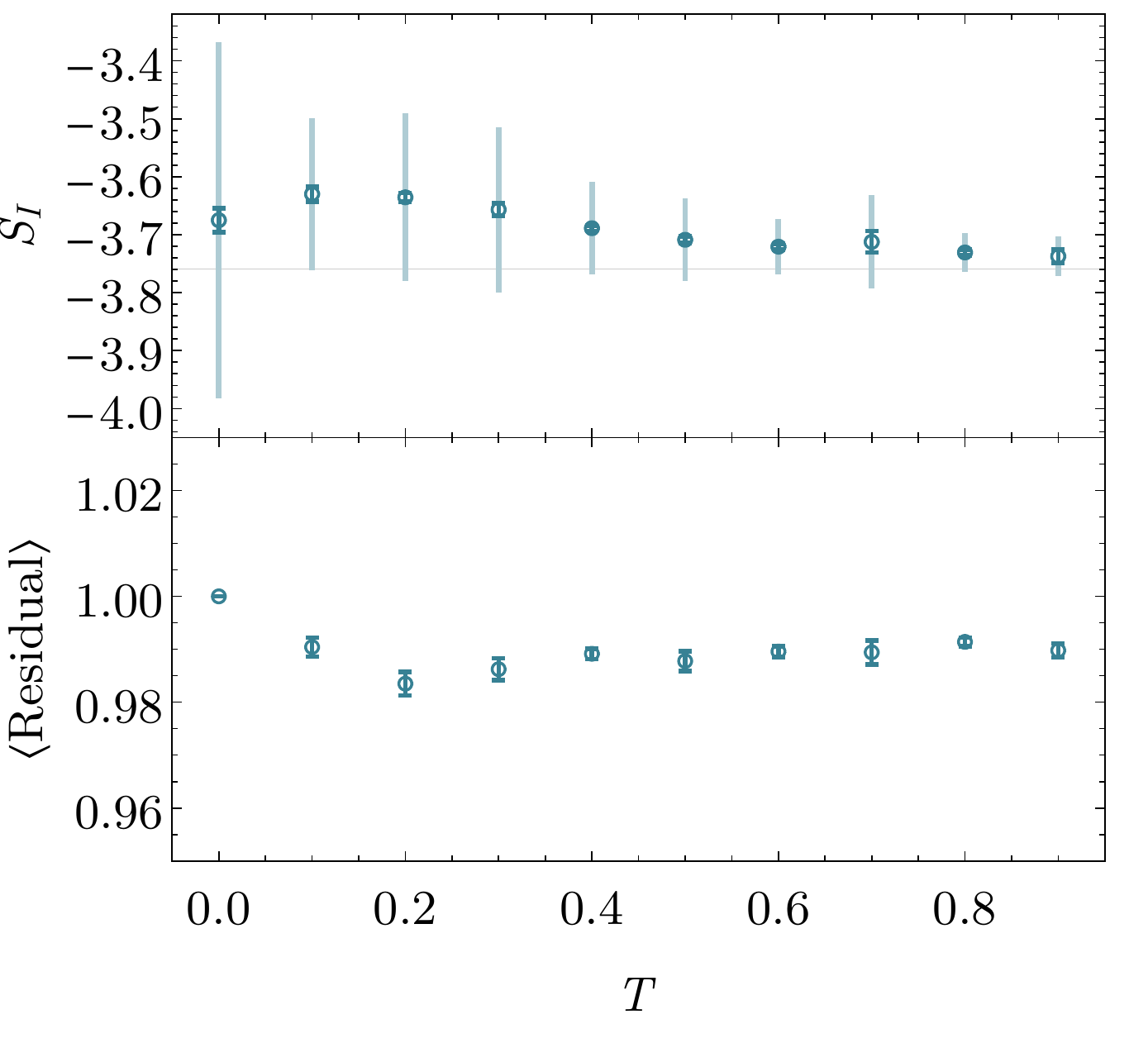}
\caption{The imaginary part of the action and the residual phase computed on $\mathcal{F}_T(T_+)$ using the contraction algorithm. The horizontal line denotes the value of $\text{Im } S(\phi_+)$.
}
\label{fig:obs-fcn-of-flow}
\end{figure}

Since our purpose is to illustrate the Contraction Algorithm, let us consider only the $\mu>\mu_c$ case. As an example, let $m=\lambda =1.0,~h = 0.1(1+i/10)$ and $\mu = 1.3$. With these choices, $\mathcal{T}_+$ contributes most to the path integral. 
The results obtained on flowed manifolds are plotted in \fig{fig:obs-fcn-of-flow}. The variance of $S_I$ decreases as a function of flow time; this demonstrates that the integral over $\mathcal{T}_+$ indeed has reduced phase fluctuations relative to $\mathbb{R}^N$. Furthermore, the convergence of observables as a function of flow time strongly suggests convergence to $\mathcal{T}_+$. 
%

\subsection{Generalized thimble method}
The main limitation of the  methods discussed so far is that they are capable of computing the integral over only one  thimble. However, the integral over the real variables is generically equivalent to the integral over a collection of thimbles. Finding these collection of thimbles is a daunting process; integrating over all of them an even harder task. Fortunately, there is a way of bypassing this difficulty based on what we learn in section \ref{sect:cauchy}: the generalized thimble method.
  
  Recall that if every point of $\R^N$ (the integration region of the path integral) is taken to be the initial condition for the  \eq{eq:flow} that is then integrated for a time $T$, we obtain a manifold $\mathcal{M}_T=\mathcal{F}_T(\R^N)$ that is equivalent to the initial $\R^N$ manifold (in the sense that the path integral over $\R^N$ and $\mathcal{M}_T$ are the same). In addition, for large enough values of $T$, $\mathcal{M}_T$ approaches exactly the combination of thimbles equivalent to $\R^N$. It is important to understand how the thimbles are approached. In the large $T$ limit an isolated set of points in $\R^N$, let us call each of them $\zeta^c$, approach the critical points $\phi^c$ of the relevant thimbles. Points near them initially approach the critical points but, when close to them, move along the unstable directions,  almost parallel to the thimble but slowly approaching it (see \fig{fig:approach-thimble}).  Points far from $\zeta^c$ run towards a point when the action diverges, either at infinity or at a finite distance (in fermionic theories thimbles meet at points where the action diverges as exemplified by the Thirring model discussed below). This means that the correct combination of thimbles equivalent to the original path integral can be parametrized by points in $\R^N$. This is an advantage over the contraction method where only one thimble at a time could be parametrized. We have then
  \begin{align}
  \int_{\R^N}d\phi\ e^{-S(\phi)} & = \int_{\mathcal{M}_T}  \!\!\!\! \!\!\! d\tilde\phi \ e^{-S(\tilde\phi)} \det J(\tilde\phi) \nonumber \\
  & =
  \int_{\R^N}\!\!\!\!  d\zeta\  e^{-S[\mathcal{F}_T(\zeta)]} \det J(\zeta).
  \end{align} 
  
     \begin{figure}[t!] 
   \includegraphics[width=.85\linewidth]{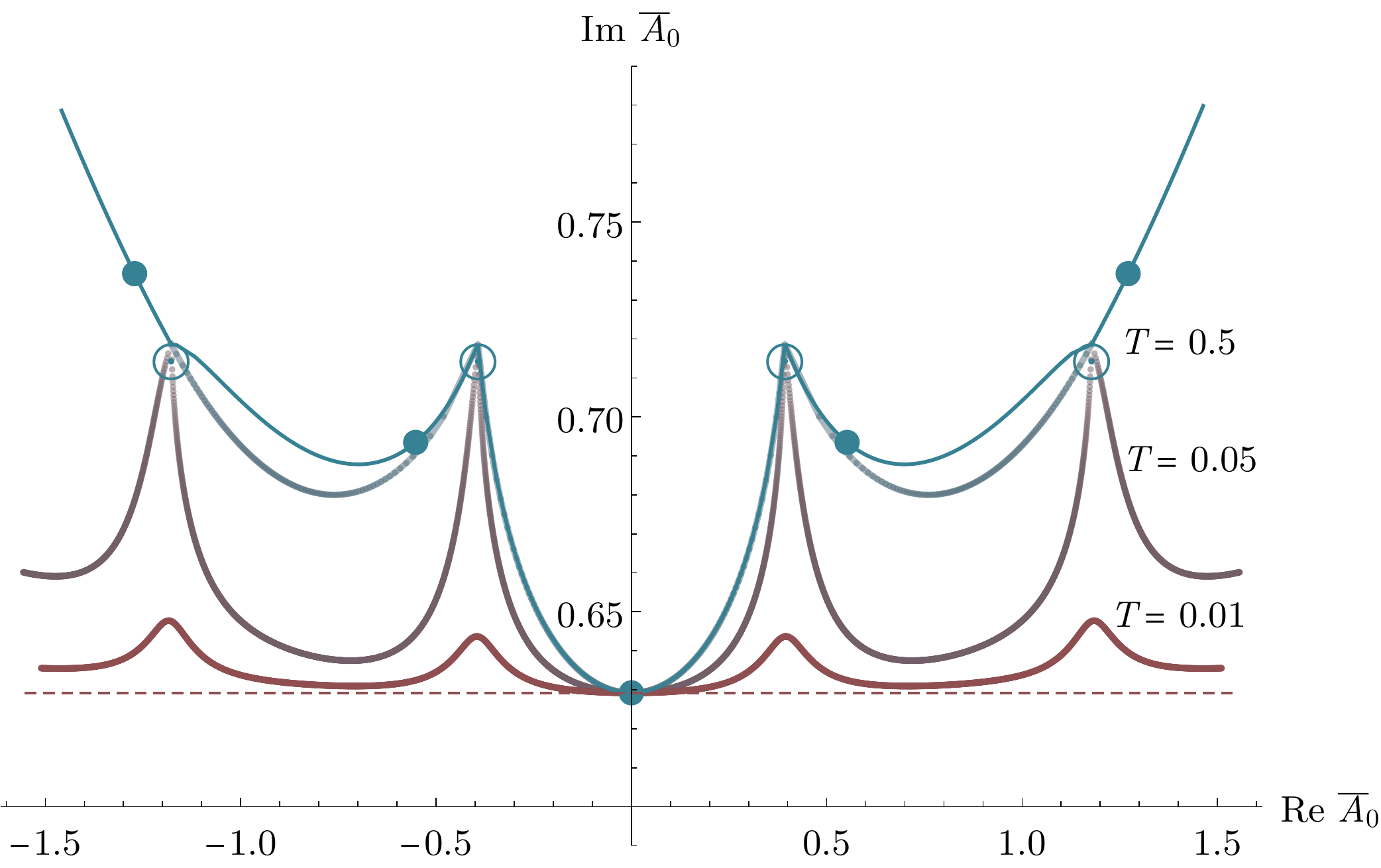}
  \caption{Complex $\overline{A}_0=1/V\sum_x A_0(x)$ plane for the Thirring model. The blue squares are critical points, the blue lines the thimbles. The dashed line is the tangent space to the ``main" thimble while the other solid lines are the manifolds $\mathcal{M}_T$ obtained by flowing the tangent space by $T=0.01, 0.05$ and $0.5$. Notice how $\mathcal{M}_T$ approaches the correct combination of thimbles as $T$ is increased.}
  \label{fig:flow-tangent-plane}
  \end{figure}
  
  The generalized thimble method consists in using a Metropolis algorithm on $\R^N$ with the action $\Re S_\text{eff}$ where $S_\text{eff}(\zeta) = S[\mathcal{F}_T(\zeta)]-(\log \det J(\zeta))$.
   \begin{center}
 {\bf Generalized Thimble Algorithm (GTA)}
 \end{center}
   \begin{enumerate}
  
  \item Start with a point $\zeta$ in $\R^N$. Evolve it  by the holomorphic flow by a time $T$ to find $\phi_f=\mathcal{F}_T(\zeta)$.
  \item Propose new coordinates $\zeta'=\zeta+\delta\zeta$, where $\delta\zeta$ is a random vector drawn from a symmetric distribution. Evolve it by the holomorphic flow by a time $T$ to find $\phi_f'=\mathcal{F}_T(\zeta') $.
 
  \item Accept $\zeta'$ with probability $P_\text{acc}=\min\{1,e^{-\Delta \Re S_\text{eff}}\}$.
  \item Repeat from step 2 until a sufficient ensemble of configurations is generated.
  \end{enumerate}
  Methods to speed up---or bypass---the frequent computation of the Jacobian $J$ are an improvement of the method and will be discussed below (see \ref{sec:algo-jacobians}).
  
  While the algorithm above is exact, the practical applicability of the GTA depends on 
  the landscape induced by $\exp(-\Re S_\text{eff})$ on $\mathcal{M}_T$. At large $T$, the points $\zeta$ that are mapped to the statistically significant parts of $\mathcal{M}_T$ lie on small, isolated regions. 
  This explains why the phase of the integrand fluctuates less on $\mathcal{M}_T$ than on $\R^N$. The imaginary part of $S[\mathcal{F}_T(\zeta)]$  on points on  $\mathcal{M}_T$ are the same as the imaginary parts of the action $S(\zeta)$ in a little region around $\zeta^c$,  the only region with significant statistical weight $\exp(-\Re S_\text{eff}[\mathcal{F}_T(\zeta)])$.

  In between the regions around the different $\zeta^c$ lie areas with small statistical weight $\exp(-\Re S_\text{eff}[\mathcal{F}_T(\zeta)])$ that are mapped to points where the action (nearly) diverges, as we discussed in \ref{sect:thimbles-and-pl}. A probability landscape of this form may trap the Monte Carlo chain in one of the high probability regions, breaking ergodicity. A  trapped Monte Carlo chain is effectively sampling only one of the thimbles contributing to the integral (more precisely, it is an approximation to a one thimble computation). This problem can be alleviated by making $T$ small. In that case $\mathcal{M}_T$ will be farther away from the thimbles, the phase oscillations are larger and the original sign problem may not be controlled. The usefulness of the GTA relies then in being able to find a value of $T$ such that the sign problem is sufficiently ameliorated while the trapping of the Monte Carlo chain is not a problem. In several examples discussed below, over a large swatch of parameter space, it is not difficult to find a range of values of $T$ for which the GTA is useful. Still, one should perform due diligence  and try to diagnose  trapping signs in every calculation, as it is always the case in Monte Carlo calculations.

\subsection*{Case study: 0+1D Thirring model}\label{sect:1Dthirring}
    
 \begin{figure}[t]
\includegraphics[scale=0.4]{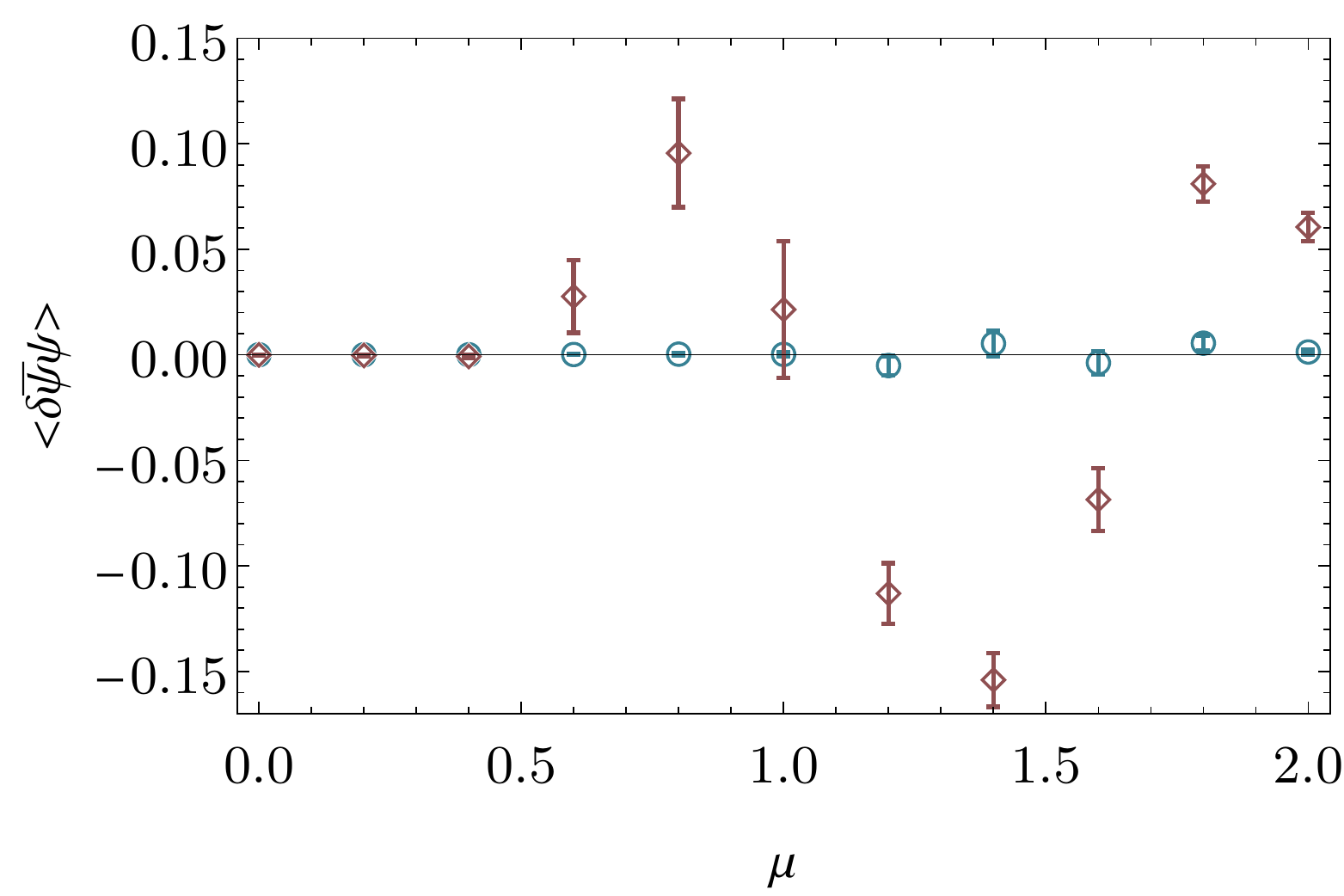}
\caption{The difference in the value of the chiral condensate between the exact result and the one obtained by the contraction method (with $T=2$) shown in red and the generalized thimble method (with $T=0$, that is, integration over the tangent space. )
The parameters are $N = 8$, $m = 1$ and $ g^2 = 1/2$ (lattice units). }
\label{fig:thirring-1D}
\end{figure}

 \begin{figure}[t]
\includegraphics[scale=0.35,angle=-0]{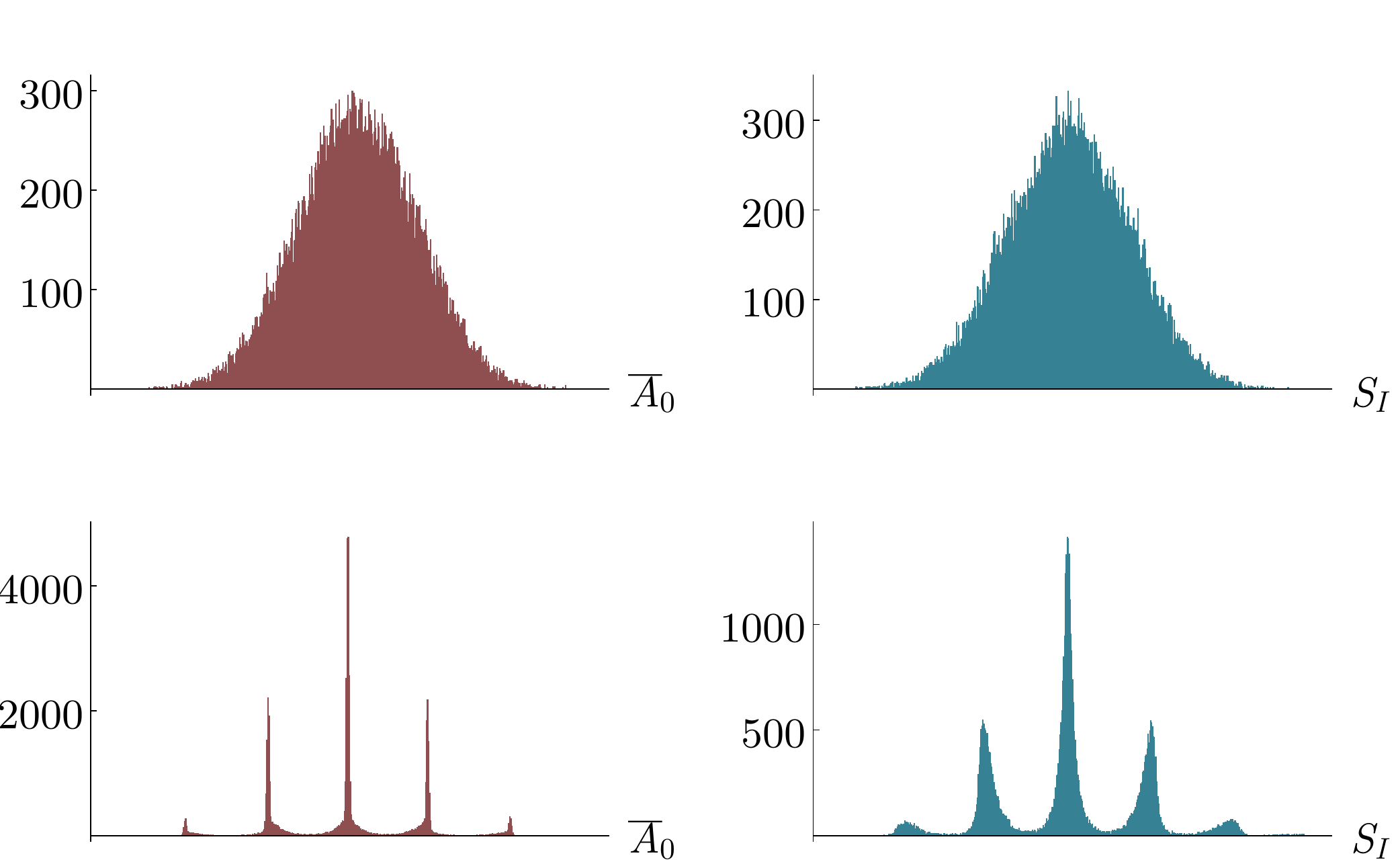}
\caption{
Histogram of  average field $\overline{A}_0$ (left) and imaginary part of the action (right) in a Monte Carlo sampling in the 1D Thirring model using the GTM with $T = 0$ (top line) and $T = 0.5$ (bottom line), $g^2a^2=1/2, N=32$ calculation.  In the $T= 0$ calculation the phase $e^{-i S_I}$ fluctuates too wildly and the result has large uncertainties. On the $T = 0.5$ calculation on the bottom line the phase fluctuates much less. It is also evident that regions on the the tangent space corresponding to several thimbles are being sampled. The multimodal distribution in the $T=0.5$ calculation indicates that larger flow values could lead to trapping of the Monte Carlo chain in a region corresponding to only one thimble.
}
\label{fig:thirring-histograms}
\end{figure}

 We will use the finite density/temperature Thirring model in $0+1, 1+1$ and $2+1$  spacetime dimensions to illustrate several of the techniques discussed in this review. The Thirring model was initially formulated as an example of solvable model in $1+1$ dimensions \cite{Thirring:1958in} and it describes fermions with a contact vector-vector interaction and it is described by the Lagrangian density
 \beq
 \mathcal{L} =\bar\psi^a (i \partialslash+m+\mu \gamma^0)\psi^a + \frac{g^2}{2N_F} \bar\psi^a\gamma_\mu\psi^a \bar\psi^a\gamma^\mu\psi^a,
 \eeq where $\phi$ is a spinor for the appropriate spacetime dimension and $a$ indexes the $N_F$ different flavors of fermions. This theory is, in $1+1$ dimensions, asymptotically free. The $N_F$ case is identical to the Gross-Neveu model and its ground state breaks a discrete symmetry spontaneously and, in this respect, resembles QCD. For $N_F>1$ the chiral condensate exhibits power law decay, the closest behavior to long-range order possible in one spatial dimension \cite{1978NuPhB.145..110W}.
 
 We will use two discretizations of the Thirring model, one using staggered fermions and the other using Wilson fermions. The lattice action in $d$ dimensions is:
 \beq\label{eq:thirring-continuum}
 S = \sum_{x,\nu}    \frac{N_F}{g^2} (1-\cos A_\nu(x)) +\sum_{x,y}\bar\psi^a(x) D_{xy}  \psi^a(y)  ,
 \eeq with
 \beqs
 D^W_{xy} = \delta_{xy}& - \kappa \sum_{\nu=0,1} \big[
 (1-\gamma_\nu) e^{iA_\nu(x)+\mu\delta_{0\nu} } \delta_{x+\nu,y}  \\
 &+ (1+\gamma_\nu) e^{-iA_\nu(x)-\mu\delta_{0\nu}} \delta_{x,y+\nu} 
 \big]
\eeqs 
with $1/\kappa = 2m+4d$ or
\beqs
 D^{KS}_{xy} = m \delta_{xy}+ \frac{1}{2} \sum_{\nu=0,1} &\big[
 \eta_\nu(x)  e^{iA_\nu(x)+\mu\delta_{0\nu}} \delta_{x+\nu,y} \\
 &-  \eta^\dagger_\nu(x)  e^{-iA_\nu(x)-\mu\delta_{0\nu} }\delta_{x,y+\nu}
 \big]
\eeqs 
with $\eta_0(x)=1, \eta_1=(-1)^{x_0},  \eta_2=(-1)^{x_0+x_1}$ and the flavor index goes from $1$ to $N_F$ in the Wilson fermion case but from $1$ to $N_F/2$ in the staggered case. Integrating over the bosonic field $A_\nu(x)$ leads to a discretized version of \eq{eq:thirring-continuum}, showing their equivalence. Integration over the fermion fields leads to  purely bosonic action more amenable to numerical calculations:
 \beq\label{eq:thirring-lattice-A}
 S= N_F \left( \frac{1}{g^2} \sum_{x,\nu} (1-\cos A_\nu(x)) - \gamma \log{\rm det} D(A)\right),
 \eeq with $\gamma=1$ (Wilson) or $\gamma=1/2$ (staggered). Both of these actions describe $N_F$ Dirac fermions in the continuum. The presence of the chemical potential $\mu$ renders the fermion determinant complex and is the origin of the sign problem in this model.
 
 The  $0+1$ dimensional case can be solved exactly with the lattice action in \eq{eq:thirring-lattice-A} and it has been used as a check on several methods designed to handle sign problems \cite{Pawlowski:2013pje,Fujii:2017oti,Li:2016srv}. 
 Its thimble structure is known. In the $A_0(x)=\text{constant}$ sector it is shown in \fig{fig:flow-tangent-plane}. There is one purely imaginary critical point that has the smallest value of the real part of the action, therefore called the ``main critical point". Therefore, in the semiclassical limit it should dominate the path integral. Thimbles touch each other at points where the fermion determinant vanishes and the effective bosonic action diverges (shown as blue squares in \fig{fig:flow-tangent-plane}). The tangent space to the main thimble ($\mathbb{T}$) is just the real space shifted in the imaginary direction (dashed red line in \fig{fig:flow-tangent-plane}). The integration over the tangent space is no more expensive than over the real space since no flowing is required and the Jacobian of the transformation is one. The tangent space, lying parallel to the real space, has the same asymptotic behavior as $\R^N$ and is equivalent to it for the computation of the integral. The figure also shows the result of ``flowing" the tangent space by different values of $T$; the larger the value of $T$, the closer the resulting manifold(s) approach the thimbles. Starting from the tangent space and using a flow time $T=2$ the manifold $\mathcal{F}_T(\mathbb{T})$ obtained is nearly indistinguishable from the thimbles.  
 
  In \cite{Alexandru:2015xva} the model was studied using the contraction algorithm. The results, shown on \fig{fig:thirring-1D} indicate that the fermion condensate, for instance, is close to the exact result but does not agree with it, in particular for certain values of $\mu$ near the transition from $\langle \bar\psi\psi\rangle =0$ to $\langle \bar\psi\psi\rangle \neq 0$. The size of the discrepancy is consistent with a semiclassical estimate of the contributions of other thimbles (besides the main thimble). Similar behavior was seen a 1-site model of fermions \cite{Tanizaki:2015rda}. The integration over the tangent space, however, gives the correct result. Of course, the average sign on the tangent space is smaller than the one obtained with the contraction method.  For not too low temperatures the sign fluctuation is, however, small enough to allow for the computation to be done on the tangent plane. But as the temperature is lowered, the sign fluctuations grow and it becomes difficult to sample the correct distribution, as predicted by general arguments (see \eq{eq:average-sign-free-energy}). One can then use the generalized thimble method and integrate on the manifold $\mathcal{F}_T(\mathbb{T})$ for a suitable value of $T$. Too small a $T$ the sign fluctuation is too large; a $T$ too large is essentially an integration over one thimble and the wrong results is obtained. It is interesting to understand how the transition between these two behaviors occur. In \fig{fig:thirring-histograms} histograms of the imaginary part of the effective action are shown for both $T=0$ and $T=0.5$. It is clear that for $T=0.5$ the fields sampled are concentrated around the pre-image of a few (five) critical points while with $T=0$ (no flow) the distribution is broader. Consequently, the values of the phase $\exp(-i \Im S)$ fluctuate less when there is flow and the sign problem is minimized. On the other hand, for large enough flow time, the probability distribution $\exp(-\Re S)$ becomes multimodal and the trapping of Monte Carlo chains can prevent  proper sampling. Thus, the GTM trades the sign problem by a the problem of sampling a multimodal distribution. This trade is not without profit: in many cases one can find values of $T$ such that the sign problem is sufficiently alleviated but trapping has not set in yet. These values of $T$ can be determined by trial and error.  As $T$ is increased trapping occurs, quite suddenly, and it is not difficult to detect it by noticing a jump on the values of the observables. Also, there are well studied ways to deal with trapping, as explained in the next section. Still trapping is a source of concern in GTM calculations and other, more general techniques, have been developed to avoid it (see \sect{sect:others}).

\subsection*{Case study: 1+1D Thirring model}
   \begin{figure}[t]
\includegraphics[scale=0.6]{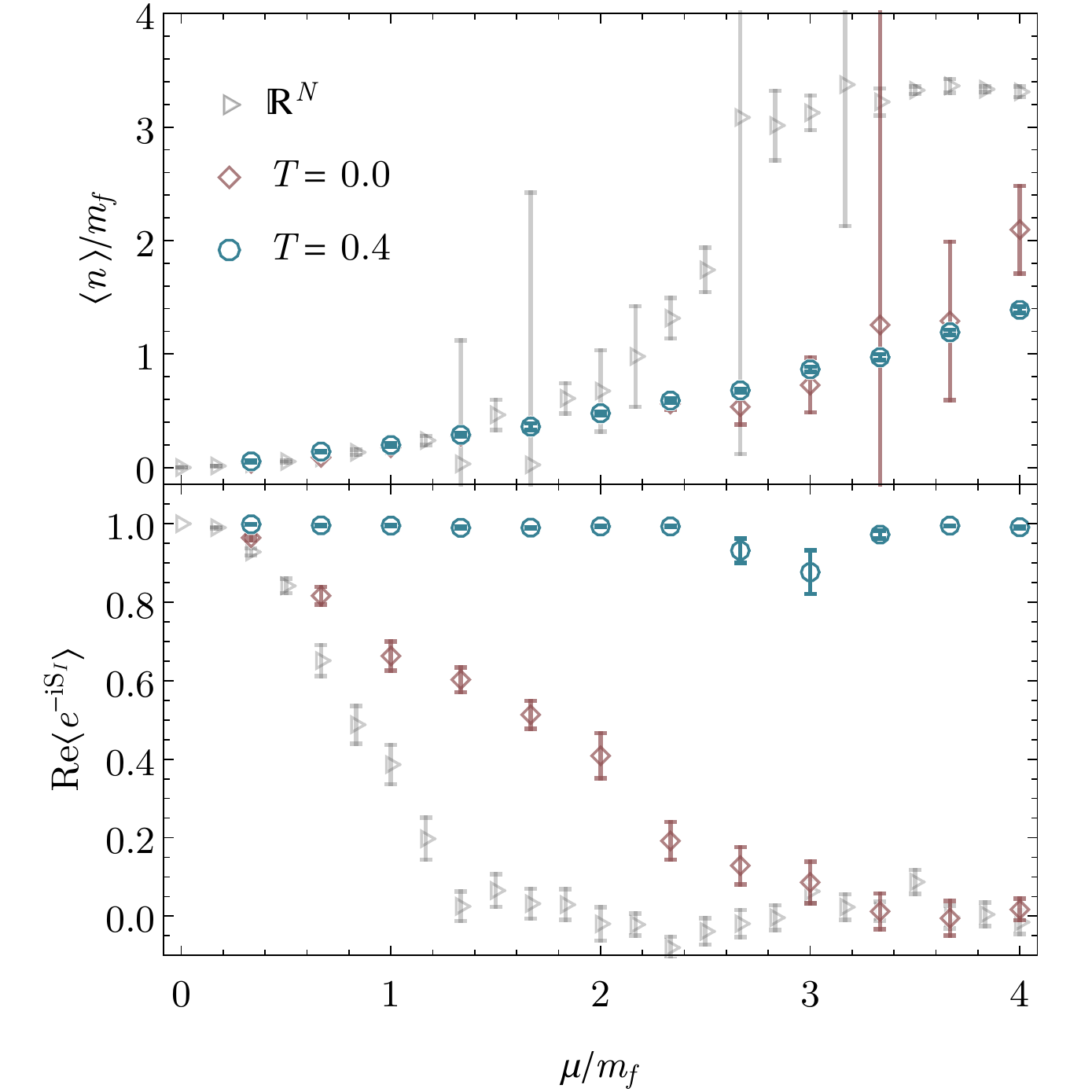}
\caption{ Fermion density (top) and average sign (bottom) of the $1+1$dimensional Thirring model
on a $10\times 10 $ lattice, $g=1, m=-0.25$ (lattice units). The sign problem is strongly suppressed and one moves the path integration from $\R^N$ to the tangent plane $\mathbb{T}$ and from that to the flowed manifold $\mathcal{F}_T(\mathbb{T})$ allowing for precise measurements of the density and other observables \cite{Alexandru:2016ejd}. }
\label{fig:2Dthirring}
\end{figure}
 The lessons learned in applying the generalized thimble method to the $0+1$ dimensional Thirring model carry on to the more interesting $1+1$ dimensional case.  Extensive calculations on the finite density/temperature $1+1$ dimensional Thirring model with two flavors were made over a range of parameters in the strong coupling region~\cite{Alexandru:2016ejd} with both Wilson and staggerred fermions. The thimble structure of the $1+1$ models is more complex than the $0+1$ case. Still, all critical points/thimbles present in the $0+1$ dimensional case have analogues in $1+1$ dimensions (which has many others without a $1+1$ dimensional analogue). It is still true that the closest critical point to the real space (the ``main critical point") is a constant shift of $A_0(x)$ by an imaginary amount and that its tangent space is just a translation of $\R^N$ by an imaginary amount (see \fig{fig:flow-tangent-plane}). The path integration over $\R^N$ has a bad sign problem for all values of the chemical potential larger than the fermion (renormalized) mass ($\mu > m_f$), that is, for all values of $\mu$ for which there is an appreciable number of fermion-antifermion unbalance\footnote{We note here that, contrary to other approaches, the thimble method trivially reproduces the ``Silver Blaze" phenomena, the fact that the system is trivial at small temperatures and chemical potentials smaler than the mass of the lightest fermionic excitation \cite{Cohen:2003kd}.}. The integration over the tangent space of the main thimble can be accomplished at no extra cost by simply shifting the variables of integration by a constant imaginary amount. This step, by itself, improves the sign problem considerably. The reason is that the tangent space is a (rough) approximation to the main thimble, specially the region near the critical point that dominates the path integral in the semiclassical regime. Still, for larger volumes, smaller temperatures and higher chemical potential, the shift to the tangent space is not enough to control the sign fluctuation. It was determined that flow times of the order of $T=0.4$ are sufficient to drastically reduce the sign fluctuation and, at the same time, not cause problems with trapping and ergodicity of the Monte Carlo chain. Some of the results are summarized in \fig{fig:2Dthirring}. In \cite{Alexandru:2016ejd} it was also demonstrated that the same method works well as the continuum and thermodynamic limits are approached.

\subsection{Trapping and tempered algorithms}

The landscape induced by $\exp(-\Re S_\text{eff})$ on the parametrization manifold changes as a function of the flow time $T$. For small $T$ the landscape is typically flat, while for larger $T$ the landscape is steeper. When the sign problem is severe enough to require large flow times, the landscape of $\exp(-\Re S_\text{eff})$ has high peaks and low valleys and the probability distribution can become multi-modal. The purpose of this section is to detail several algorithms addressing this difficulty. 


We first discuss the method of \emph{tempered transitions} \cite{Neal:1996fk}.
Designed to combat trapping, a tempered proposal is a composite proposal assembled from small steps which, taken together, more rapidly cover phase space than a standard proposal. A tempered proposal is constructed as follows. First, let $p_0(\phi),p_1(\phi),..., p_n(\phi)$ be a sequence of increasingly relaxed probability distributions such that $p_0(\phi) \equiv p(\phi)$ is the distribution of interest and $p_n(\phi)$ is significantly more uniform. Next, for every $i$, let $\hat T_i$ be a transition probability satisfying detailed balance with respect to $p_i$, that is
\beq
p_i(\phi) \hat T_i(\phi \rightarrow \phi') = p_i(\phi') \hat T_i(\phi' \rightarrow \phi) ~.
\eeq
Then a tempered update $\hat T$ is executed by first generating a sequence of $2n$ configurations
\beq
\phi_0 \rightarrow \phi_1 \rightarrow ...\rightarrow \phi_n \equiv \phi_n' \rightarrow \phi_{n-1}' \rightarrow ...\rightarrow \phi_0'~,
\eeq
using transition probabilities $\hat{T}_1,\hat{T}_2,\ldots,\hat{T}_n,\hat{T}_n,\ldots,\hat{T}_1$,
followed by an accept/reject step with probability:
\beq
P_\text{acc}(\phi_0 \rightarrow ... \rightarrow \phi_0') 
  = \min \{1, {F(\phi)}/{F(\phi')}\} \,.
\eeq
where
\beq
F(\phi) \equiv
\frac{p_1(\phi_0)}{p_0(\phi_0)} \frac{p_2(\phi_1)}{p_1(\phi_1)} \cdots \frac{p_{n-1}(\phi_{n-2})}{p_{n-2}(\phi_{n-2})}\frac{p_n(\phi_{n-1})}{p_{n-1}(\phi_{n-1})} \,.
\eeq
What is gained by using tempered proposals is enhanced ergodicity. Since the distributions $p_i$ are increasingly uniform, the corresponding transition probabilities $\hat T_i$ may grow in support without decreasing the acceptance probability. To apply this general framework to simulations trapped by holomorphic gradient flow, suppose the flow time $T$ is large enough that the probability distribution of interest
\beq
p(\zeta ) = p_0(\zeta) = \frac{e^{-\Re S_\text{eff}(\zeta)}}{Z}
\eeq
is multi-modal. Consider a sequence of flow times $T_0<T_1<...<T_n$ such that $T_0 = T$ and $T_n \ll T_0$. This defines a sequence of probability distributions $p_0(\zeta), p_1(\zeta), \ldots,p_n(\zeta)$ which are decreasingly multi-modal; we use this sequence to perform tempered proposals. 

Applying this method to the (0+1) dimensional Thirring Model at finite density \cite{Alexandru:2017oyw}, severely trapped simulations have been liberated. Certain thermodynamic parameters exist for which at least five thimbles contribute non-negligibly to the path integral. Trapping to a single thimble, however, can become arbitrarily severe: for example, at $T=0.5$, the multi-modality of $p_0(\zeta)$ is so severe that over the course of a Metropolis with $10^7$ steps not a single transition occurred. Tempered proposals free these trapped MCs however; this is demonstrated in Fig.~\ref{fig:tempered-proposal-history} where proper sampling of the $T=0.5$ probability distribution is achieved. In this case, five separate thimbles are sampled over the course 2000 tempered proposals. Even though tempered proposals cost more than standard proposals, the improvement in ergodicity renders the added effort worthwhile.

\begin{figure}[t!]
\includegraphics[scale=0.4]{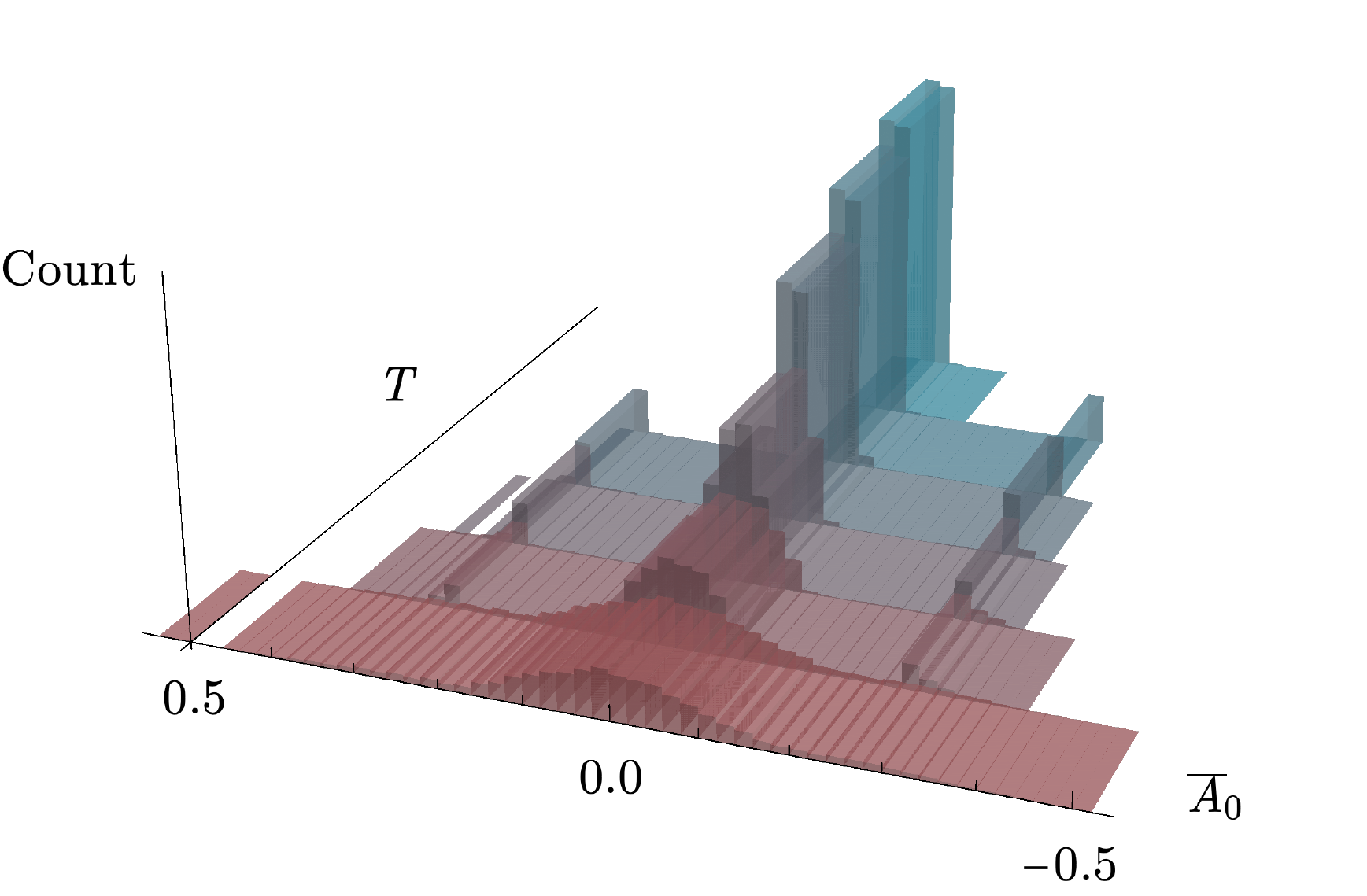}
\caption{Here we show the evolution of field space sampled in the (0+1) dimensional Thirring model as a function of flow time. At small flow times 
the distribution is relatively uniform and much of phase space is sampled. The distribution sharpens as the flow time increases, and at sufficiently large flow times, the shoulder thimbles centered about $\pm 0.3$ cease to be sampled.  }
\label{fig:trapping-process}
\end{figure}

\begin{figure}[t!]
\includegraphics[scale=0.55]{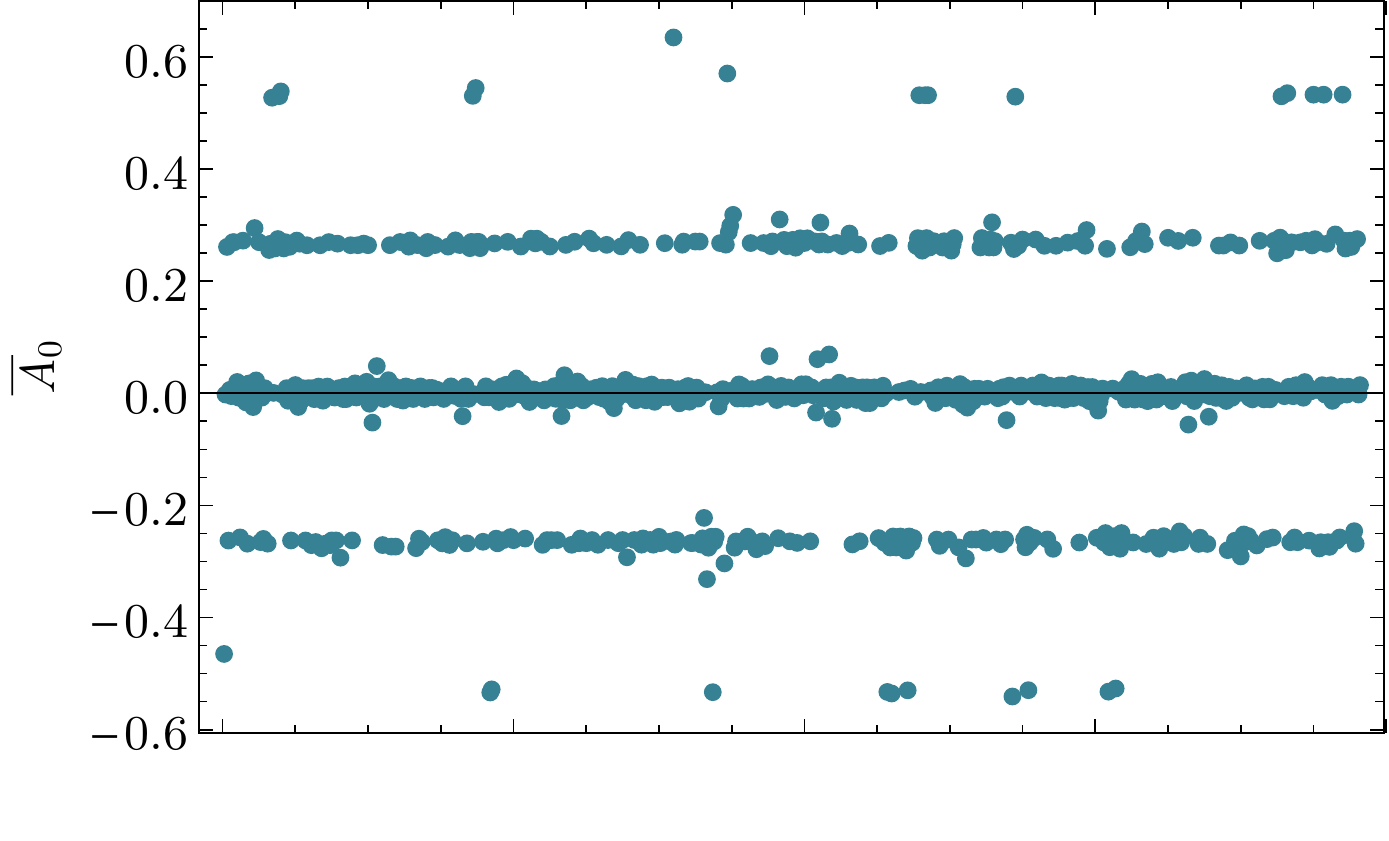}
\caption{Plotted is $\overline{A}_0 = N_t^{-1} \sum_t{A_{0,t}}$ at after each tempered transition. The  five heavily visited positions in field space correspond to five thimbles contributing to the path integral. This distribution is to be compared with the sharpest distribution of Fig. \ref{fig:trapping-process}, where only one thimble is sampled.}
\label{fig:tempered-proposal-history}
\end{figure}

A similar method, \emph{parallel tempering}, was proposed to help sample such from multi-modal
distributions~\cite{PhysRevLett.57.2607,geyer91,Earl_2005}. Parallel tempering involves simulating 
$n$ replicas of the system of interest, each having a particular value of the tempering parameter. 
Each stream evolves separately and swaps between replicas are added satisfying detailed balance. The swapping of configurations between adjacent replicas leads to enhanced ergodicity relative to the single chain case. 
Fukuma et. al. have developed the ``Tempered Lefschetz Thimble Method" (TLTM),  an application of parallel tempering to multi modal distributions generated by flow~\cite{Fukuma:2017fjq}. As with tempered transitions, in this method the flow time is chosen as a tempering parameter. 
The TLTM method has been successfully applied to the (0+1) Thirring model \cite{Fukuma:2017fjq} where trapping due to flow times as large as $T=2.0$ have been solved \footnote{Because the thermodynamic parameters used in \cite{Fukuma:2017fjq} do not match those in \cite{Alexandru:2017oyw} it is currently not possible to compare the efficacy of tempered transitions and the TLTM. A comparison would, however, be useful.}. The authors also studied how to pick the flow times optimally and devised a geometric method for this optimization~\cite{Fukuma:2018qgv}. More recently, the TLTM has been applied to the Hubbard model away from half filling on small lattices \cite{PhysRevD.100.114510}.

\input algo-G

%% file: algo-G.tex
\subsection{Algorithms for the Jacobian}\label{sec:algo-jacobians}

The most computationally expensive part of  many algorithms involving deformation of contours in field space -- like the contraction or the generalized thimble method -- is the calculation of the Jacobian $J$ related to the parametrization of the manifold of integration. For bosonic systems where the Hessian can be computed efficiently the calculation time is dominated by the matrix multiplication in the flow equation and its computation complexity is $\CO(N^3)$ where $N$ is proportional to the spacetime volume of the theory. The calculation of $\det J$ has also similar computational complexity. This prohibitive cost prevents the study of all but the smallest models.

Fortunately, there are ways of bypassing this large cost.
In Ref.~\cite{Cristoforetti:2014gsa} a stochastic estimator was introduced to compute the {\em phase}, 
$\Phi(\phi_n)=\arg \det J(\phi_n)$. The main idea stems from the observation that the Jacobian 
can be expressed as $J=UR$ for some unitary matrix~$U$ and some real, upper-triangular matrix~$R$, 
a property that follows from the fact that $J^\dagger J\in\R$; therefore $\arg\det J=\arg\det U$. 
Note that since $J$ and $U$ are related by a real matrix, this corresponds to a change in basis in the 
tangent plane, so the columns of $U$ form a basis of the tangent space too, an orthonormal basis. 
Moreover $U$ satisfies ${d \log\det U(t)/ dt }=-i \Im \Tr\big(U^T(t) H(t) U(t)\big)$. 
The trace can be estimated stochastically by using random vectors $\xi \in \R^N$ with
$\av{\xi_i \xi_j}=\delta_{ij}$, where the average is taken over the random source; if
we generate $N_R$ vectors we have
\beq
\Tr \big(U^T(t) H(t) U(t)\big) \approx \frac1{N_R} \sum_{r=1}^{N_R} (\xi^{(r)})^T U^T(t) H(t) U(t) \xi^{(r)} \,.
\eeq
Now $\eta^{(r)}=U\xi^{(r)}$ is a random vector in the tangent plane, isotropically distributed and 
its length, with respect to the real Euclidean metric, satisfies $\av{\braket{\eta^{(r)}}{\eta^{(r)}}}=N$.
We can generate such vectors without computing $U$: we generate a random vector $\tilde\eta$ isotropically in $\C^N$ with 
$\av{\tilde\eta^\dagger \tilde\eta} = 2N$, for example using a gaussian distribution $P(\tilde\eta)\propto\exp(-\tilde\eta^\dagger \tilde\eta/2)$, and then project it to the tangent space 
$\eta=P_\parallel(\phi_f) \tilde\eta$ 
using the same procedure presented when we discussed the HMC algorithm.
Using $i\Phi(t) = \log\det U(t)$, the phase can then be estimated from
\beq
\Phi(T)\approx  \Phi(0) -\Im \int_{0}^{T} dt \frac1{N_R}\sum_{r=1}^{N_R}\Tr\left( \eta^{(r)}(t)^T H(t) \eta^{(r)}(t)\right)\,,
\eeq
  whose computational cost scales as $\CO(N\times N_R)$. By comparing this stochastic estimation algorithm by explicit computation for a complex $\phi^4$ theory, Ref. \cite{Cristoforetti:2014gsa} presented numerical evidence that this algorithm indeed provides a nontrivial speedup for the computation of the residual phase in relatively large systems. However, its applicability is limited to the phase of the Jacobian; the GTM requires the magnitude also. 
 
For methods that require the Jacobian, we can substitute them with computationally cheap {\em estimators}.
The idea is to use the estimators during the generation of configurations and correct for the difference
when computing the observables.
Two estimators for $\log\det J$ have been introduced in~\cite{Alexandru:2016lsn}. They are given by 
\beq\label{eq:wrongians}
  W_1=\int_0^{T} dt \sum_i \rho^{(i) \dagger} \overline {H(t) \rho^{(i)}}  \,,
\quad
W_2=\int_0^{T} dt \Tr \overline H(\tau)
\eeq
where $\rho^{(i)}$ are the Takagi vectors of $H_{ij}(0)$ with positive 
eigenvalues. The first estimator, $W_1$, is equal to $\log\det J$ for quadratic actions. The second estimator is equal to $\ln\det J$ when the Jacobian is real along the flow. As such, it is expected to be a good estimator for Jacobians which are mostly real. The bias introduced by the use of estimators instead of the Jacobian is corrected by reweighting the difference between them when computing observables with the help of:
\beq\label{eq:wrongian-reweighting}
\av{\mathcal{O}} = \frac{\av{\mathcal{O} e^{-\Delta S}}_{\Re S'_\text{eff}}}
{\av{ e^{-\Delta S}}_{\Re S'_\text{eff}}}
\eeq 
where $S'_\text{eff} = S- W_{1,2}$ and $\Delta S=S_\text{eff}-\Re S'_\text{eff}$. The estimator is useful when $\Delta S $ has small fluctuations over the sampled the field configurations, that is, if $W_{1,2}$ ``tracks" $\log\det J$ well. 
  
For theories where the Hessian can be computed efficiently, for example for bosonic theories with local actions,
$W_1$ estimator has computational cost of $\CO(N^2)$ and $W_2$ has $\CO(N)$ complexity, a significant improvement
over $\CO(N^3)$ for the full Jacobian. 
In order to use~\eq{eq:wrongian-reweighting} the correct Jacobian $J$ needs to be computed. This has to be done, however,  only on field configurations used in the average in~\eq{eq:estimator}. Typically, configurations obtained in subsequent Monte Carlo steps are very correlated and only one configuration out of tens or hundreds of steps are used in~\eq{eq:wrongian-reweighting}. The idea is then to use the cheaper Jacobian estimators, like $W_1, W_2$ during the collection of configurations and to compute the expensive Jacobian $J$ only when make measurements, which cheapens the calculation by orders of magnitude. This strategy was used, for instance, in the $\phi^4$ model in $3+1$ dimensions~\cite{Alexandru:2016san} and the Thirring model in $1+1$ dimensions~\cite{Alexandru:2016ejd}, both at finite density. However for other class of problems, such as real time systems, the estimators $W_1, W_2$ do not provide a significant improvement.

 
A rather more robust algorithm for the Jacobian have been introduced in Ref.~\cite{Alexandru:2017lqr}. 
The key idea is to modify the proposal mechanism in such a way as to incorporate the Jacobian as part 
of the effective action. As an added bonus, the procedure leads to isotropic proposals on the integration manifold.  
As in the contraction algorithm, the goal is to generate a distribution on the parametrization manifold with
probability proportional to $\exp[-\Re S_\text{eff}(\phi_n)]$.
This is a Metropolis method, so we need to make a proposal and then accept/reject it. For update proposals,
we generate a random complex vector in the tangent plane at $\phi_f$, uniformly distributed with normal distribution
$P(\eta)\propto \exp(-\eta^\dagger \eta/\delta^2)$. The parameter $\delta$ controls the step-size and is tuned
to optimize the acceptance rate. The vector $\eta$ is generated using the projection discussed earlier:
a $\tilde\eta \in \C^N$ sampled from a Gaussian distribution and then $\eta=P_\parallel(\phi_f)\tilde\eta$
using the vector flow projection. The update in the parametrization space is $\phi_n' = \phi_n + \epsilon$ where
$\epsilon=J^{-1}(\phi_n) \eta$ is a vector in the tangent space at $\phi_n$.
Here we take advantage of the fact that the parametrization space is flat and $\phi_n'$ does not need to
be projected.


Since the proposals are not symmetric, the accept/reject step has to be slightly modified to satisfy detailed balance.
The added factor does not cancel the Jacobian, unless the proposal satisfies an implicit equation that is not easy to
solve. A better alternative is based an algorithm by Grady~\cite{Grady:1985fs}: the ratio of Jacobians is taken into
account implicitly using a stochastic generated vector.
The vector is generated with probability $P(\xi)\propto \exp[-\xi^\dagger(J'^\dagger J')\xi]$, where $J'=J(\phi_n')$
and the proposal is accepted with probability~\cite{Alexandru:2017lqr}:
\beq
P_\text{acc} = \min\{ 1, e^{-\Re [S'-S]+\xi^\dagger\Delta J\xi -\epsilon^\dagger\Delta J\epsilon} \} \,,
\eeq
where $\Delta J = (J'^\dagger J')-(J^\dagger J)$.
We stress that $\xi$ is a {\em complex} random vector with $2N$ independent components, whereas $\epsilon$ has only
$N$ independent components.
 
The highlight of this method is that by construction is samples the probability distribution $e^{-\Re S} |\det J|$ 
without an explicit computation of $|\det J|$. It only requires the computation of $J^{-1} \eta$ and 
$J\epsilon$ both of which scale as $\CO(N)$ for most bosonic theories.

A simplified algorithm that may lead to further computational speedup can be achieved when instead 
of $J(\phi_n)$ we approximate it with $J(\phi^c)$. The Jacobian is then only required to compute the
displacements $\epsilon$ and the accept/reject is done simply based on the change of the action since $\Delta J=0$.
For this method $J(\phi^c)^{-1}$ can be computed once at the start of the simulation.
Of course, the difference between $J(\phi_n)$ and $J(\phi^c)$ has to be included by reweighting the observables 
as it was done with $W_{1,2}$. This method should work well when the fluctuations 
of $J(\phi_n)$ are mild. In Ref.~\cite{Alexandru:2017lqr} this was to shown to be the case for the real time study 
of a 1+1 dimensional $\phi^4$ theory even in the strongly coupled regime. 

\subsection*{Case study: real time field theory}\label{sec:realtime}
  
  The generalized thimble method and the whole machinery used in dealing with the computational cost of the Jacobian was applied to one of the most challenging sign problems: the calculation of real time correlators in field theory. These correlators are the building blocks for the computation of transport coefficients like diffusivity, conductivity, viscosities, etc., and are of great importance in a variety of physical contexts. Similar methods can also be used in fully non-equilibrium situations. At the same time the available theoretical tools to study this problem are limited. Even perturbation theory requires complicated resummations  and in the strongly coupled regime the conventional lattice methods are not applicable as detailed below. Alternatively, stochastic quantization (or ``complex Langevin") have been utilized but it seems to converge to the wrong result if the time separation $t-t'$ is more than the inverse temperature $\beta$ \cite{Berges:2006xc}.
  
\begin{figure}[t!]
\includegraphics[scale=0.5]{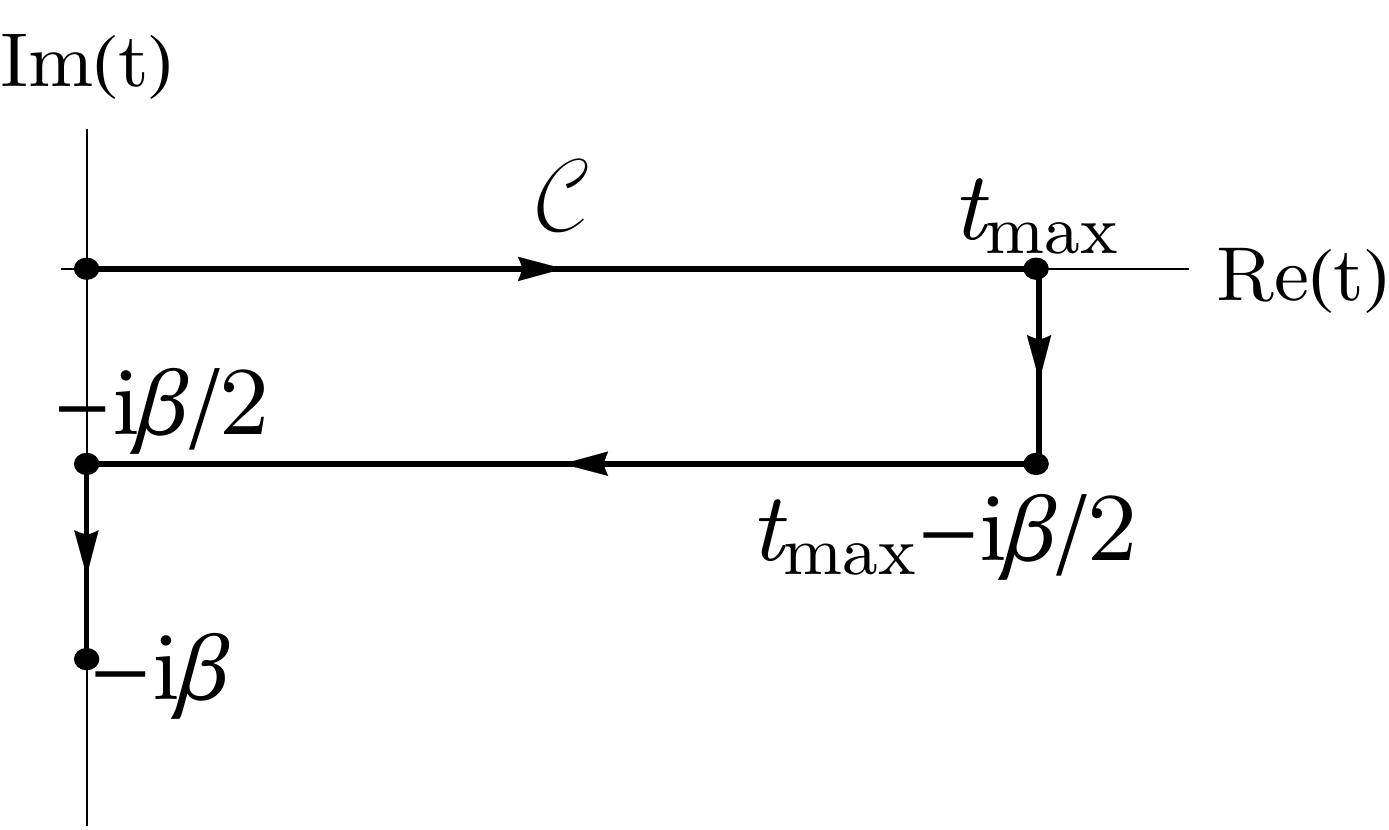}
\caption{The Schwinger-Keldysh contour in complex time plane. The real part corresponds to forward and backwards time evolution and the imaginary part corresponds to the insertion of the equilibrium density matrix. }
\label{fig:sk}\end{figure}

\begin{figure}
\includegraphics[scale=0.53]{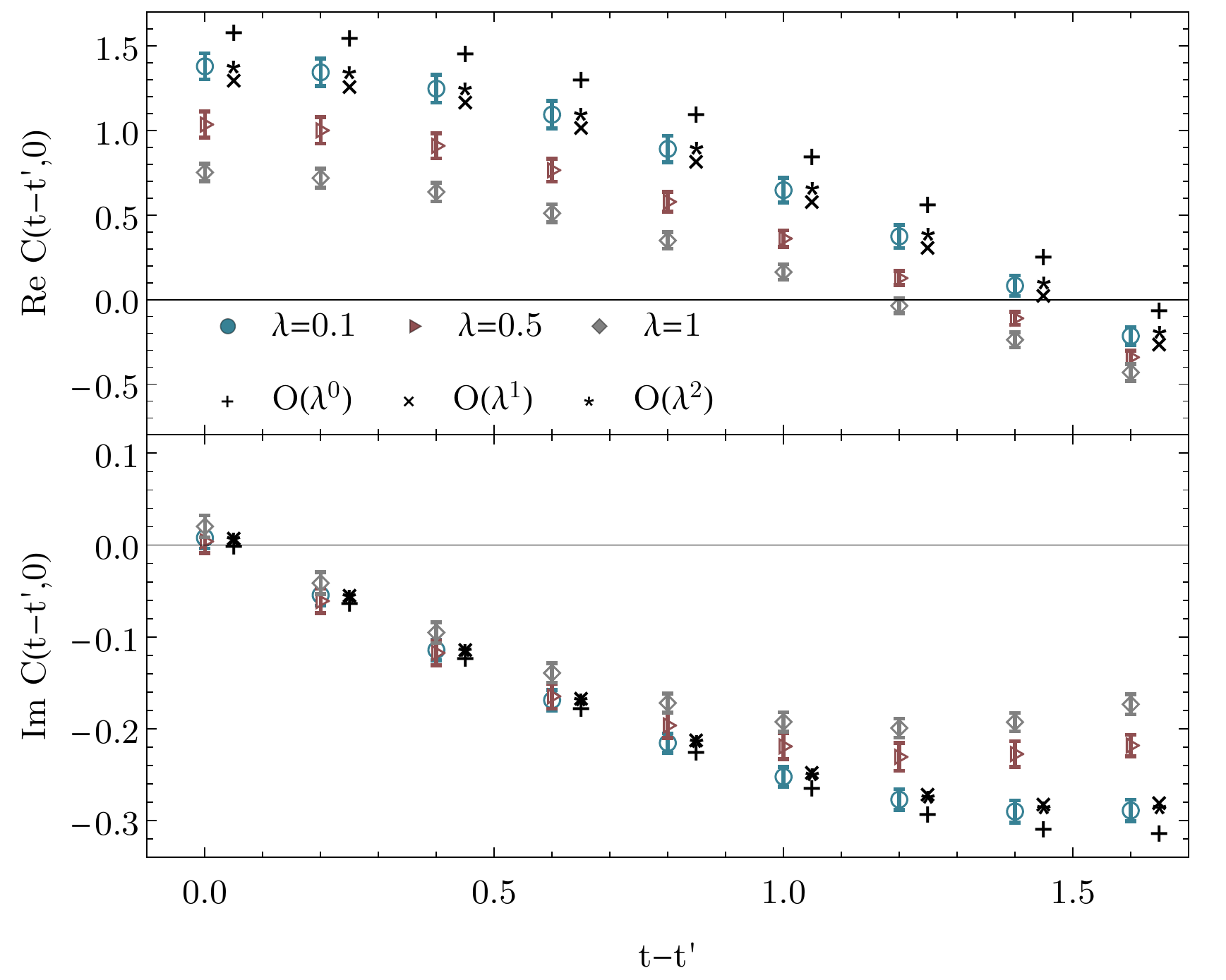}
\caption{The Monte Carlo computation of time order correlation function defined in Eq. \eqref{eq:rt-c} for $p=0$ and $\lambda=0.1,0.5,1$. The $\lambda=0.1$ result is compared with the analytical perturbation theory calculations at ${\cal O}(\lambda^0)$,${\cal O}(\lambda^1)$ and ${\cal O}(\lambda^2)$ which are offset in the $x$ axis for visual clarity. }
\label{fig:rt-c0}
\end{figure}

\begin{figure}
\includegraphics[scale=0.53]{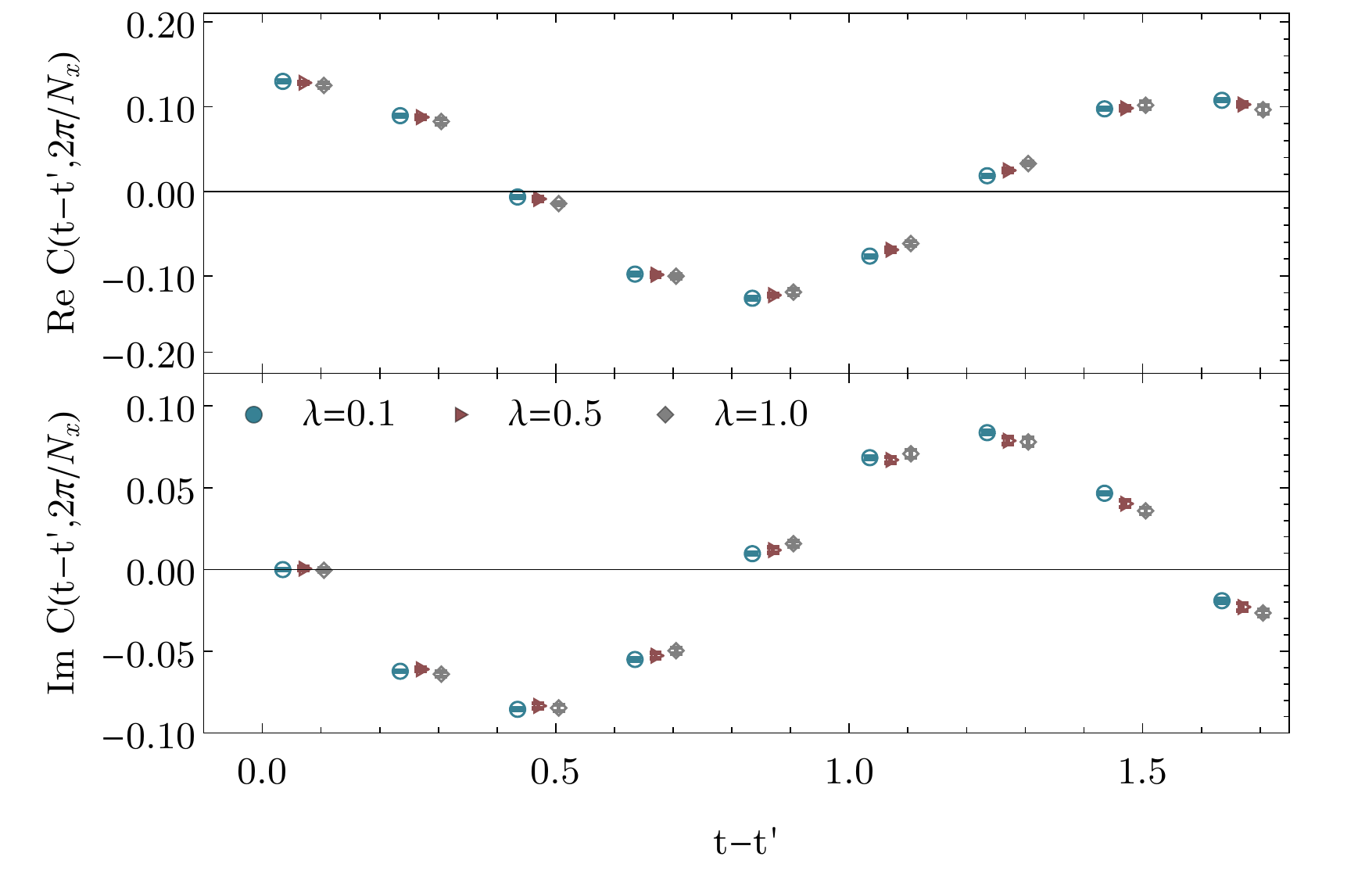}
\caption{The Monte Carlo computation of time order correlation function defined in Eq. \eqref{eq:rt-c} for $p=2\pi/N_x$ and $\lambda=0.1,0.5,1$.}
\label{fig:rt-c1}
\end{figure}

\begin{figure}
\includegraphics[scale=0.6]{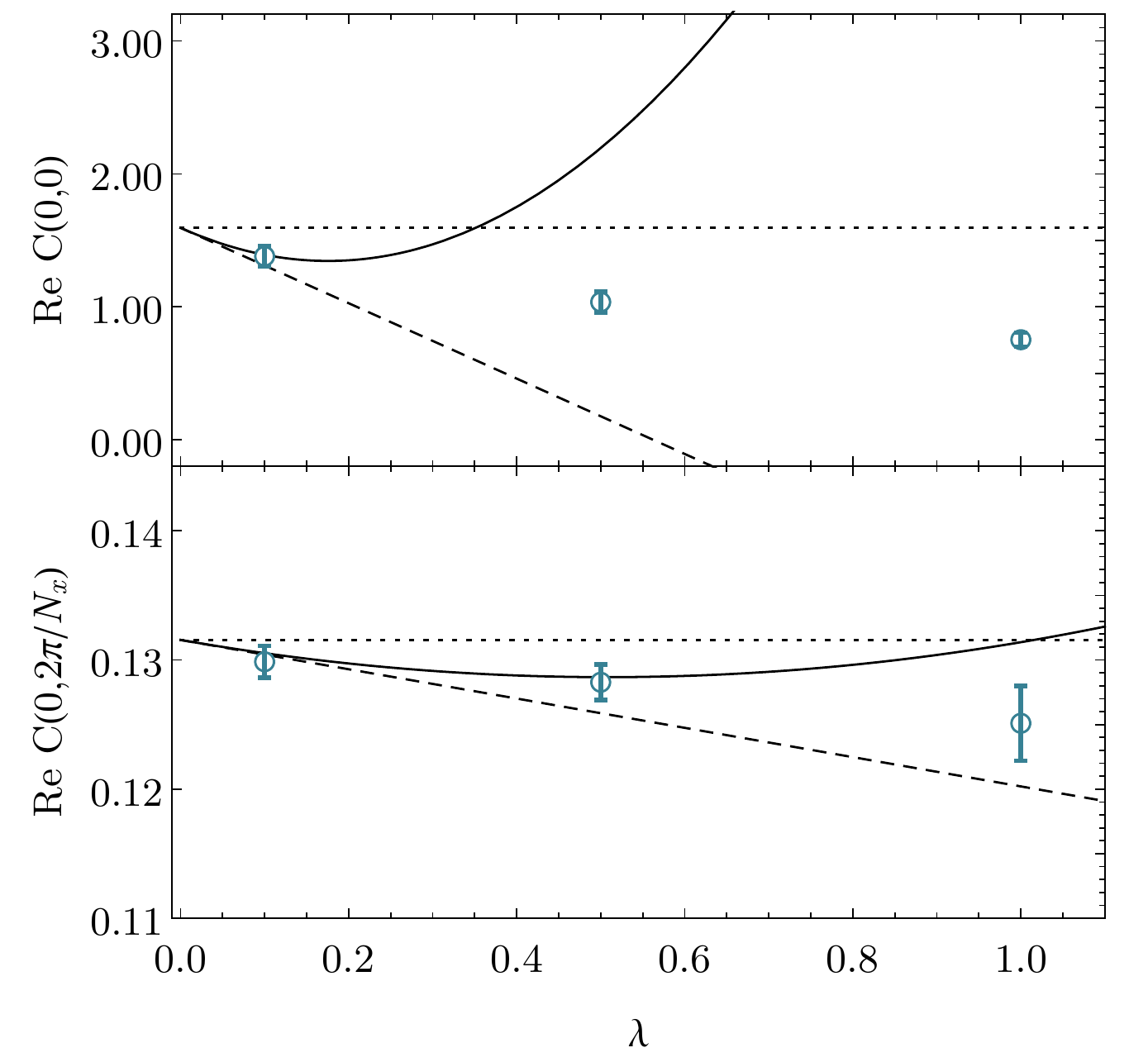}
\caption{The comparison between the perturbative calculation and the Monte-Carlo result.The dotted, dashed and solid lines denote ${\cal O}(\lambda^0)$,${\cal O}(\lambda^1)$ and ${\cal O}(\lambda^2)$ calculations respectively }
\label{fig:rt-pert}
\end{figure}

  The central objects of interest here are \textit{time dependent} correlation functions  of the form  
  \bea\label{eq:corr1}
\av { \CO_1(t) \CO_2(t') }_\beta &=& \Tr (\hat\rho \,\CO_1(t) \CO_2(t')  )
\eea
where $\hat\rho$ is the density matrix which reduces to the familiar Boltzmann factor, $e^{-\beta H}/Tr(e^{-\beta H})$, in equilibrium. 
 Time dependent correlation functions can be generated from the Schwinger-Keldysh (SK) path integral \cite{Schwinger:1960qe,Keldysh:1964ud},
  \bea\label{eq:corr2}
\av{ \CO_1(t) \CO_2(t')}_\beta
&=& \Tr[\CO_1(0)\,e^{-iH(t-t^\prime)}\,\CO_2(0) \,e^{iH(t-t^\prime+i\beta) }]\nn\\
&=&{1\over Z} \int {\cal D}\phi\, e^{i S_{SK}[\phi]}\CO_1(t)\CO_2(t^\prime),\\
\eea
where the SK action is obtained by integrating the Lagrangian over a complex contour, shown in Fig. \ref{fig:sk}. 
The real part corresponds to forward and backward time evolution and the imaginary part corresponds to the insertion of the equilibrium density matrix, $e^{-\beta\hat H}/\Tr e^{-\beta \hat H}$. 
For instance, a discretized Schwinger-Keldysh action for a scalar theory reads:
\footnote{For simplicity we consider the bosonic case but the formalism can be generalized to the fermionic case in a straightforward fashion. We also include an overall factor $i$ so that the associated Boltzmann weight is $e^{-S}$.},
\bea
\label{eq:sk_discrete}
S(\phi)
&=& \sum_{t,\vec x} a_t a   \left[
 \frac12\frac{(\phi_{t+1,\vec x}-\phi_{t,\vec x})^2}{a_t^2} 
+ \frac12 \sum_{\hat i} \frac{(\phi_{t,\vec x+\hat i}-\phi_{t,\vec x})^2}{a^2} \right.\nn\\
&& \qquad \qquad \qquad \qquad \qquad \qquad \qquad \qquad
+ V(\phi_{t,\vec x})  \Bigg], \nn\\
 a_ t &=&\begin{cases}ia\,\text{ for } 0\leq t < N_t  \\ -ia\,\text{ for }  N_t\leq t < 2N_t \\ a\,\text{ for }  2N_t\leq t < 2N_t+N_\beta \end{cases}
\eea
from which the correlators follow
\bea\label{eq:skcorr}
\av{\phi_{t_1,\vec x_1} \phi_{t_2,\vec x_2} } = {\int \big(\prod_{t,\vec x}d\phi_{t,\vec x} \big) e^{-S(\phi)} \phi_{t_1,\vec x_1} \phi_{t_2,\vec x_2}\over \int \big(\prod_{t,n}d\phi_{t,\vec x} \big) e^{-S(\phi)}} \,.
\eea
The Boltzmann weight of the Minkowski part of the SK contour, $0\leq t < 2N_t$, is pure imaginary as expected from the real time evolution and leads to a severe sign problem. In fact, due to the fact that its pure phase with no damping term, it is impossible to define a ``phase quenched" measure and reweigh the phase. For this reason conventional lattice methods do not work not  for real time problems     even if  unlimited computational power is available\footnote{In principle, it is possible to extract the real time correlator \eqref{eq:corr2} from a purely Euclidean time correlator by analytic continuation.  The extrapolation is, however, numerically unstable and requires exponentially accurate precision in Euclidean time.}. By contrast, on any manifold ${\CM}$ that is obtained by flowing from $\R^N$ by some fixed flow time, $\Re S>0$ and the action provides a damping factor making the real time path integral well defined.
The generalized thimble method has been successful in computing time dependent correlation functions in 0+1 dimensional (quantum mechanics) \cite{Alexandru:2016gsd,Mou:2019gyl} and 1+1 dimensional bosonic field theories with $V(\phi)=\lambda \phi^4/4!$ potential. 
In Figs. \ref{fig:rt-c0} and \ref{fig:rt-c1} the two lowest spatial Fourier modes of the time-ordered correlator
\begin{equation}
\label{eq:rt-c}
C(t-t^\prime,p)= \text{T} \langle \phi(t,p) \phi(t^\prime,p)^\dagger \rangle_\beta
\end{equation}
where
\begin{equation}
 \phi(t,p)=\frac1{N_x}\sum_{x=0}^{N_x-1}e^{ipx}\phi_{tx}
\end{equation}
are plotted for different values of $\lambda$ \cite{Alexandru:2017lqr}. To ensure the validity of the method the weak coupling ($\lambda=0.1$) Monte-Carlo result is compared with the zeroth, first and second order perturbation theory calculations performed analytically. In the strong coupling regime which lies outside of the domain of perturbation theory (see Fig. \ref{fig:rt-pert}) the method works as well as it does in the weak coupling regime without any problems. In the quantum mechanical case a similar cross check has been performed which showed agreement between the Monte-Carlo results and the exact result obtained from numerically solving the Schr\"odinger equation \cite{Alexandru:2016gsd}. In Refs. \cite{Mou:2019gyl,Mou:2019tck} the 1+1 dimensional model was studied with a non-equilibrium density matrix.   

The sign problem in the real time problem gets more severe when the time interval between the operators, $|t-t^\prime|$, in units of inverse temperature is increased. This is because the real part of $\CC$ that generates the pure phase contribution to the path integral becomes larger. Therefore a larger flow time is needed to handle larger $|t-t^\prime|$. Currently with the help of the algorithms for the Jacobian described above, $|t-t^\prime|=4 \beta$ has been achieved on a lattice with $N_t=8$, $N_\beta=2$, and $N_x=8$.  Extending to larger time separations seem to be hindered by trapping in a local minima in the Monte-Carlo (Metropolis-Hastings) evolution which calls for alternative sampling methods to be utilized.    
 

\subsection{Gauge theories}

  The thimble structure of gauge theories is more complicated due to the fact that critical points are not isolated points but they are continuous manifolds formed as gauge orbits. The thimbles attached to the critical points carry the same degeneracy due to the gauge symmetry. One might envision instead of working with the degenerate  field space $\CM$ fixing the gauge and working with the quotient space $\CM/\CG$ where the critical points would be isolated and Picard-Lefschetz theory can be used as usual. As we discuss below, this is possible for an abelian gauge fields, however not for non-abelian gauge fields. The reason for that is that some critical points have nontrivial stabilizers and become singular points on $\CM/\CG$ \footnote{For example the zero field configuration is such a critical point stabilized by the whole gauge group.}. In this case Picard-Lefschetz theory has to be modified to accommodate these complications which has been discussed in the context of Chern Simons theory in \cite{Witten:2010cx}

 Lattice gauge theory remains largely unexplored from the perspective of Picard-Lefschetz theory at the moment\footnote{Though exploratory studies exist \cite{DiRenzo:2017igr,Pawlowski:2020kok}.}. We review a few exploratory examples from the literature in the next two sections. Before doing so we first discuss some generalities. In lattice gauge theory the fundamental degrees of freedom are gauge links, $U_i$ where $i\equiv(x,\mu)$ collective index for the link variable $U_{x\mu}\equiv U(x+\hat\mu,x)\equiv U_i$. The derivative with respect to the link variable is defined as 
   \bea\label{eq:groupder}
  \CaD_i^a f(U)\equiv  {\partial \over \partial t} \left. f \left(e^{it T^a} U_i\right) \right|_{t=0}\,.
  \eea
 As usual we consider the complexification of the Lie group where the link variables can be parameterized as $U=e^{i \xi_a T^a}$ where $T^a$ are the group generators and $\xi_a$ are complex variables. For example the complexification of $SU(N)$ leads to $SL(N)$. The holomorphic flow equation reads
\bea\label{eq:gaugeflow}
{d U_i\over d\tau} =i \sum_{a} (T^a \overline{ \CaD_i^a S(U)} ) U_i 
\eea 
and it satisfies the desired properties $d\Re S(U)/dt=|{ \CaD_i^a S(U)}|^2>0$ and $d\Im S(U)/dt=0$. Unlike ordinary derivatives, the groups derivatives do not commute,
  \bea\label{eq:commutators}
  [\CaD_i^a, \CaD_j^b] &=&-f^{abc}\delta_{ij} \CaD^{c}_j ,\quad[\bar\CaD_i^a, \bar\CaD_j^b] =-f^{abc} \delta_{ij}\bar\CaD^{c}_j,
  \nonumber \\
  \quad [\CaD_i^a, \bar\CaD_j^b] &=&0,
  \eea
 where $f^{abc}$ are the structure constants of the gauge group such that $[T^a,T^b]=if^{abc}T^c$. Therefore the flow equation for the tangent space generated by $e^a_i=\overline{\CaD^a_i S}$ is modified as
 \bea\label{eq:group}
{d e_i^a \over d\tau}= \overline{e_j^b\CaD_j^b\CaD_i^a S }-f^{abc}e_i^b\overline{\CaD_i^a S}\, .
 \eea
 
\subsection*{Case study: Heavy-Dense QCD}

  Some exploratory work towards implementing the thimble method in QCD has been done within the so-called ``heavy-dense QCD'' that is QCD with heavy quarks in the high density limit \cite{Zambello:2018ibq}. As opposed to the heavy mass limit where the quarks decouple from the theory, in the simultaneous high-mass, high-density limit
  \begin{equation}
   m_0\rightarrow \infty, \,\,\mu\rightarrow \infty, \,\,e^{\mu}/m_0:\text{fixed}, 
    \end{equation}
  the quarks remain in the picture and the theory has a nontrivial phase structure controlled by $\mu$.  Just like QCD, heavy-dense QCD also exhibits a sign problem. At the same time it is not as computationally demanding as full QCD which makes it a fruitful arena for testing new approaches to the sign problem  \cite{Aarts:2016qrv, Zambello:2018ibq}. 
 In this limit the fermion determinant simplifies quite dramatically as \cite{Bender:1992gn, Blum:1995cb}
 \begin{equation}
 \label{eq:Dirac-hd}
 \det D_f\rightarrow \prod_{\vec x}\det\left(1+\gamma P_{\vec x}\right)^2 \det \left(1+\tilde \gamma P^{-1}_{\vec x}\right)^2
 \end{equation}
 where  $\gamma\equiv(2e^\mu/m_0)^{N_t}$ and $\tilde\gamma\equiv(2e^{-\mu}/m_0)^{N_t}$ and  $ P_{\vec x}= \prod_{t=0}^{N_t-1} U_0(\vec x, t)$ is the Polyakov loop. Eq. \eqref{eq:Dirac-hd} has a simple physical interpretation: in the infinite mass limit, quarks are pinned to their spacial location and do not move. Therefore a quark (anti-quark) at a spatial point $\vec x$ is simply described by the Polyakov loop $P_{\vec x}$ ($P^{-1}_{\vec x}$). Furthermore due to the high density limit, the anti-quark contribution is negligible  (i.e. $\gamma\gg \tilde \gamma$) and one can neglect the second determinant in the right hand side of Eq. \eqref{eq:Dirac-hd}. 
 Since the fermion determinant has no dependence on the spatial links, $U_{\mu\neq0}$, one can obtain the effective action for heavy-dense QCD by integrating out the spatial degrees of freedom in the QCD path integral,
   \bea\label{eq:heavydensereduce}
Z_{QCD}&=&\int DU_\mu e^{{\beta\over 2N_c}\sum_{p} \left( \Tr U_p+\Tr U^\dagger_p\right)}\prod_{f=1}^{N_f} \det D_f\nn\\
&\rightarrow & \int DU_0 e^{-S_{HD}(U_0)}
\eea
where we assumed all $N_f$ quarks are heavy and has identical chemical potentials for simplicity. The effective action for the heavy-dense QCD in this case is 
 \bea\label{eq:heavydenseL}
S_{HD}&\equiv &S_{gauge}-2N_f \sum_{\vec x} \log \left[\det\left(1+\gamma P_{\vec x}\right) \right],\nonumber \\
S_{gauge}&\approx&-\left({\beta\over18}\right)^{N_t} \sum_{\langle \vec x \vec y \rangle}\left(\Tr P_{\vec x}\Tr P^{-1}_{\vec y} +\Tr P_{\vec y}\Tr P^{-1}_{\vec x} \right)\nn\\
\eea
The leading order pure gauge action, including the coefficient $(\beta/18)^{N_t}$, follows from the character expansion of the original gauge action \cite{Langelage:2014vpa}.  In the low temperature limit where $N_t\gg1$ and $\beta \sim \CO(1)$ the pure gauge contribution, $S_{gauge}$, can further be neglected and $S_{HD}$ simply reduces  to 
\bea\label{eq:heavydenseS}
&&S_{HD}\approx-2N_f\sum_{\vec x}\log\det\left(1+\gamma P_{\vec x}\right)
\eea
In particular for $N_c=3$ the determinant over the gauge group reduces to
\bea
\det\left(1+\gamma P_{\vec x}\right)&=&1+\gamma\Tr P_{\vec x}+\gamma^2\Tr P^{-1}_{\vec x}+\gamma^3 
\eea
Higher order corrections to $S_{HD}$ is given in powers of the hopping parameter, $\kappa\equiv 2/m_0$, and can be found in Ref. \cite{Zambello:2018ibq}. Furthermore Ref. \cite{Zambello:2018ibq} focuses on $\mu\approx\mu_c\equiv m=-\log(2\kappa)$ where nuclear phase transition occurs at zero temperature. It is possible and convenient to work in the temporal gauge which eliminates all the links in $P_{\vec x}$ but one for fixed $\vec x$ (say $t=0$), so that $P_{\vec x}=U_0(\vec x,t=0)\equiv U_{\vec x}$ . The holomorphic gradient flow equation \eqref{eq:gaugeflow} in the temporal gauge reads
\bea\label{eq:hdflow}
{d U_{\vec x}\over dt} &=i \sum_{a} (T^a \overline{ \CaD^a S_{HD}[U]} ) U_{\vec x} = -2\gamma \sum_a T^a \overline{\Tr(T^aU_{\vec x})\over\det(1+\gamma  U_{\vec x})} U_{\vec x}.\nn\\
&
\eea 
 The critical points satisfy $\Tr(T^aU^{cr}_x)=0$ and therefore are elements of the center:
 \begin{equation}
 U_{\vec x}^{cr}=e^{ i \omega_{\vec x}}, \quad \omega_x\in \left\{\left. \frac{2\pi n}{ N_c}\right|n=0,\dots,N_c-1 \right\} \, .
 \end{equation}
Since $\omega_x$ can take one of these three values at each lattice site, the number of critical points exponentially grows with the volume as $(N_c)^V$. However they contribute to the path integral with different weights. Ref.  \cite{Zambello:2018ibq} studied this model with $N_c=3$  in small spatial volumes up to $3^3 - 4^3$ and in a parameter range where only a few critical points, hence thimbles, contribute significantly to the path integral and estimated their semiclassical weights, $e^{-S_{HD}[U^{cr}]}$,  by importance sampling. Furthermore they performed the Monte-Carlo computations of the charge density $\av{n}$ and the Polyakov loop $\av P = 1/N \sum_{\vec x}\Tr \av{ U_{\vec x}}$ over the thimbles with one and two lattice sites. The results show the expected behavior in the cold limit near $\mu=\mu_{cr}$, namely $\av{ n} $ sharply changing from $0$ to $1$ \footnote{The saturation of the fermion density is due to finite volume.} (i.e. the Silver Blaze behavior \cite{Cohen:2003kd}) and $\av{ P}$ having a narrow peak around $\mu_{cr}$. Furthermore the contribution of three thimbles is necessary to obtain this expected result.

\subsection*{Case study: 2D QED}
  
  Another example of a gauge theory, two-dimensional QED with the lattice action
  \beq
	S = \frac 1 {g^2} \sum_r \left(1 - \cos P_r\right) - \sum_a \ln \det D^{(a)}, 
	\eeq
	with 
	\bea
	D^{(a)}_{xy} = m_a \delta_{xy} + \frac 1 2 &&\sum_{\nu\in\{0,1\}}
	\big[
		\eta_\nu e^{i Q_a A_\nu(x) + \mu \delta_{\nu 0}}\delta_{x+\hat\nu,y} \nn\\
		&&- \eta_\nu e^{-i Q_a A_\nu(x) - \mu \delta_{\nu 0}} \delta_{x,y+\hat\nu}
	\big]
	\text.
\eea
was studied in Ref. \cite{Alexandru:2018ngw} by using the generalized thimble method. Here $D^{(a)}$ denotes the (Kogut-Susskind) fermionic matrix for flavor $a$\footnote{In order to have neutral excitations a three flavor model with charges $(2,-1,-1)$ have been studied in Ref. \cite{Alexandru:2018ngw}. The two flavor model with equal charges has no sign problem due to charge conjugation symmetry.}, and $P_r$ denotes the plaquette,
\bea
P_r \equiv A_1(r) + A_0(r+\hat x) - A_1(r+\hat t) - A_0(r),
\eea
and $\hat{t}$ and $\hat{x}$ are the unit vectors in time and space direction. In the case of abelian gauge theories it is convenient to work with the complexified gauge field, $A_\mu(x)\in \mathbb C^{2N}$, where $N$ is the number of lattice sites, instead of the gauge links. This way, degeneracies to due gauge redundancy can be addressed in a straightforward fashion. For any point $x$, the gauge orbit is generated by  gauge transformations, $A_{\mu}(x) \rightarrow A_{\mu}(x) + \alpha(x+\hat\mu) - \alpha(x)$, which forms an $N-1$ dimensional\footnote{Note that $\alpha$=constant is not a gauge transformation.} subspace. The original real field space can therefore be locally expressed as a direct product  $\mathbb R^{2N} = \mathcal M_0 \times \CG$ where $\CM_0$ is the space of gauge inequivalent field configurations and $\CG$ is the gauge orbit. The flow leaves $\CG$ invariant because the gradient $\overline{\frac{\partial S}{\partial A}}$ is orthogonal to the gauge orbits, and it commutes with gauge transformations. Therefore the middle-dimensional manifold (i.e. a manifold with real dimension $N$) obtained by flowing $\R^{2N}$ by an amount $T$, can be decomposed as $ \CM_T \times \CG$ where $\CM_{T}$ is the result of flowing $\CM_0$ by $T$. Furthermore, the critical points on the gauge fixed slice are now isolated and the thimble decomposition follows straightforwardly as the limit $T\rightarrow \infty$. This argument illustrates conceptually how the generalized thimble method works in the presence of an abelian gauge field. For the actual lattice computations, however,  there is no need to fix the gauge as the Markov chain will randomly sample $\CG$ which has no effect on the results as long as only gauge invariant observables are evaluated. 

In Ref. \cite{Alexandru:2018ngw} the generalized thimble method has been put into action in a $U(1)$ gauge theory with three fermion species with charges $1,1$ and $-2$. Their charges are chosen in such a way that a state with finite fermion density does not necessarily have a net charge, which would render the energy of the state infinite in the thermodynamics limit. 
It is shown that a computational speedup can be achieved in computing the equation of state compared to conventional real space computation. Even though the cold limit, which shows the Silver Blaze behavior (on a $14\times 10$ lattice), can be achieved this way, faster algorithms are needed to go to larger lattices. Finally, two-dimensional QED in the continuum has been studied in the mean field approximation in \cite{Tanizaki:2016xcu} where the exact thimble decomposition has been worked out for $N_f=1,2,3$ and it was shown that the sign problem can be eliminated by deforming the path integral into the thimble decomposition.

%% file: glory.tex
\subsection{Well beyond thimbles}

We have seen above that deforming the integration from $\R^N$ to a proper combination of thimbles is not always desirable from the numerical point of view. The generalized thimble method, for instance,  uses a rough approximation of thimbles that, while having a smaller average sign, has better ergodic properties. There is no reason, however, to be limited to manifolds close to the thimbles. The condition that $S_I$ is constant is only one constraint in a $2N$ dimensional space and, presumably, there are many manifolds of integration where the sign of the integrand is fixed.
In this section we consider a few methods  to search for other manifolds unrelated to thimbles that both alleviate the sign problem and are numerically convenient. 

An ideal manifold of integration would i) ameliorate the sign problem significantly both because the action is nearly real on it and because the residual phase is small, ii) be computationally cheap to find and iii) be parametrized in such a way that the associated Jacobian is also computationally cheap. All these restrictions are hard to satisfy at the same time and only the first steps in this direction were taken. It seems that insight into particular models will be essential to exploit this general idea profitably. We will show below that, in cases where the thimble method generate some of this insight, it is not difficult to improve it by allowing for more general manifolds. In other theories it is an open problem to find a way to capitalize on the freedom of picking more general manifolds.

\subsection{Learnifolds}
\label{sec:learnifold}


Suppose a number of points on the thimble(s)---or some approximation of it---are obtained using the computationally costly holomorphic flow equations. It is reasonable to expect that the sign fluctuations on a manifold that interpolates between the original manifold, $\R^N$, and the thimble will be small. 
A point found using the flow (in reality a complex field configuration or an element of $\C^N$) can be viewed as a map connecting its real part to the corresponding imaginary part, that is for $\phi\in{\cal M}_T$ the map $f$ takes $\Re \phi\to f(\Re \phi)= \Im\phi$. Here we assume that the manifold does not ``fold"; i.e. every real value of the field corresponds to an unique imaginary part.
We seek to find a manifold with a simple parametrization that ``interpolates'' the configurations sampled on ${\cal M}_T$. 
The result of the interpolation problem can be thought as a map that approximates $f$, the map that connects the given real parts of coordinates to their imaginary parts. Thus, finding the interpolation of the points obtained by the flow can be formulated as learning a general rule from a set of examples. This is a typical problem studied by the artificial intelligence community and we can borrow some of their techniques to bear on it.


\def\tset{{\cal S}}

\begin{figure}
  \includegraphics[width=.8\linewidth]{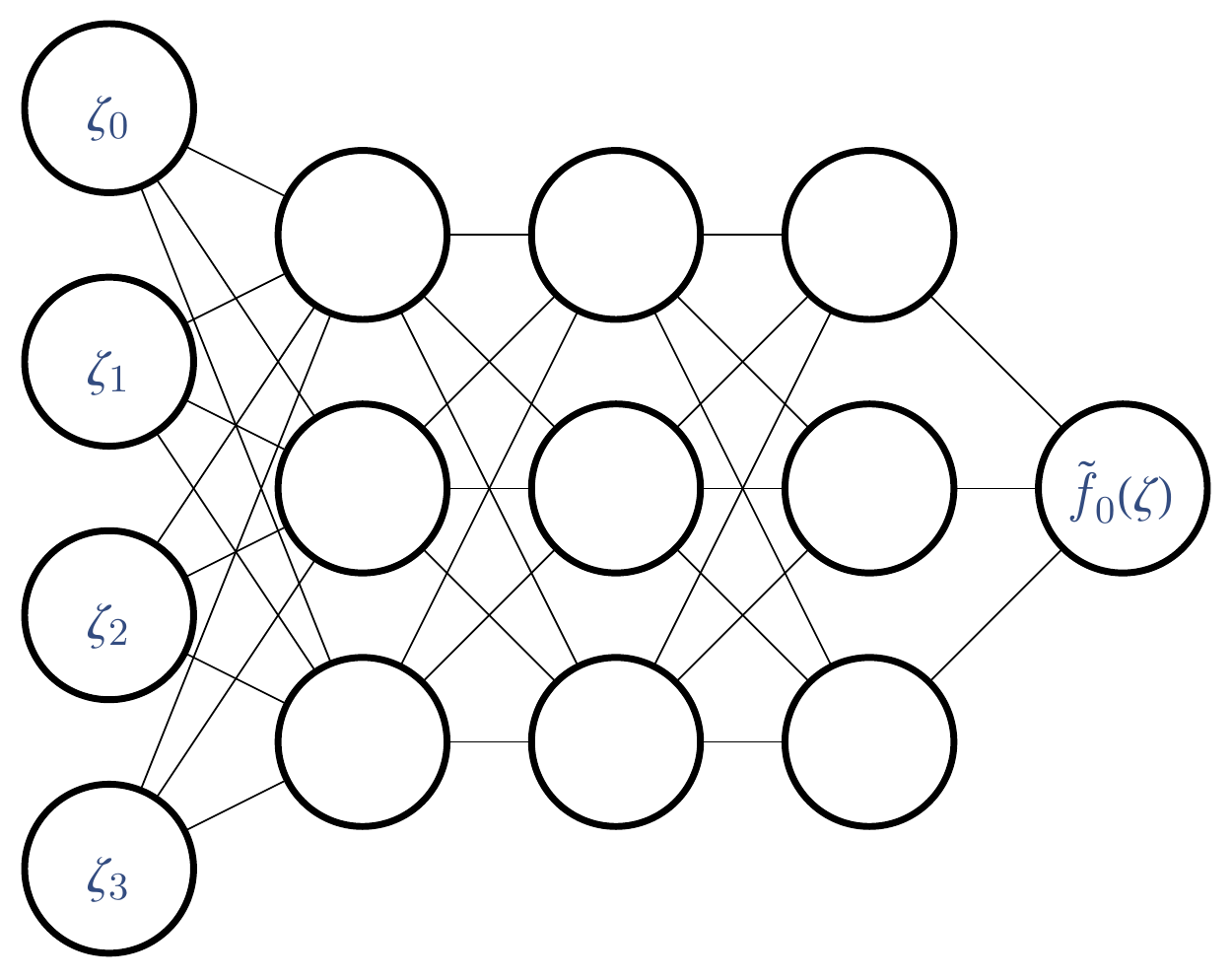}
  \caption{Topology of a feed-forward network with~5 layers: one input layer with~4 nodes, three intermediate layers with~3 nodes each and one output layer with one node. The inputs in the incoming layer (shown on the left) are the (real) values of the field. The output is the imaginary value of the coordinate of the first point of the lattice~$\tilde f_0(\zeta)$. 
}\label{fig:net}
\end{figure}

More concretely, suppose we have a ``training set", $\tset$, that is a number of field configurations $\phi^{a}_i$  lying on the manifold $\mathcal{M}_T$, where $i=1,\dots, N$ indexes their components and $a=1, \dots, |\tset|$, with $|\tset|$ being the size of the training set.
We parametrize the interpolating manifold $\mathcal{L}_{\mathcal{S}}$, an approximation to $\mathcal{M}_T$, as
\beq\label{eq:learn_parametrization}
\phi_i = \zeta_i + i \tilde f_i(\zeta),
\eeq where $\zeta_i\in\R^N$  and $\tilde f_i$ is a real function approximating $f$. The function $\tilde f_i$ is represented by a feed-forward network of the type depicted in~\fig{fig:net}. The nodes on the left layer represent the input values, in our case the values of $\zeta_i$.
The results are combined on the second layer by making linear combinations of them, adding a bias and  feeding them to a nonlinear  function $\sigma(x)$ that we will take to be the of the form $\sigma(x) = \log(1+e^x)$.  The result is
\beq
v_j = \sigma\left( b_j + \sum_i w_{ij} \zeta_i \right),
\eeq where $j$ indexes the nodes of the second layer.
These results are then combined again, piped through $\sigma(x)$ and fed to the next layer. At the end all results are combined in a single number which represents $\tilde f_{i=0}(\zeta)$. By translation invariance, the values of $\tilde f_i(\zeta)$ for other $i\neq 0$ can be obtained by translating the inputs $\zeta_i$. The feed-forward network is parametrized by the weights ($w$'s) of every link and biases $b$'s of every node. These parameters are chosen in order to minimize the discrepancy between the training set and the results of the network:
\beq
C(w,b) = \frac{1}{|\tset|} \sum_{a=1}^{|\tset|} \left | \tilde f_{w,b}(\Re\phi^{a}) -  \Im(\phi^a )   \right|,
\eeq 
where $\tilde f_{w,b}(\Re\phi^a)$ is the result of applying the network, with parameters $w_{ij}$ and $b_j$ to $\Re\phi^a$.\footnote{Other cost functions can be used instead of $C(w,b)$ used here.} In order to minimize $C(w,b)$ a gradient descent algorithm is used. The computation of the gradient is efficiently done using the back-propagation algorithm. In fact, the existence of this simple algorithm is an important motivation to use feed-forward networks as opposed to networks with more complicated topologies. The minimization process is sped up tremendously by using the Adaptive Moment Estimate Algorithm (ADAM)\cite{2014arXiv1412.6980K} (other methods are discussed in \cite{2016arXiv160904747R}), another borrow from the artificial intelligence literature. Since the manifold $\mathcal{L_{\mathcal{S}}}$ is defined by a network which learned how to approximate $\mathcal{M}_T$ we call $\mathcal{L_{\mathcal{S}}}$ the ``learnifold".
An example of the practical use of this method will be described in the next section. 

Three comments are worth making at this point. Firstly, while the usefulness of this method can be gauged on a case by case basis, its correctness is guaranteed  by construction. In fact, any network will define a manifold of integration in the same homology class as $\R^N$ since the mapping
\beq
\phi_i(\zeta) = \zeta_i + i s \tilde f_i(\zeta),
\eeq for $0\leq s \leq 1$ define a one-parameter family of manifolds interpolating between $\R^N$ and 
$\mathcal{L_{\mathcal{S}}}$.  Care must be taken on the asymptotic behavior of $\mathcal{L_{\mathcal{S}}}$, determined the function $\tilde f_i(\zeta)$, in order for $\mathcal{L_{\mathcal{S}}}$ to be in the same class as $\mathcal{M}_T$ but, in the case of periodic $\zeta$ as in the applications below, this is automatic.

\begin{figure}[t!]
\centering
\begin{subfigure}[t]{.5\textwidth}
  \centering
  \includegraphics[width=0.8\linewidth]{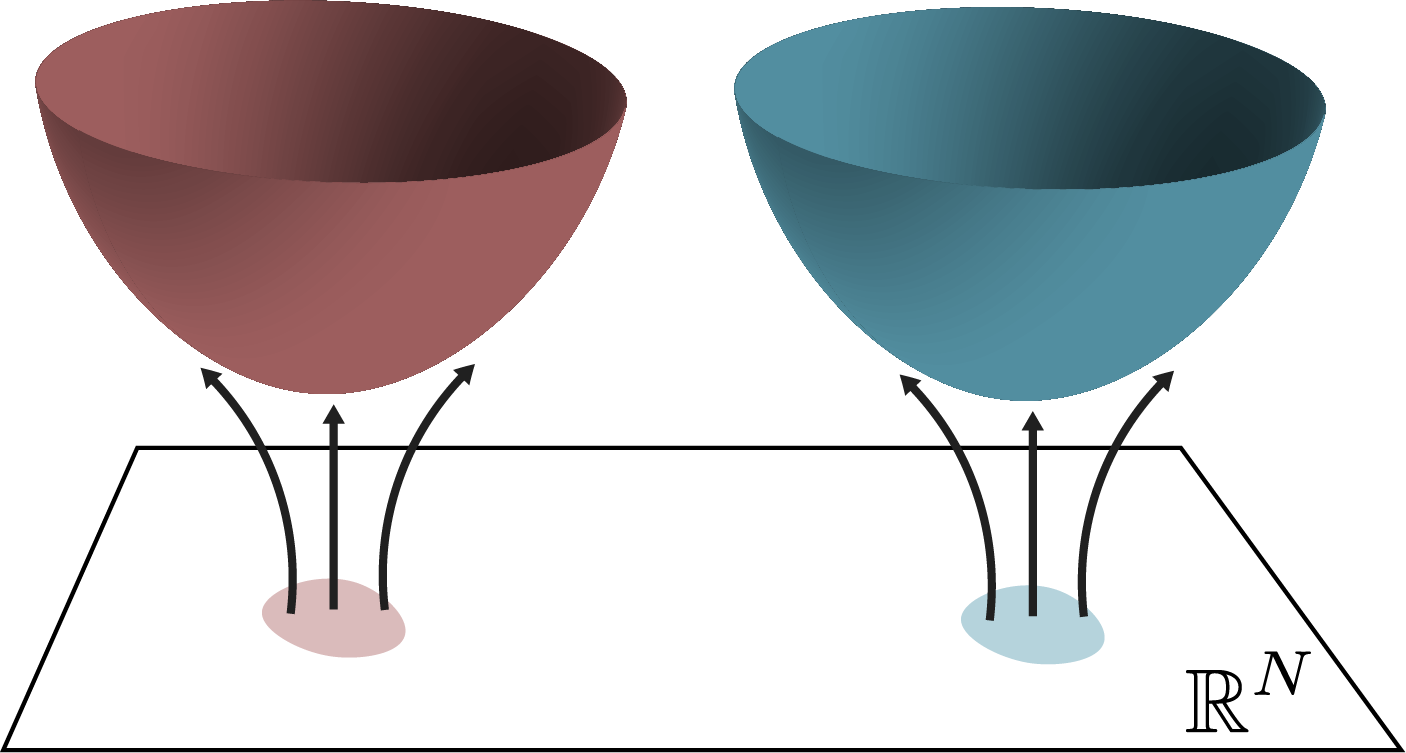}
  \caption{ Parametrization using the holomorphic flow: small regions of $\R^N$ map into large regions of thimbles (or $\mathcal{M}_T$).
  }
  \label{fig:flow-para}
\end{subfigure}%
\vskip 0.8cm
\begin{subfigure}[b]{.5\textwidth}
  \centering
  \includegraphics[width=0.8\linewidth]{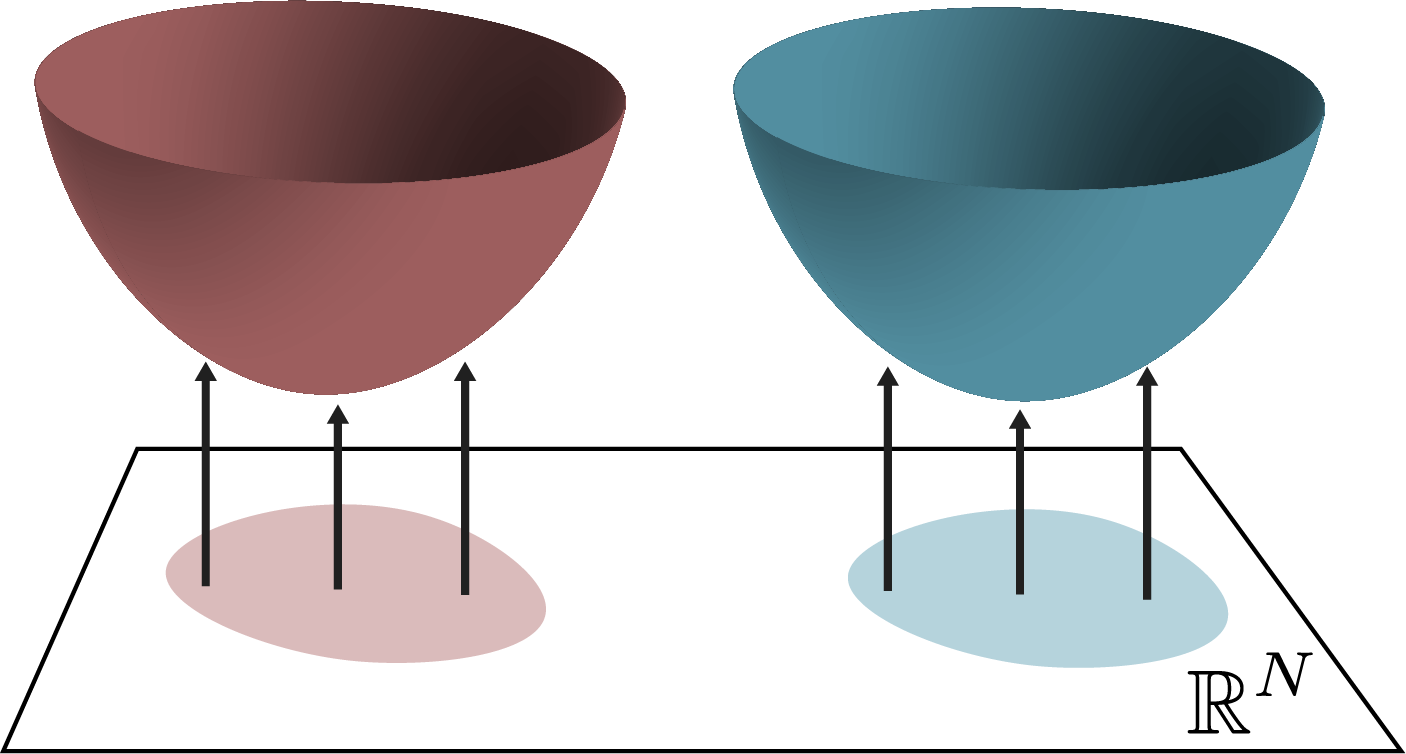}
  \caption{  
 Parametrization of \eq{eq:learn_parametrization}: regions of $\R^N$ mapping into large regions of  
  $\mathcal{L_{\mathcal{S}}}$ are larger and with smaller gaps between them, which helps to prevent trapping of the Monte Carlo's Markov chain. }
  \label{fig:learn-para}
\end{subfigure}
\caption{}
\end{figure}

Secondly, the parametrization \eq{eq:learn_parametrization} helps avoiding ``trapping" of Monte Carlo chains as compared to the parametrization through the flow used in the (generalized) thimble method. This is explained by the~\fig{fig:learn-para} where we can see that the same manifold parametrized by a small region of $\R^N$ (in the generalized thimble method) or a much larger region using \eq{eq:learn_parametrization}. Regions of $\R^N$ with large statistical weights are then less separated by low probability regions, which facilitates the Monte Carlo sampling. Relatedly, the Jacobian of the parametrization~\eq{eq:learn_parametrization} fluctuates less than the Jacobian of the flow parametrization.

Finally, it is unlikely that the learnifold $\mathcal{L_{\mathcal{S}}}$ is better at controlling the sign problem than the manifold obtained by flow (of which the training set is taken). The usefulness of the method relies in the possibility of sampling the learnifold at a cost orders of magnitude cheaper than flowing (and computing/estimating the Jacobian). The hope is that the increase in statistics compensates allowed by the speed of the process compensates for the smaller average sign.

\subsection*{Case study: 1+1D Thirring model revisited}
  
 
The application of the learnifold method to the $1+1$ dimensional Thirring model is discussed in \cite{Alexandru:2017czx}.  The first step is to collect a number of complex configurations $\phi^a=\mathcal{F}_T(\zeta^a)$ obtained by evolving real configurations $\zeta^a$ by a ``time" $T$ according to the flow equations  \eq{eq:flow}  in order to form the training set $\tset$. This  step is identical to what is done in the generalized thimble method.
One would like to approximate the manifold $\mathcal{M}_T$ particularly well in the region that is going to be sampled the most. 
We also want to sample some of the field configurations on ${\cal M}_T$ toward the large field value region to make sure that the 
profile of the learnifold matches $\mathcal{M}_T$ in this region too.
Therefore some configurations are collected running a Metropolis chain with weight $e^{-S}$ and some others with weight $e^{-S/\tau}$, with $\tau>1$. This way  $\mathcal{M}_T$ is sufficiently sampled so the interpolation manifold  $\mathcal{L_{\mathcal{S}}}$ approximates it well in the statistically important regions and it is not radically wrong at asymptotically far away regions.
Notice that there is no particular reason for the Monte Carlo chain to be thermalized while collecting these configurations; al that is required is to have a good enough sampling of the  statistically important regions of  $\mathcal{M}_T$ and some sampling of the other regions. After these configurations are generated they are used as the training set for the feed-forward network. In order to enforce translation invariance, all translations of them are added to the training set, resulting in a larger set, typically of the order of $10^5$ elements. This set is too large to be used in the minimization process so different subsets (``minibatches") are used at different steps of the gradient descent (or ADAM step). 
Details can be found in \cite{Alexandru:2017czx}.   A comparison of the computational cost between the learnifold and the generalized thimble methods is not straightforward because the cost of the learnifold method is divided into a ``fixed" cost related to the generation of the training set and the minimization of the cost function on one hand and the running of the Monte Carlo given the optimal manifold. The second part is faster than the flowing required by the thimble method by orders of the magnitude but the first part may dominate the total costs. Calculations of the kind done in \cite{Alexandru:2016ejd} can be done more effectively using the learnifold method.

Similar methods have recently been applied to the solution of the Hubbard model away from half-filling \cite{Ulybyshev:2020aa}

 \begin{figure} [t] 
\centering
  \includegraphics[width=.95\linewidth]{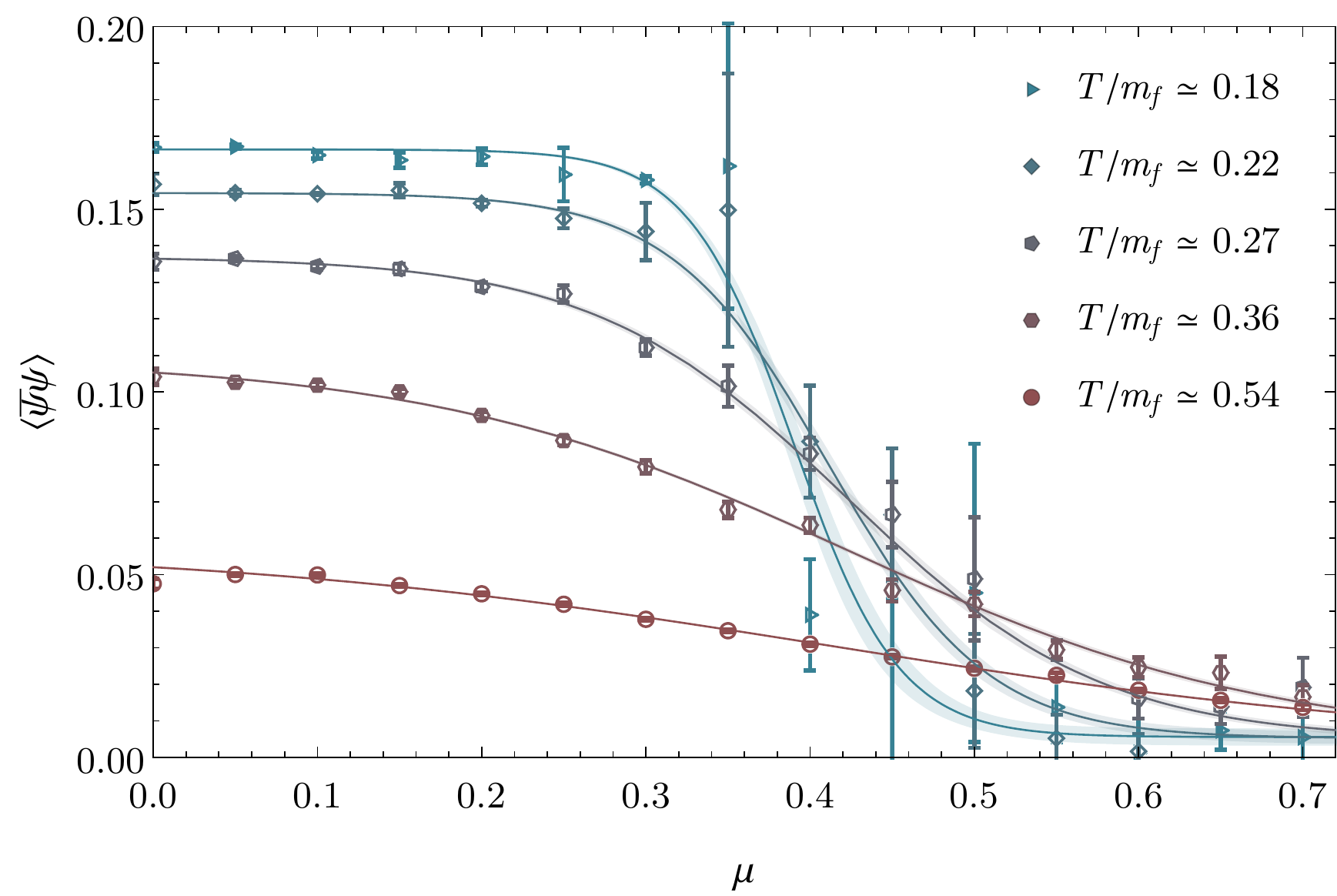}
  \caption{$\langle \bar\psi \psi\rangle$ as a function of $\mu$ for the $2+1$ dimensional Thirring model  in a $\beta\times 6^2$ lattice showing the melting of the chiral condensate as the density is increased. The solid lines are fit to the functional form $\langle \bar\psi \psi \rangle=A\ {\rm tanh}(\beta(\mu-\mu_c))$}
  \label{fig:learnifold-thirring}
\end{figure}

  \subsection{Path optimization }
  
  An even more radical departure from thimbles is embodied in the path optimization method.
  The idea here is to consider a family of manifolds $\mathcal{M}_\lambda$ parametrized by a set of parameters (collectively denoted by $\lambda$) and to maximize
  the average sign within that family. The result of the maximization defines a manifold $\mathcal{M}_{\lambda_0}$ that is the best, within that family, at ameliorating the sign problem. The viability of the method rests on the observation that the gradient of the average sign in parameter space can be calculated with a (usually very short) {\it sign-problem free} Monte Carlo calculation. Indeed, the average sign on a manifold $\mathcal{M}$ is given by
\beq\label{eq:<sign>}
  \langle \sigma \rangle_\lambda =
  \frac{
  \int_\mathcal{M_\lambda} d\phi \  e^{-S(\phi)}  }
  {   \int_\mathcal{M_\lambda} |d\phi|  \  e^{-\Re S(\phi)}   }
  =
  \frac{
  \int_{\R^N} d\zeta\  e^{-S_\text{eff}(\zeta)}  }
  {   \int_{\R^N} d\zeta  \  e^{-\Re S_\text{eff}(\zeta)}   },
\eeq 
where $S_\text{eff}(\zeta) = S[\phi(\zeta)] - \ln \det J(\zeta)$ includes the determinant of the Jacobian of the $\mathcal{M}_\lambda$ parametrization  $\phi_i = \phi_i(\zeta)$. The numerator of~\eq{eq:<sign>} is independent of $\lambda$, due to Cauchy's theorem. The denominator, however, being an integral of a non-holomorphic function,  does depend on $\lambda$. In fact,
\beq\label{eq:grad<sign>}
\frac{\nabla_\lambda   \av{\sigma} }{ \av{\sigma} }
=
\frac{   {   \int_{\R^N} d\zeta  \  e^{- S_\text{eff}(\zeta)} 
\left(-\nabla_\lambda \Re S + \Re \Tr (J^{-1} \nabla_\lambda J)  \right)  }  }
  {   \int_{\R^N} d\zeta  \  e^{-\Re S_\text{eff}(\zeta)}   }.
\eeq 
The average sign does not affect the direction of the vector $\nabla_\lambda \av{ \sigma } $ and can be neglected during the maximization process while the right hand side term in~\eq{eq:grad<sign>}, the average  $\av{   -\nabla_\lambda \Re S + \Re \Tr (J^{-1} \nabla_\lambda J)  }_{\Re S_\text{eff}}$,  can be computed by the Monte Carlo method {\it without} encountering a sign problem. Knowledge of the gradient $\nabla_\lambda \av{ \sigma } $ (to be more precise, knowledge of its direction in $\lambda$-space) allows a maximization routine, like ADAM \cite{2014arXiv1412.6980K} to find the values of $\lambda$ leading to the largest possible sign within the parametrized family of manifolds. 
  
  A few facts make this scheme practical. First, a rough computation of the gradient is usually enough; some stochastic noise is actually useful in avoiding local minima that could otherwise trap the maximization process. Second, the last configurations obtained with one value of $\lambda$ is close to being thermalized as $\lambda$ is changed to a nearby value during the maximization process, bypassing the need for long thermalization periods at each step of the minimization. Finally, it is imperative that the family of manifolds considered i) includes only manifolds in the same homology class as $\R^N$, ii) contain manifolds where the sign problem is sufficiently ameliorated and iii) are parametrized in such a way the computation of the Jacobian $J$ is cheap.  Condition i) is relatively easy to satisfy but there is tension between conditions ii) and iii). Theoretical insight into the specific model of interest is required for the successful application of this method and it is currently sorely missed in most theories of physical significance.
  
We pause to note that contour deformations can be applied to any theory with a holomorphic path integrand. This includes theories with real actions but complex observables. This fact, combined with path optimization, has been used to tame the \emph{signal to noise} problem encountered in the calculation of correlation function in simple field theories \cite{detmold2020path}. 
   
\subsection*{Case study: 2+1D Thirring model}
  
 \begin{figure} 
\centering
  \includegraphics[width=.8\linewidth]{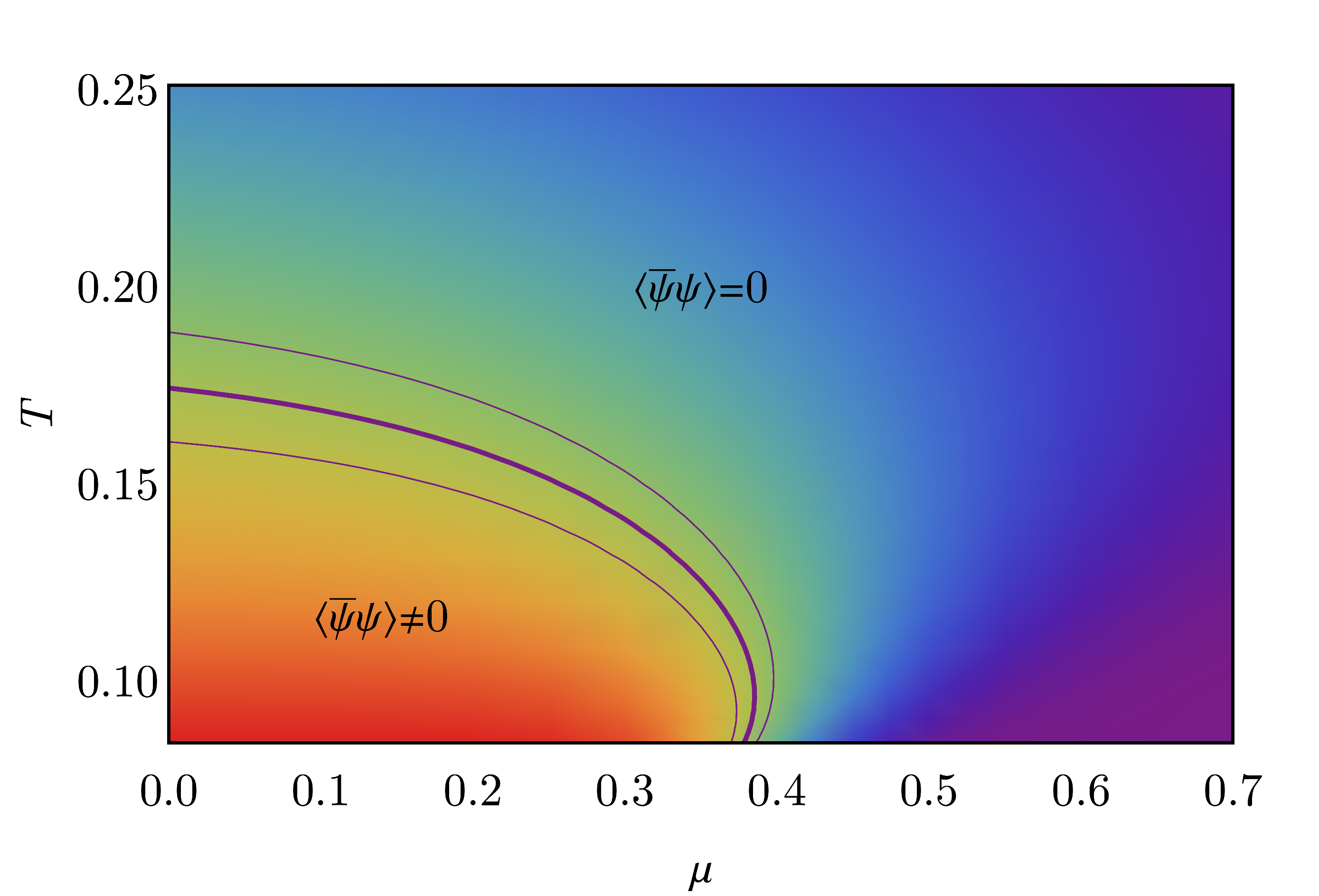}
  \caption{Phase diagram of the 3D Thirring model \cite{Alexandru:2018ddf}. The thick central line shows the location where $\langle \bar\psi\psi\rangle_{\mu,T} = 0.5 \langle \bar\psi\psi\rangle_{0,0}$ and its width the statistical errors. The thinner lines indicate  $\langle \bar\psi\psi\rangle_{\mu,T} = (0.5\pm 0.05) \langle \bar\psi\psi\rangle_{0,0}$  to help gauge the sharpness of the transition. }
   \label{fig:3D-thirring}
\end{figure}
  
  The path optimization method was applied to a one-dimensional integral in \cite{Mori:2017pne}, to the 1+1D Thirring model at finite density  in \cite{Alexandru:2018fqp}, to the 1+1D $\phi^4$ model in \cite{Mori:2017nwj} (with a neural network parametrization of the manifolds similar to the one discussed in section \ref{sec:learnifold} ), the PNJL model in 0+1D  \cite{Kashiwa:2018vxr,Kashiwa:2019lkv}, 0+1D QCD \cite{Mori:2019tux} and 1+1D $\phi^4$ \cite{Bursa:2018ykf}.  Here we discuss its application to the 3D Thirring model \cite{Alexandru:2018ddf}.
  
The action defining the 2+1D Thirring model is in \eq{eq:thirring-continuum}. The family of manifolds considered is given by
\beqs
  A_0(x) &= \zeta_0(x) + i(\lambda_0 + \lambda_1 \cos \zeta_0(x) + \lambda_2 \cos(2\zeta_0(x)))\,, \\
  A_1(x) &= \zeta_1(x)\,, \\
  A_2(x) &= \zeta_2(x)\,, 
\label{eq:thirring-3D-ansatz}
\eeqs 
where $\lambda_0, \lambda_1$ and $\lambda_2$ are real numbers parametrizing the manifolds. The ansatz in \eq{eq:thirring-3D-ansatz} is motivated by the following considerations. Firstly, the determinant of the Jacobian $J=(\partial A_\nu/\partial \zeta_i)$ is trivial to compute (the cost scales with the spacetime volume $V$, as opposed to $V^3$) since the value of $A_\mu(x)$ depends only on $\zeta(x)$ evaluated at the same spacetime point $x$:
\beq
\det J=\prod_x  \left(1-\lambda_1 \sin\zeta_0(x) -2 \lambda_2 \sin(2\zeta_0(x))  \right).
\eeq 
Secondly, in the limit $\mu\rightarrow\infty$ the partition function
\beq
\lim_{\mu\to\infty}  Z \approx \left[ \int d^3A\   e^{\frac{1}{g^2}(\sum_\nu  \cos A_\nu) + \frac{i}{2} A_0}\right]^{\beta V},
\eeq 
factorizes into a separate integral at every spacetime point and the sign problem arises entirely from $A_0$. The ansatz in~\eq{eq:thirring-3D-ansatz} reflects that. Thirdly, in the weak coupling limit~$g^2\rightarrow 0$ we expect the functional integral to be dominated by the saddle point with the smallest action that, as discussed in~\sect{sect:1Dthirring} has the form~$A_0 = i \alpha, A_1=A_2=0$, for some real constant~$\alpha$. The ansatz in~\eq{eq:thirring-3D-ansatz} contains manifolds that approach this thimble near its critical point. Finally, the variables~$A_\nu$ are periodic variables with a period~$2\pi$ (so they belong to $(S^1)^N$, not $\R^N$) and the question of whether the manifolds defined by~\eq{eq:thirring-3D-ansatz} have a different asymptotic behavior is not present. Furthermore, by varying $s$ from $s=1$ to $s=0$ in $A_0(x) = \zeta_0(x) + is(\lambda_0 + \lambda_1 \cos \zeta_0(x) + \lambda_2 \cos(2\zeta_0(x)) $ we see that every member of the family of manifolds can be smoothly deformed to $(S^1)^N$, guaranteeing the applicability of the Cauchy theorem. 
  
This method was used in \cite{Alexandru:2018ddf} in lattices of sizes up to~$10^3$ and action parameters near the continuum. It is interesting to examine how the maximization  process proceeds.  The parameter $\lambda_0$ acquires very quickly a non-zero value very close to the position of the critical point~$A_0=i \alpha, A_1=A_2=0$. The corresponding manifold does not go through exactly through the critical point and has a larger average sign than the space tangent to that critical point. 
Afterwards, $\lambda_1$ and $\lambda_2$ settle on their preferred values, giving a little curvature to the manifold. 
 More complicated functions of $\zeta_0$ in \eq{eq:thirring-3D-ansatz} do not seem to improve the average sign. It seems that one is required to go beyond the factorized form in \eq{eq:thirring-3D-ansatz} for further improvements.

The results obtained by this method show a clear transition (technically a crossover as the fermion mass breaks chiral symmetry explicitly) between a phase with large chiral condensate to another, at higher temperatures and densities, where chiral symmetry is restored and the chiral condensate is small (see~\fig{fig:learnifold-thirring}). The resulting phase diagram is shown in \fig{fig:3D-thirring}.

%% file: conclusions.tex
A bird's eye view of the developments described here reveal some broad lessons that should not be lost amidst the technical details. The first is that the main idea the thimble approach to the sign problem is based on is sound and no fundamental flaw has been revealed, either conceptual or practical. This is not to say one can currently use the method to solve any sign problem. But the set of ideas explored in this review provide a new setting where the simulation of many models can and should be attempted.
This is not a trivial statement. There is a common perception among non-practitioners that there is a ``conservation of difficulty" and that any approach to solve the sign problem will reveal, at closer inspection, a simulation cost as large as the naive attempts. This is demonstrably untrue, as the examples discussed in this review show. 

The second foundational lesson is that the integration over all the relevant thimbles is both necessary and possible, with several algorithms already proposed and tested, usually in small scale simulations. In fact, the rapid algorithmic development in the last few years generated a problem -- or an opportunity -- as most calculations were aimed at demonstrating the algorithm correctness and scaling properties and not focused on the Physics of the problem. For instance, current technology should be able to clarify the phase diagram of a variety of $1+1$ dimensional models at finite temperature/density. At a larger computational cost, $1+2$ dimensional models can also be studied now. In fact,  recent papers began setting up the path towards a solution to the repulsive Hubbard model away from half-filling,  a result that would be a game-changer in the field \cite{PhysRevD.100.114510,10.2307/2414761,Mukherjee:2014hsa,Ulybyshev:2019fte,PhysRevB.40.506,doi:10.7566/JPSJ.86.093001,Ulybyshev:2019hfm,Fukuma:2020lny,ostmeyer2020semimetalmott}

An important insight arising from the research on thimble-related methods was that deformation of the integration manifold to other manifolds is both possible and profitable. This observation, as simple as it is, has vast consequences. Indeed, the condition that the imaginary part of the (effective) action to be constant is only one constraint in a $2N$ dimensional space. This leaves a $2N-N-1=N-1$  parameters family of possible direction of the tangent space of the integration manifold to choose from while still solving the sign problem. This freedom is not explored by holomorphic flow methods to deform contours of integration \footnote{Notice that contrary to the multidimensional case, in the familiar case of a single  complex variable, the condition $\Im S_{\text{eff}}=0$ defines a unique contour.}. A few ideas exist on how to explore this newfound freedom. One is to use information about the model obtained elsewhere to devise parametrized families of integration manifolds suitable for that particular model. This approach provides a way of bringing physical insight into a the Monte Carlo calculation that is sometimes characterized as a brute force method. Whatever insight is brought to the model, obtained by rigorous or intuitive, approximate methods, can then be used to speed up a calculation, hopefully exponentially,  that is guaranteed to converge to the correct answer by the Monte Carlo method. A surprising recent development is that the physical insight into a model can be substituted by systematic machine learning techniques. We expect the near future to bring much more developments in this direction.

%% file: acknowledgement.tex
The authors would like to thank Hank Lamm, Scott Lawrence, Greg Ridgway and Sohan Vartak, who collaborated with us on this topic. We also thank Tom Cohen, Gerald Dunne,  Tarun Grover, Sergei Gukov, Jay Sau, Christian Schmidt, Luigi Scorzato, and Mithat \"Unsal for conversations on this topic over the last few years. This work was supported in part by the US DoE under contracts No. DE-FG02-93ER-40762, DE-FG02-95ER40907. AA gratefully acknowledges the hospitality of the Physics Department at the University of Maryland where part of this work was carried out.

%% file: appendix-jacobian.tex
\appendix
\section{Computation of the Jacobian}\label{app:jacobian}

The evolution by the holomorphic flow  \eq{eq:flow}  by time $T$ maps initial conditions $\zeta$ into $\Phi(\zeta)$. The Jacobian of this transformation is derived in this appendix.

\begin{figure}[t!]
  \includegraphics[scale=0.8]{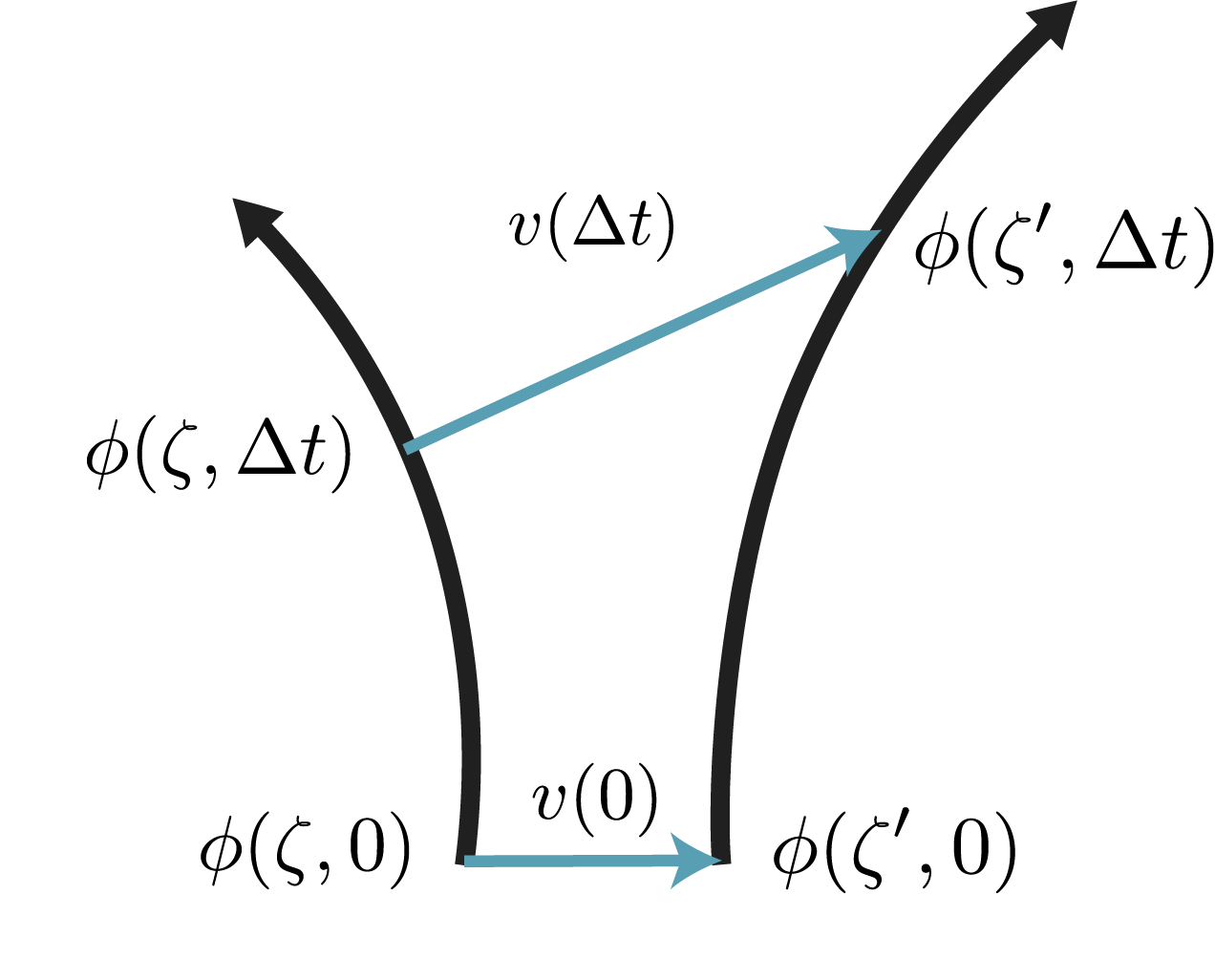}
  \caption{Two nearby points evolving by the holomorphic flow. Their difference vector, shown in blue, evolves according to \eq{eq:vector-flow}.}
  \label{fig:vector-flow}
  \end{figure}

Begin by considering two infinitesimally close coordinates $\zeta$ and $\zeta'$, and let $v$ 
   \beq
   v = \phi(\zeta',0)-\phi(\zeta,0)
   \eeq
   denote the corresponding difference vector between them. Flowing for time $\Delta t$, both $\phi(\zeta',0)$ and $\phi(\zeta,0)$ move, changing the difference vector; we denote this time dependent difference as $v(\Delta t)$ (see  \fig{fig:vector-flow}). In the limit $\Delta t\rightarrow 0$:
  \begin{align}
  v_a(\Delta t) & \equiv  \phi_a(\zeta',\Delta t) -  \phi_a(\zeta,\Delta t)
  \nn\\
  & = \Big[\phi_a(\zeta',0)+\Delta t \overline{\frac{\partial S}{\partial \phi_a}(\phi_a(\zeta',0))}\Big] 
 \nn \\
&  -  \Big[\phi_a(\zeta,0)+\Delta t \overline{\frac{\partial S}{\partial \phi_a}(\phi_a(\zeta,0))}\Big] 
\nn\\
  & = \big[\phi_a(\zeta',0) - \phi_a(\zeta,0)\big] 
 \nn \\
  &+ \Delta t \overline{\frac{\partial^2 S}{\partial \phi_a \partial \phi_b}(\phi_a(\zeta,0))\big[\phi_b(\zeta',0) - \phi_b(\zeta,0)\big]} 
  \nn\\
  & = v_a(0) + \Delta t \overline{H_{ab}(\phi_a(\zeta,0)) v_b(0)}.
  \end{align}
  In other words, a vector evolves along a flow trajectory according to the differential equation
  \beq\label{eq:vector-flow}
  \frac{d v_a}{d t} = \overline{H_{ab}(\phi(\zeta,t)) v_b(t) }~.
  \eeq We can use the equation above to evolve a set of $N$ vectors forming an orthonormal basis. Packaging these vector in the columns of a matrix $J(0)=\openone$,
  we see that $J(t)$ obeys
   \beq
  \frac{d J}{dt} = \overline{H J}.
  \eeq 
    
%

%% file: appendix-thimbles.tex
\section{Another definition for thimbles}\label{app:thimble-alt}

In this appendix we give a different perspective on Lefschetz thimbles. We begin with focusing on the stationary points of the flow, namely the critical points of the action $\phi^c$, such that $\partial S/\partial\phi_i |_{\phi^c }=0$. Around a critical point\footnote{In our analysis we consider only isolated, quadratic (non-degenerate) critical points. A degenerate critical point where the Hessian determinant of $\phi$ vanishes can be split into $\mu$ number of non-degenerate critical points with a small deformation with $\mu$ being the Milnor number of the critical point and a similar analysis presented in this section follows \cite{pham}.} it is always possible to find local coordinates $\{z_i=x_i+i y_i \}$ with $i=1,\dots ,N$ such that
\bea\label{eq:morse}
S(\phi)-S(\phi^c)&=& z_1^2+\dots +z_N^2
\nn
\\
&= &(x_1^2+\dots +x_N^2)- (y_1^2+\dots +y_N^2)
\nn\\ 
&&+ 2i (x_1y_1+\dots+x_Ny_N)
\eea 
whose existence is guaranteed by the Morse lemma. 
Now consider the $N-1$ dimensional surface $v(s)$ defined by $x_1^2+\dots+x_N^2=s,y_1=\dots=y_N=0$. This surface is known as the \textit{vanishing cycle} as it vanishes at the critical point. It can be viewed as the level set of the action around the critical point, $S^{-1}(s+s_c)$ where $s_c=S(\phi^c)$. We now ``move" the vanishing cycle by varying $s$. This can be done by taking the vanishing cycle around the critical point $v(\epsilon)$ and then flowing it. As $s$ runs from $0$ to $\infty$, the vanishing cycle sweeps an $N$ (real) dimensional surface.  This $N$ dimensional surface,  defined as the union of vanishing cycles on the half line  $0 \leq s < \infty$, ${\cal T}=\cup_s v(s)$, is known as the \textit{Lefschetz thimble} associated with the critical point $\phi^c$. 
Similarly, we define an $N-1$ dimensional ``dual'' cycle, $v^D(s)$ by $x_1=\dots=x_N=0, y_1^2+\dots+y_N^2=s$.  We shall call the union of these dual cycles on the half line $0 \leq s < \infty$, ${\cal K}=\cup_s v^D(s)$, the \textit{dual thimble}\footnote{Note that what we call the thimble and the dual thimble are referred as downward/upward cycles referring to the fact that the weight, $e^{-\Re S}$, monotonically decreases/increases over them with flow\cite{Witten:2010cx}.}. 
